\documentclass[11pt,a4paper]{article}

\usepackage[T1]{fontenc}
\usepackage[utf8]{inputenc}
\usepackage[condensed]{roboto}

\usepackage[a4paper,margin=2.5cm]{geometry}
\usepackage{xcolor}
\usepackage{setspace}
\usepackage{fancyhdr}
\usepackage{titlesec}
\usepackage{graphicx}
\usepackage{caption}
\usepackage{booktabs}
\usepackage{ragged2e}
\usepackage{booktabs}     
\usepackage{graphicx}     
\usepackage{amssymb}      
\usepackage{array}        
\usepackage{multirow}     
\usepackage{caption}      
\usepackage[numbers]{natbib}
\usepackage{tabularx}
\usepackage{hyperref}
\usepackage{eso-pic}

\newcommand{\kapaksayfasi}[1]{%
	\clearpage
	\thispagestyle{empty}
	\AddToShipoutPictureBG*{%
		\AtPageLowerLeft{%
			\includegraphics[
			width=\paperwidth,
			height=\paperheight
			]{#1}%
		}%
	}%
	\null
	\clearpage
}

\providecommand{\bibcommenthead}[1]{}
\definecolor{maintext}{RGB}{0,0,60}
\color{maintext}

\justifying

\titleformat{\section}
{\bfseries\fontsize{16pt}{18pt}\selectfont\color{maintext}}
{\thesection.}{0.5em}{}

\titleformat{\subsection}
{\bfseries\fontsize{14pt}{16pt}\selectfont\color{maintext}}
{\thesubsection.}{0.5em}{}

\titleformat{\subsubsection}
{\bfseries\itshape\fontsize{11pt}{12pt}\selectfont\color{maintext}}
{\thesubsubsection.}{0.5em}{}

\begin{document}
\pagenumbering{gobble}
\kapaksayfasi{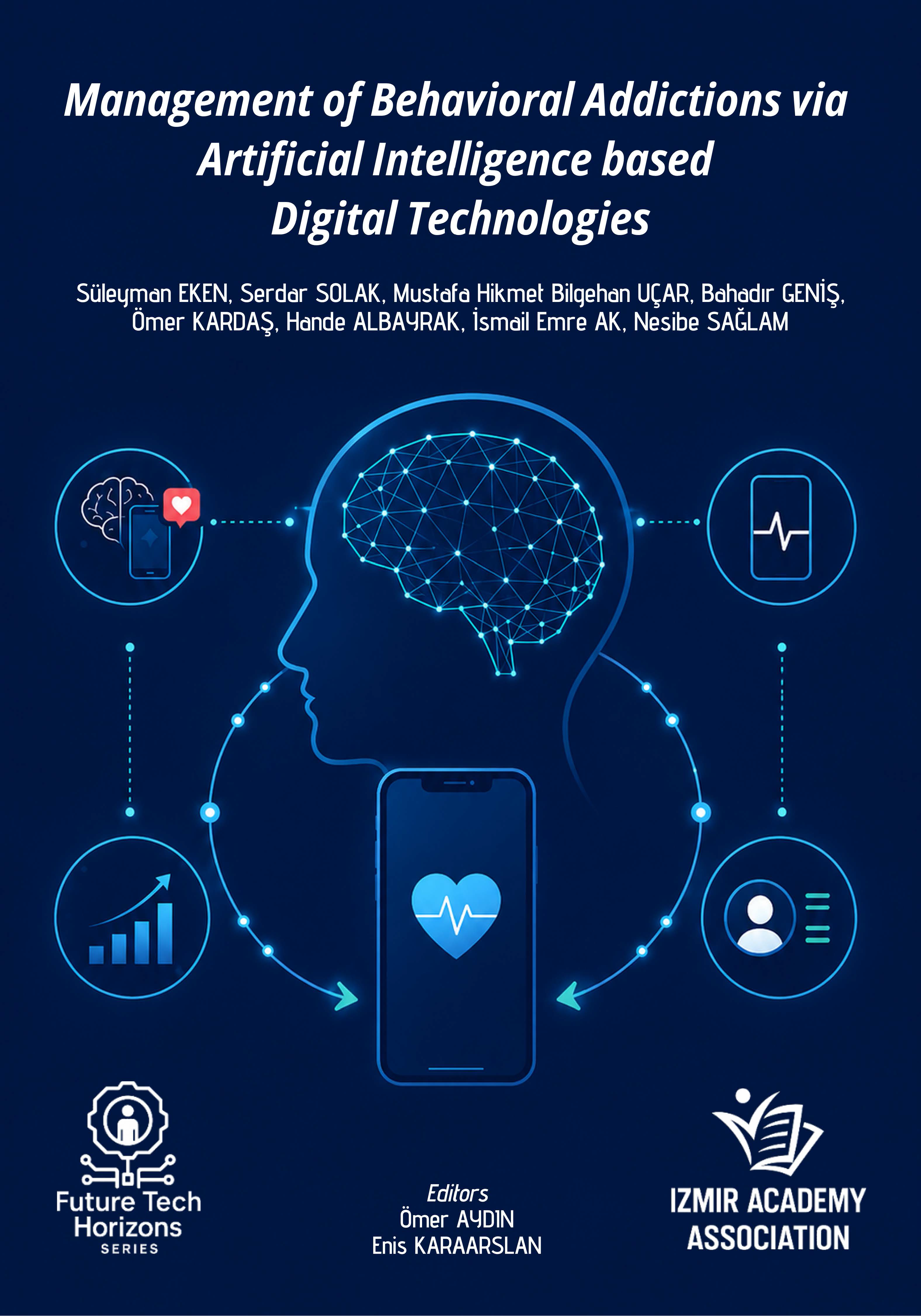}

\section{Introduction}\label{intro}
Addiction is widely recognized as a chronic and relapsing disorder involving persistent engagement in rewarding behaviors despite harmful consequences. It is typically associated with impaired behavioral control and dysregulation of neural reward and control systems \cite{koob2016neurobiology,volkow2016neurobiologic}. Behavioral addictions, also known as process addictions, involve persistent engagement in rewarding behaviors despite negative consequences and diminished behavioral control \cite{grant2010introduction,brand2019interaction}. 

Epidemiological evidence indicates rising prevalence rates for gambling disorder, internet gaming disorder, problematic smartphone use, and online behavioral addictions, particularly among adolescents and young adults \cite{j2014internet,panova2021internet}. These disorders impose substantial societal burdens through academic decline, productivity loss, psychiatric comorbidity, and healthcare utilization. Traditional treatment approaches, including cognitive behavioral therapy (CBT), motivational interviewing, and relapse-prevention protocols, demonstrate efficacy but remain constrained by limited accessibility and engagement challenges \cite{cowlishaw2012psychological}. 

Behavioral addictions are inherently context-dependent and temporally dynamic, with craving intensity, cue exposure, and self-regulation capacity fluctuating across everyday environments \cite{nahum2025just}. Consequently, episodic clinic-based interventions are often insufficient for sustained relapse prevention \cite{mestre2023features, volkow2023substance}. In parallel, the same digital environments that facilitate addictive behaviors also generate fine-grained behavioral traces, enabling continuous monitoring and real-time assessment. Leveraging these data streams, recent advances in digital phenotyping, mobile sensing, and artificial intelligence have enabled a shift toward precision mental health and adaptive, context-aware intervention strategies \cite{mohr2013continuous, onnela2016harnessing,insel2017digital,  jacobson2022digital,baumel2025more}. 

These developments support AI-driven DTx and just-in-time adaptive interventions (JITAIs), which enable personalized, real-time intervention delivery. However, despite rapid technological advances, existing literature remains fragmented across modality-specific reviews, AI-centric prediction studies, and disorder-specific surveys, highlighting the need for a unified and comprehensive framework for digital management of behavioral addictions.

\subsection{Definition and taxonomy of addictions}
Building on the general addiction model, behavioral addictions are defined as non-substance-related behaviors characterized by persistent engagement despite adverse consequences and diminished behavioral control \cite{grant2010introduction,brand2019interaction}. These conditions involve repetitive, reinforcement-driven behaviors that provide short-term reward or relief but lead to long-term functional impairment. They share core features with substance use disorders, including compulsive engagement, escalation of behavior, withdrawal-like symptoms, cue-reactivity, and elevated relapse risk \cite{grant2010introduction,brand2019interaction}.

Despite the absence of exogenous substances, similar neurocognitive and behavioral mechanisms are implicated, particularly those related to RL and impaired inhibitory control \cite{koob2016neurobiology,volkow2016neurobiologic}. Theoretical frameworks such as the Interaction of Person–Affect–Cognition–Execution (I-PACE) model conceptualize behavioral addictions as dynamic and interactional processes \cite{brand2019interaction}. These processes emerge from the interplay between individual predispositions, affective states, cognitive biases, and executive control mechanisms \cite{brand2019interaction}. Within this perspective, addictive behaviors are understood as failures of self-regulation maintained through conditioned cue-response associations and maladaptive learning processes \cite{brand2019interaction}. Importantly, behavioral addictions are increasingly conceptualized as dimensional phenomena characterized by fluctuating vulnerability states rather than fixed diagnostic categories.

\begin{figure*}[h!]
	\centering
	\includegraphics[width=0.95\textwidth]{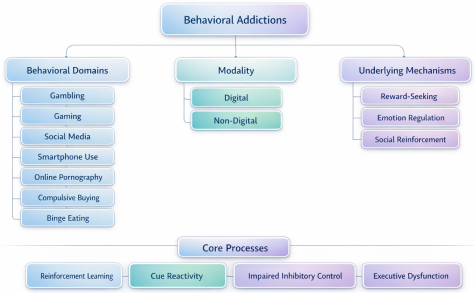}
	\caption{Multidimensional taxonomy of behavioral addictions.}
	\label{fig:taxonomy_behavioral_addictions}
\end{figure*}

From a taxonomical perspective, behavioral addictions can be categorized along multiple dimensions \cite{brand2019interaction}, including (i) the primary behavioral domain (e.g., gambling, gaming, compulsive buying), (ii) modality (digital versus non-digital), and (iii) underlying reinforcement mechanisms (e.g., reward-seeking, emotion regulation, social reinforcement). Gambling disorder is formally recognized as a behavioral addiction in DSM-5-TR and ICD-11, while gaming disorder is included in ICD-11 and listed as a condition for further study in DSM-5-TR \cite{american2013diagnostic, bandyopadhyay2014american,world2024clinical}. In addition, problematic internet use (PIU), excessive social media engagement, and smartphone overuse are widely studied forms of digital behavioral addictions \cite{panova2021internet,montag2023digital, tomczyk2020problematic, moreno2026social}. Other forms, including compulsive online pornography consumption and binge-eating behaviors, are also discussed within the broader behavioral addiction spectrum \cite{brand2019interaction,grant2010introduction}. 
A schematic overview of this multidimensional taxonomy is presented in Fig.~\ref{fig:taxonomy_behavioral_addictions}, providing a structured representation of domains, modality, and underlying mechanisms, along with shared core processes.

\begin{figure*}[h!]
	\centering
	\includegraphics[width=0.95\textwidth]{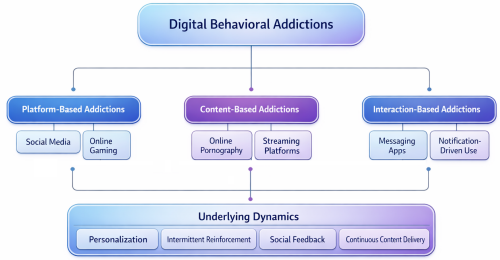}
	\caption{Taxonomy of digital behavioral addictions based on platform type, content characteristics, and interaction patterns.}
	\label{fig:digital_taxonomy}
\end{figure*}

Digital behavioral addictions constitute a rapidly expanding subgroup, driven by ubiquitous connectivity, algorithmically optimized engagement systems, and continuous access to personalized content \cite{montag2023digital}. These digital environments amplify reinforcement cycles through mechanisms such as variable reward schedules, social feedback loops, and real-time content personalization \cite{montag2023digital,brand2019interaction}. This multidimensional taxonomy provides a conceptual foundation for understanding behavioral addictions and supports the development of computational models, digital phenotyping approaches, and adaptive intervention systems discussed in subsequent sections. A detailed taxonomy of digital behavioral addictions is presented in Fig.~\ref{fig:digital_taxonomy}. The figure provides a structured classification across platform-based, content-based, and interaction-based categories, and delineates the underlying dynamics that drive and sustain these behaviors.

\subsection{Prevalence and societal impact}

Behavioral addictions have shown a marked increase in prevalence over the past decade, particularly in the context of rapid digitalization and widespread access to internet-enabled devices \cite{panova2021internet,montag2023digital,stevens2021global,dadi2024problematic}. Epidemiological studies suggest that PIU affects approximately 6--7\% of the global population, with higher prevalence observed among adolescents and young adults \cite{panova2021internet, tomczyk2020problematic, pan2020systematic, gentile2017internet}. Internet gaming disorder (IGD) has been reported with prevalence rates ranging from 1\% to 10\% depending on population characteristics and diagnostic criteria \cite{singh2021prevalence, stevens2021global, ratan2022prevalence, kim2022prevalence, ko2024prevalence, zhou2024meta, gisbert2025prevalence, chew2025prevalence,fraiwan2025internet}. Excessive social media use and smartphone overuse have also emerged as significant concerns, with elevated prevalence reported particularly in younger populations \cite{olson2011diffusion, van2015modeling, tateno2019internet, alotaibi2022smartphone, montag2023digital, moreno2026social}.

Beyond prevalence, behavioral addictions impose substantial individual and societal burdens. At the individual level, these conditions are associated with impaired academic and occupational functioning, reduced quality of life, and increased risk of comorbid psychiatric disorders, including depression, anxiety, and attention-related conditions \cite{kuss2016internet,paulus2018internet,restrepo2020problematic, ratan2022prevalence}. Their chronic and relapsing nature further contributes to long-term functional impairment and diminished psychosocial well-being \cite{chakrabarti2014s}.

At the societal level, behavioral addictions contribute to productivity loss, increased healthcare utilization, and broader economic costs. Emerging evidence also suggests that excessive digital engagement may negatively impact social relationships through mechanisms such as social withdrawal, reduced face-to-face interaction, and altered communication patterns \cite{christensen2018social, alavi2021effects, montag2023digital}. Moreover, the algorithmic design of digital platforms—optimized for engagement and retention—can amplify addictive usage patterns at scale, raising concerns regarding ethical design, user autonomy, and public health implications \cite{yao2025governing, de2025social, li2026algorithm}.

Notably, vulnerability to behavioral addictions is not uniformly distributed. It is shaped by demographic factors (e.g., age, gender), psychological traits (e.g., impulsivity, sensation-seeking), and environmental conditions (e.g., accessibility, social context) \cite{schlosser1994compulsive, durkee2012prevalence, koc2013facebook, ainin2017exploring}. Adolescents and young adults are consistently identified as high-risk groups due to developmental sensitivity and intensive exposure to digital environments \cite{panova2021internet, sussman2022substance,montag2023digital, imrani2026interventions}.

Taken together, the increasing prevalence and multifaceted societal impact of behavioral addictions highlight the need for scalable, adaptive, and technology-driven approaches to prevention, monitoring, and intervention \cite{insel2017digital, montag2020discussing, mestre2023features, dadi2024problematic, mohr2013continuous}. These trends further motivate the integration of digital technologies and artificial intelligence into addiction management frameworks, as discussed in subsequent sections.

\subsection{Motivation for digital approaches in prevention, treatment, and recovery}

The increasing prevalence and dynamic nature of behavioral addictions necessitate innovative approaches that extend beyond traditional, clinic-based treatment models. Conventional interventions, including CBT, motivational interviewing, and relapse-prevention strategies, have demonstrated effectiveness; however, they remain constrained by limited accessibility, high cost, social stigma, and the inability to provide continuous monitoring and real-time support \cite{cowlishaw2012psychological,oesterle2020substance,edinoff2022addiction}.

Behavioral addictions exhibit substantial variability across real-world contexts, with fluctuations in craving, cue exposure, emotional states, and self-regulation capacity. This complexity reduces the effectiveness of episodic, clinic-based interventions and underscores the need for continuous, context-aware, and adaptive support for relapse prevention \cite{brand2019interaction,nahum2025just}. Digital technologies offer a transformative opportunity to address these limitations by enabling continuous, context-aware, and scalable intervention strategies. The widespread adoption of smartphones, wearable devices, and internet-connected platforms has facilitated the emergence of digital phenotyping approaches, which enable passive and real-time monitoring of behavioral, physiological, and contextual data \cite{onnela2016harnessing,montag2023digital,torous2021growing}. These data streams provide unprecedented insights into user behavior and risk states, forming the basis for personalized and adaptive intervention systems.

In parallel, DTx have evolved from static content delivery systems into intelligent, data-driven platforms capable of tailoring interventions to individual needs. Artificial intelligence (AI) and machine learning techniques enable predictive modeling, user profiling, and dynamic decision-making, thereby supporting precision mental health frameworks \cite{jacobson2022digital,baumel2025more,isa2024exploring,suva2024artificial}. These approaches facilitate the development of adaptive treatment pathways that respond to changes in user behavior and context over time.

A particularly promising paradigm in this domain is the concept of JITAIs, which aim to deliver the right type of support at the right time and in the right context. JITAIs leverage real-time data and sequential decision-making models to optimize intervention delivery, improving both engagement and treatment effectiveness \cite{nahum2016just,nahum2025just,henry2025just,dowling2022gambling,rodda2022smartphone}.

Furthermore, the integration of multimodal data sources—including behavioral logs, sensor data, and contextual signals—has enhanced the ability of AI systems to model complex addiction dynamics. Multimodal learning approaches enable more accurate risk prediction, early detection of relapse, and personalized intervention strategies \cite{kamrul2025multimodal,alsaad2024multimodal,kato2023computational,wyant2024machine}.

Recent studies have also demonstrated the feasibility of AI-based prediction models for identifying addiction risk and behavioral escalation using digital behavioral traces \cite{gezgin2025using,auer2026using}. At the same time, digital mental health ecosystems—including mobile health applications, conversational agents, and telehealth platforms—continue to expand access to care while introducing new challenges related to engagement, regulation, and ethical considerations \cite{torous2025evolving,refolo2022ethics,saeidnia2024ethical}.

These developments highlight the critical role of digital technologies and AI-driven systems in advancing the prevention, treatment, and long-term management of behavioral addictions.

\subsection{Previous works and our contributions}

The growing interest in behavioral addictions and digital health technologies has led to a substantial body of review literature spanning multiple disciplines, including psychology, psychiatry, computer science, and digital health. Existing works can be broadly categorized into several thematic groups.

First, theoretical and clinical reviews have focused on conceptual frameworks and the underlying mechanisms of behavioral addictions. Notably, the I-PACE model and its extensions provide a comprehensive theoretical basis for understanding the interaction between individual vulnerabilities, affective processes, cognitive biases, and executive control mechanisms \cite{brand2019interaction}. More recent work has further advanced this perspective by integrating insights from neuroscience and clinical practice. Despite these efforts, existing studies remain fragmented, and no work provides a unified framework integrating these dimensions.

Second, intervention-oriented reviews have examined the effectiveness of traditional and digital treatment approaches. Systematic reviews of web-based interventions highlight the potential of online platforms for delivering scalable treatments, particularly in the context of behavioral addictions such as problematic internet use and gaming disorder \cite{park2022content}. Meta-analyses on relapse prevention strategies further emphasize the importance of continuous support and adaptive intervention mechanisms in improving long-term outcomes \cite{tabugan2025novel}. Early digital platforms further illustrate the feasibility of technology-assisted addiction management systems \cite{balhara2019behavior}.

Third, computational and artificial intelligence–driven studies have explored the role of data-driven modeling in addiction research. Reviews in this domain highlight the growing use of machine learning and computational approaches to model addictive behaviors, predict risk, and support decision-making \cite{kato2023computational,suva2024artificial}. These studies demonstrate the potential of AI in advancing personalized and predictive frameworks for addiction management.

Fourth, digital mental health and e-health reviews have investigated the role of emerging technologies in transforming mental health care delivery. Studies in digital psychiatry and mobile health emphasize the increasing adoption of digital platforms, the potential for scalable interventions, and the challenges associated with engagement, regulation, and ethical considerations \cite{ferreri2018addictology, torous2021growing,torous2025evolving, ali2025artificial}.

Fifth, recent work has also examined the broader research ecosystem, including data sharing practices and ethical considerations in addiction research. These studies highlight critical issues related to data governance, privacy, transparency, and fairness in AI-driven systems \cite{corwin2025comprehensive,refolo2022ethics,saeidnia2024ethical}.

Existing studies provide valuable insights into specific aspects of behavioral addiction research. Theoretical models establish a strong conceptual foundation, while intervention-focused reviews and meta-analyses demonstrate the effectiveness of treatment strategies. Digital platforms and AI-driven studies further highlight the feasibility and potential of technology-assisted addiction management systems. However, these contributions remain largely isolated and do not fully support an integrative perspective.

A comparative overview of existing works is presented in Table~\ref{tab:survey_comparison}, highlighting the limitations of prior work and the lack of an integrative framework. Despite these contributions, the current body of literature remains inherently fragmented across conceptual, technological, and clinical dimensions. Existing works typically address isolated components, such as theoretical models, specific intervention modalities, or AI-based prediction, without providing a unified perspective that bridges these domains.

\begin{table*}[ht]
	\centering
	\caption{Comparison of existing works on behavioral addictions and digital approaches.}
	\label{tab:survey_comparison}
	\footnotesize
		\begin{tabular}	{p{1.5cm} p{5cm} p{1cm} p{1cm} p{1cm} p{1.5cm} p{1.5cm}}
			\toprule
			\textbf{Study} & \textbf{Focus} & \textbf{Digital} & \textbf{AI/ML} & \textbf{JITAI} & \textbf{Multimodal} & \textbf{Unified} \\
			\midrule
			
			\cite{brand2019interaction} & Theory (I-PACE) & $\times$ & $\times$ & $\times$ & $\times$ & $\times$ \\
			
			\cite{brand2025current} & Clinical advances & $\times$ & $\times$ & $\times$ & $\times$ & $\times$ \\
			
			\cite{park2022content} & Web-based interventions & $\checkmark$ & $\times$ & $\times$ & $\times$ & $\times$ \\
			
			\cite{tabugan2025novel} & Meta-analysis (relapse) & $\times$ & $\times$ & $\times$ & $\times$ & $\times$ \\
			
			\cite{balhara2019behavior} & Digital platform & $\checkmark$ & $\times$ & $\times$ & $\times$ & $\times$ \\
			
			\cite{ali2025artificial} & AI in mental health & $\checkmark$ & $\checkmark$ & $\times$ & $\checkmark$ & $\times$ \\
			
			\cite{suva2024artificial} & AI in addiction & $\checkmark$ & $\checkmark$ & $\times$ & $\times$ & $\times$ \\
			
			\cite{kato2023computational} & Computational models & $\times$ & $\checkmark$ & $\times$ & $\times$ & $\times$ \\
			
			\cite{torous2025evolving} & Digital mental health & $\checkmark$ & $\checkmark$ & $\times$ & $\times$ & $\times$ \\
			
			\cite{nahum2025just} & JITAI framework & $\checkmark$ & $\checkmark$ & $\checkmark$ & $\times$ & $\times$ \\
			
			\cite{corwin2025comprehensive} & Data/ethics & $\checkmark$ & $\times$ & $\times$ & $\times$ & $\times$ \\
			
			\midrule
			\textbf{This book} & \textbf{Behavioral addiction management} & $\checkmark$ & $\checkmark$ & $\checkmark$ & $\checkmark$ & $\checkmark$ \\
			
			\bottomrule
		\end{tabular}%
\end{table*}

In particular, prior work lacks (i) an integrated framework linking behavioral addiction theory with digital intervention technologies, (ii) systematic incorporation of adaptive and real-time intervention paradigms such as JITAIs, (iii) comprehensive utilization of multimodal data for continuous monitoring and personalization, and (iv) a clear mapping between computational modeling approaches and clinically actionable intervention strategies.

Furthermore, although recent studies acknowledge the potential of AI and DTx, they rarely operationalize how these components can be integrated into a coherent, end-to-end addiction management framework, limiting their translational impact.

To address these limitations, this book presents the management of behavioral addictions via digital technologies, with a particular focus on integrative and adaptive approaches. The main contributions of this work are summarized as follows:

\begin{itemize}
	\item A unified and multidimensional taxonomy of behavioral addictions, incorporating behavioral domains, modality (digital vs.\ non-digital), and underlying mechanisms, along with a refined taxonomy for digital behavioral addictions are provided.
	
	\item Digital approaches for addiction management, including DTx, mobile health systems, AI-driven models, and just-in-time adaptive interventions are reviewed systematically.
	
	\item A structured synthesis of computational and machine learning techniques applied to behavioral addiction modeling, risk prediction, and intervention personalization is presented.
	
	\item The role of multimodal data integration and digital phenotyping in enabling real-time monitoring and adaptive intervention strategies is analyzed.
	
	\item Key research gaps and challenges, including issues related to data quality, model generalizability, user engagement, and ethical considerations are identified.
	
	\item Future research directions toward scalable, explainable, and clinically integrated AI-driven addiction management systems are outlined.
\end{itemize}

\subsection{Book organization}

The remainder of this book is organized as follows. Section~\ref{sec2} presents the clinical and psychological foundations of behavioral addictions, including neurobiological mechanisms, psychological models, comorbidity, and current treatment paradigms. 

Section~\ref{sec3} reviews digital technologies used in behavioral addiction management, covering mobile applications, wearable and IoT-based monitoring systems, web-based intervention platforms, and AI-powered DTx.

Section~\ref{sec4} introduces the functional capabilities and use-cases enabled by digital systems, including digital screening, diagnostics and risk stratification, early detection and relapse prediction, behavior tracking and digital phenotyping, personalized and adaptive interventions, harm reduction strategies, and continuous monitoring frameworks such as just-in-time adaptive interventions.

Section~\ref{sec5} describes the data ecosystem underlying digital addiction systems, including self-reported data, physiological signals, behavioral and usage logs, contextual information, and social interaction traces, along with challenges in multimodal data fusion.

Section~\ref{sec6} presents artificial intelligence and machine learning approaches for addiction management, including risk prediction models, recommendation systems, RL strategies, NLP methods, and issues related to explainability, bias, and generalization.

Section~\ref{sec7} discusses ethical, legal, and societal considerations, including privacy, data protection, security, equity, regulatory frameworks, and healthcare system integration.

Section~\ref{sec8} examines economic and ecosystem dimensions, including business models, healthcare integration, and public health implications.

Section~\ref{sec9} outlines key challenges and future research directions, including limitations in validation, engagement, scalability, multimodal integration, and emerging directions such as generative AI, hybrid human-AI systems, and preventive digital health strategies.

Finally, Section~\ref{sec10} concludes the book.

\newpage
\section{The Clinical and Psychological Foundations of Addiction}\label{sec2}

\subsection{Neurobiological mechanisms and reward systems}
For a long time, addiction was viewed solely as a disease resulting from the direct stimulation of the brain’s reward systems by psychoactive substances such as alcohol, cocaine, or heroin \cite{ross2009neurobiology}. In this classical approach, it was accepted that substances activate the mesolimbic dopaminergic pathways, leading to increased dopamine in the nucleus accumbens (NA), and that this forms the basis of addiction \cite{kalivas2005neural}. However, it is now clear that the process is more complex than this. While dopamine elevation is significant, it is believed that the primary mechanisms sustaining addiction are neuroplastic changes that develop over time and the reorganization of neural networks \cite{suckling2017neurobiology}.

The functioning of the reward system is crucial at this point. The mesocorticolimbic system operates as a circuit connecting the ventral tegmental area (VTA), the NA, and the pre-frontal cortex. Dopamine released from the VTA enhances reward anticipation in the NA, while the prefrontal cortex evaluates this information to determine whether to continue the behavior. It should be particularly emphasized here that dopamine is not merely a molecule associated with pleasure. Rather, it serves as a signal that guides learning through “reward prediction errors.” Better-than-expected outcomes ensure that the behaviors leading to these outcomes are reinforced via dopaminergic mechanisms \cite{clark2019neuroimaging, schultz1997neural}. From a clinical perspective, this mechanism helps explain why behavior can become so persistent, particularly under variable and unpredictable reward conditions such as in gambling or online gaming.

Beyond dopamine release, changes at the receptor level are equally important. The D1 and D2 receptors located in the striatum have distinct functions; while D1 receptors primarily facilitate the reinforcement of reward-related behaviors, D2 receptors are associated with inhibitory control \cite{volkow2016neurobiologic}. It has been demonstrated that chronic substance use leads to a decrease (down regulation) in D2 receptors in particular, and that this is associated with impaired impulse control \cite{volkow2010addiction, kim2011reduced}. This phenomenon can be considered the neurobiological basis for the clinically common state of “being unable to stop despite knowing one should not.” The fact that similar findings have been demonstrated in behavioral addictions suggests that these disorders share a common biological basis \cite{grant2010introduction, leeman2013targeted}. Table \ref{tab:nurobiological} shows key neurobiological mechanisms in behavioral addictions and their clinical implications.

\begin{table}[h!]
	\centering
	\footnotesize
	\caption{Key neurobiological mechanisms in behavioral addictions and their clinical implications (adapted from \cite{leeman2013targeted}).}
	\label{tab:nurobiological}
	\setlength{\tabcolsep}{4pt}
	\renewcommand{\arraystretch}{1.1}
	\begin{tabular}{p{4.5cm} p{5.5cm} p{5.5cm}}
		\toprule
		\textbf{Mechanism} & \textbf{Neurobiological basis} & \textbf{Clinical implication} \\
		\midrule
		
		Reward system dysfunction 
		& Changes in frontostriatal circuits and reward processing 
		& Continuing the behavior despite negative consequences \\
		
		Cue reactivity / increased salience 
		& Increased sensitivity to addiction-related cues in reward pathways 
		& Craving and cue-triggered behavior \\
		
		Dopamine dysregulation 
		& Altered dopamine signaling and receptor function 
		& Impaired reward evaluation and increased risk-taking \\
		
		Reduced prefrontal control 
		& Decreased activity in brain regions responsible for control and decision-making 
		& Difficulty stopping the behavior and impaired control \\
		
		Habit formation
		& Shift from reward-driven to habit-based behavior 
		& Automatic and repetitive behavior patterns \\
		
		Serotonin dysfunction 
		& Changes in the serotonin system (mixed findings) 
		& Impulsivity and emotional instability \\
		
		Glutamate involvement 
		& Changes in glutamate pathways affecting learning and control 
		& Persistence of behavior and relapse \\
		
		White matter changes 
		& Reduced connectivity between key brain regions 
		& Problems with behavioral regulation \\
		
		Genetic vulnerability 
		& Familial and genetic contributions 
		& Increased susceptibility to addiction \\
		
		\bottomrule
	\end{tabular}
\end{table}

This process is also linked to the development of tolerance. Stimuli that initially elicit a strong dopaminergic response begin to have a weaker effect over time, and the individual repeats the behavior more frequently or intensely to achieve the same effect \cite{koob2014addiction, kalivas2009glutamate}. 

At the same time, the “reward sensitivity” model provides an important framework for understanding addiction. According to this model, over time, cues associated with the reward come to the forefront rather than the reward itself, and these cues create a powerful motivational pull. In this process, craving behavior becomes pronounced, and the individual often continues the behavior not to seek pleasure but to reduce this intense craving \cite{robinson2022incentive}. This distinction is clinically significant as it shows that the processes of craving and pleasure can be dissociated from one another.

As addiction progresses, the role of dopamine diminishes relatively, while the glutamatergic system becomes more prominent. In particular, changes in glutamatergic projections extending from the pre-frontal cortex to the NA play a significant role in behavior shifting from goal-directed to habit-based and compulsive \cite{kalivas2005neural}. Therefore, addiction can be viewed not merely as a process driven by reward-seeking, but also as a habit that becomes established over time and a disorder of behavioral regulation \cite{everitt2005neural}.

Although dopamine plays a central role, the contribution of other neurotransmitter systems cannot be overlooked. The serotonin system influences impulsivity and behavioral inhibition, and it is thought that disruptions in this system may increase the tendency toward short-term rewards \cite{dalley2012dopamine}. While glutamate plays a decisive role in learning processes and the maintenance of behavior, the endogenous opioid system modulates the reward experience \cite{kalivas2009glutamate, reed2017endogenous}. The endocannabinoid system, meanwhile, functions as a high-level modulator that regulates the overall sensitivity of this network \cite{lopez2008pharmacology}.

In recent years, evidence that behavioral addictions also develop through similar neurobiological mechanisms has broadened the scope of the concept of addiction. It is known that behavioral addictions such as gambling, online gaming, and compulsive shopping also activate the reward circuit and are maintained in a similar manner \cite{grant2010introduction}. This has also been reflected in the classification changes made in the DSM-5 and ICD-11 \cite{edition2013diagnostic}. Current data indicate that behavioral addictions overlap significantly with substance addictions from both neurobiological and clinical perspectives \cite{leeman2013targeted}.

The role of the prefrontal cortex is particularly important in terms of behavioral control. While the dorsolateral prefrontal cortex is involved in impulse control and working memory, the orbitofrontal cortex plays a role in the evaluation of reward value, and the anterior cingulate cortex plays a role in error monitoring processes \cite{feil2010addiction}. Functional changes in these regions contribute to the maintenance of compulsive patterns by reducing an individual’s capacity to regulate behavior. Neuroimaging studies have also reported significant changes in frontostriatal circuits associated with addiction \cite{suckling2017neurobiology}.

It has been shown that similar neural circuits are functionally affected in behavioral addictions as well. In neuroimaging studies, particularly in individuals with gambling disorder and internet gaming addiction, increased activity in striatal regions associated with reward processing and decreased activation in prefrontal areas associated with executive control have been reported \cite{clark2019neuroimaging}. Additionally, it has been shown that an increased neural response to addiction-related cues (“cue reactivity”) is prominent in these individuals and is associated with craving. These findings support the notion that common neurobiological mechanisms—involving heightened reward anticipation and weakened inhibitory control that contribute to the maintenance of behavior are also relevant in behavioral addictions.

In conclusion, addiction is a condition that arises from the interaction of a broad neural network involving reward processing, learning, and executive control processes. While the dopaminergic system is a key component of this process, the contribution of glutamatergic and other neurotransmitter systems results in a more complex structure \cite{koob2014addiction, kalivas2009glutamate}. Changes in dopamine receptor levels and functional differences in frontostriatal circuits provide an important framework for understanding the clinical characteristics of addiction.

\subsection{Psychological models}
In understanding behavioral addictions, the psychosocial model offers a more holistic framework that goes beyond explanations limited to biological mechanisms, instead addressing individual, interpersonal, and sociocultural levels together \cite{griffiths2005components, karim2012behavioral}. In this approach, addiction is viewed not as a condition reducible to a single cause, but rather as a dynamic process resulting from the individual’s developmental history, psychological characteristics, learning processes, and interaction with their environment \cite{brand2016integrating}. Behavioral addictions are characterized by repetitive behavioral patterns that are maintained despite short-term rewards and long-term negative consequences. It is known that these behaviors share classic addiction components such as salience, tolerance, withdrawal, conflict, and relapse. All of this suggests that treating addiction solely as a neurobiological process is insufficient; it must be evaluated within a multidimensional network of interactions.

In this context, the I-PACE model developed by Brand et al. \cite{brand2016integrating} provides an important theoretical framework for explaining the development and maintenance of behavioral addictions. According to the model, specific internet use disorders such as gaming, gambling, pornography, shopping, and social media use arise from the interaction of the individual’s neurobiological and psychological predispositions (genetic traits, early life experiences, depression, social anxiety, and impulsivity), coping strategies, cognitive biases regarding internet use, and the interaction of affective-cognitive responses to situational triggers. Within this framework, impulsivity, reward sensitivity, and personality traits form the individual foundation; meanwhile, emotional responses and cognitive processes (particularly craving and cognitive distortions) emerge as the primary mechanisms that trigger and sustain behavior. Executive functions play a role in regulating behavior, particularly through impulse control and decision-making processes; environmental factors and the availability of stimuli, meanwhile, shape the contextual dimension of this process. Additionally, the model emphasizes that classical and operant conditioning in the addiction process reinforce cue-response and craving responses; conversely, executive functions and inhibitory control weaken over time. From a clinical perspective, it is observed that these components are often not independent of one another but rather operate in a dynamic and reciprocal interaction.

In the updated version of the model, Brand et al. \cite{brand2019interaction} argued that the I-PACE model is not limited to internet use disorders; it is also applicable to other behavioral addictions such as gambling disorder, gaming disorder, shopping disorder, and compulsive sexual behavior disorder. The updated model addresses the addiction process in terms of early and advanced stages. In the early stage, general inhibitory control plays a regulatory role between emotional responses and behavioral decisions; in the advanced stage, cue-reactivity and craving, reinforced by conditioning processes, weaken stimulus-specific inhibitory control and lead to the behavior gradually becoming habitual. At the neurobiological level, findings suggest that in the early stages, activity in the ventral striatum associated with reward anticipation is more pronounced; as the process progresses, a shift toward the dorsal striatum occurs.

Individual psychological characteristics also play a significant role in the development of addiction. In particular, traits such as impulsivity and low self-control have been shown to be associated with addictive behaviors \cite{everitt2005neural}. In these individuals, short-term rewards are more likely to become appealing, and long-term consequences are more likely to be disregarded. Additionally, it has long been known that adverse childhood experiences and traumatic events are associated with addictive behaviors \cite{felitti1998relationship}. It is thought that such experiences may contribute to the emergence of addictive behaviors as a coping mechanism through difficulties in emotional regulation \cite{munno2016addictive, alimoradi2024associations}. Furthermore, it has been proposed that early-life schemas, particularly themes of rejection and disconnection, provide a cognitive backdrop for this process \cite{de2023psychosocial}.

The social and relational context also plays a significant role in the development of addiction. The family environment is a fundamental setting that shapes an individual’s emotional development and coping skills; in particular, it has been demonstrated that attachment patterns are associated with addictive behaviors \cite{thorberg2006attachment, schindler2005attachment}. Insecure attachment styles are more strongly associated with behavioral addictions \cite{d2019addiction}. In addition, it is thought that environmental factors such as the family environment and level of social support may influence the risk of addiction \cite{kuss2011online}. Peer influence and social learning processes, in particular, play a significant role in the maintenance of behaviors among adolescents and young adults.

It is well established that behavioral addictions are closely linked to learning and reinforcement processes. In particular, reward-based learning mechanisms facilitate the repetition of behavior and its reinforcement over time. The reward structures observed in gambling and digital games emerge as a significant factor reinforcing this process. Behaviors that are initially more goal-directed gradually acquire a habit-based and more compulsive nature over time \cite{everitt2005neural}. During this process, the individual may continue the behavior not only to obtain pleasure but also to avoid negative emotions \cite{koob2014addiction}. Additionally, cognitive distortions—particularly the overestimation of reward expectations and the illusion of control—play a significant role in the maintenance of the behavior.

In conclusion, the psychosocial model demonstrates that behavioral addictions cannot be explained by a single-dimensional approach. The interaction of individual characteristics, cognitive and emotional processes, learning mechanisms, and environmental factors collectively plays a role in the development and maintenance of addictive behaviors. This holistic approach provides an important framework for both clinical assessment and treatment planning.

\subsection{Comorbidity}
Behavioral addictions constitute a clinically heterogeneous group that is frequently observed both within the category itself and in conjunction with other psychiatric disorders \cite{grant2010introduction, leeman2013targeted}. This comorbidity not only increases the burden of disease but also complicates the understanding of etiopathogenesis and treatment planning \cite{dkabrowska2022factors}. Indeed, various studies report a high prevalence of additional psychiatric diagnoses in individuals with behavioral addictions.

It is well-known that rates of psychiatric comorbidity are quite high in gambling addiction. In a meta-analysis by Lorains et al. \cite{lorains2011prevalence} based on 11 studies using general population samples, it was shown that the most common comorbid condition among problem and pathological gamblers was nicotine dependence (60.1\%). This is followed, in order, by any substance use disorder (57.5\%), mood disorders (37.9\%), anxiety disorders (37.4\%), and antisocial personality disorder (28.8\%).

However, it is also worth noting that these rates may vary depending on the methods used. For example, the prevalence of alcohol use disorder can reach as high as 41.2\% in interviews conducted by clinicians, while it can drop as low as 18.2\% in self-report scales. Similarly, lifetime prevalence rates (33.4\%) are reported to be higher than past-year prevalence rates (15.7\%). These findings suggest that comorbidity rates are highly heterogeneous and that interpreting them based on a single rate may be misleading.

IGD and PIU often co-occur with other psychiatric conditions. Indeed, studies report that the most common comorbid conditions associated with IGD are anxiety (92\%), depression (89\%), attention-deficit/hyperactivity disorder (ADHD; 85\%), and social problems (75\%). A study conducted on a large sample in South Korea also demonstrates that PIU is not limited to such internalizing problems; it is also associated with nicotine and alcohol use disorders, somatoform disorders, pathological gambling, sleep problems, and suicidal ideation \cite{kim2016prevalence}.

The clinical picture becomes even more severe, particularly in cases where IGD co-occurs with alcohol use disorder. It has been shown that this group exhibits higher levels of impulsivity, difficulty with self-control, and depression; they also spend more money on online games \cite{na2017comorbidity}. On the other hand, functional imaging studies in adolescents with IGD have revealed increased connectivity between the salience network and executive control networks, and it has been suggested that this may be associated with comorbid psychiatric disorders \cite{han2017brain}.

Data on causality, however, point to a different conclusion. A longitudinal study found that existing psychopathologies, particularly depression and anxiety, predict the development of IGD; conversely, IGD does not increase psychopathology over time \cite{leo2021depression}. This suggests that internalizing disorders may often be a precursor rather than a consequence of IGD. However, it should not be forgotten that comorbidity rates can vary depending on the measurement tools used. Ferguson et al.’s meta-analysis demonstrates that different scales can yield different results \cite{ferguson2011meta}.

High rates of psychiatric comorbidity have also been reported in other behavioral addictions. Clinical studies have found that conditions such as kleptomania, compulsive buying, and compulsive sexual behavior disorder frequently co-occur with substance use disorders \cite{grant2024compulsive}. In compulsive buying disorder, rates of ADHD, borderline personality disorder, and binge eating disorder were found to be significantly higher compared to the control group. In compulsive sexual behavior disorder, comorbidity with mood disorders has been reported at 36–81\%, and with anxiety disorders at 46–96\% \cite{puszcz2025neurobiological}. In general, psychiatric diagnoses such as major depressive disorder, bipolar disorder, OCD, and ADHD frequently co-occur with behavioral addictions; however, it should be noted that the vast majority of these findings are based on clinical samples, and their generalizability to the general population remains uncertain.

When exercise addiction and eating-related behavioral addictions are evaluated together, a significant overlap with eating disorders in particular is observed \cite{mueller2023update}. It has been reported that excessive exercise behavior can be observed in 30–80\% of patients with anorexia nervosa and bulimia nervosa. Obsessive-compulsive traits and perfectionism are also more frequently observed in this group. Additionally, it is noted that anxiety and depression are more common in individuals with exercise addiction.

Behavioral addictions are frequently observed not only alongside substance use disorders but also in combination with one another. Shaffer et al. \cite{shaffer2004toward} noted that individuals who engage in compulsive shopping and pathological gambling exhibit similarly high rates of addiction, explaining this pattern of co-occurrence with the concept of the “addiction syndrome.” The authors also presented the fact that individuals recovering from one addiction often transition to another behavioral addiction (for example, from gambling to excessive shopping or vice versa) as one of the key pieces of evidence for this syndrome. Grant et al. \cite{grant2010introduction} also supported this view, reporting that other behavioral addictions such as compulsive shopping, compulsive sexual behavior, and kleptomania are frequently found co-occurring with pathological gambling in clinical samples. Similarly, Chamberlain and Grant \cite{grant2016pharmacotherapy} highlighted the comorbidity and familial overlap reported in the literature between compulsive buying disorder and gambling disorder. All these findings suggest that different behavioral addictions are not independent conditions but rather different manifestations of a syndrome sharing common neurobiological reward pathways, and thus often occur together or sequentially \cite{shaffer2004toward, grant2010introduction, grant2024compulsive}.

The relationship between behavioral addictions and personality disorders is also clinically significant; in particular, significant associations have been reported with antisocial personality traits. Impulsivity and difficulties with emotion regulation are thought to play a significant role in this relationship. Furthermore, the higher prevalence of addiction and other psychiatric disorders among first-degree relatives of individuals with behavioral addictions suggests that these conditions may exhibit familial clustering.

\subsection{Current clinical treatment paradigms and limitations}
The treatment of behavioral addictions relies on evidence-based yet individualized approaches rather than standardized algorithms, due to the clinical diversity of these disorders and the relatively recent nature of their classifications \cite{leeman2013targeted, grant2010introduction, brand2025current}. The current literature indicates that evidence regarding pharmacological treatment options is limited, that no approved pharmacological treatment exists for any behavioral addiction, and that standard clinical treatment protocols have not yet been developed. Therefore, treatment planning is based not only on the diagnosis but also on the symptom pattern, the function of the behavior, disease severity, and comorbid psychiatric conditions \cite{starcevic2017relationships}.

In treatment approaches, the focus is typically on targeting the mechanisms that sustain the problematic behavior rather than the behavior itself. In behavioral addictions, processes such as increased reward sensitivity, decreased tolerance for delayed rewards, poor impulse control, cognitive distortions, and avoidance of negative emotions are thought to be key factors \cite{brand2019interaction}. While these processes create short-term soothing or rewarding effects, they contribute to the maintenance of the behavior in the long term \cite{brand2025current}. Therefore, interventions are aimed not only at reducing the behavior but also at targeting these processes to develop more functional coping strategies.

From a treatment perspective, the I-PACE model emphasizes that it is more effective to target variables that play a mediating and regulatory role in the process (coping styles, expectations regarding internet use, attentional biases, cue-reactivity and craving, and executive functions) rather than predisposing factors that are difficult to change. In this vein, it is suggested that CBT-IA for internet addiction including intervention methods such as cognitive restructuring, cue exposure therapy, attention retraining, and inhibition training can target cognitive biases and functional impairments in the addiction process. The updated model also supports this theoretical framework; by emphasizing the need to target mediating variables (expectations, emotional and cognitive responses, inhibitory control) in treatment, it offers empirically testable hypotheses for future intervention studies.

In clinical practice, treatment often follows a stepwise approach. In the first phase, the goal is to increase the patient’s awareness of the problem and reduce ambivalence toward change using motivational interviewing techniques \cite{miller2012motivational}. Following this, cognitive-behavioral therapy focuses on identifying triggers, defining high-risk situations, and developing alternative behaviors. Techniques such as stimulus control, exposure, cognitive restructuring, and impulse management are frequently used in this process. In the relapse prevention phase, the focus is particularly on emotional triggers, high-risk situations, and dysfunctional thought patterns.

The presence of comorbid psychiatric disorders is quite common in behavioral addictions and can directly influence the treatment process. It is known that conditions such as depression, anxiety disorders, and ADHD not only serve as factors that sustain addictive behavior but also influence the response to treatment. Therefore, it is important to assess comorbid disorders during treatment planning and prioritize them when necessary. Dell’Osso et al. \cite{dell2006epidemiologic} emphasized that the pattern of comorbidity in internet addiction could guide the selection of pharmacological treatment and that treatment should be planned not only based on the addictive behavior itself but also on the co-occurring psychopathological dimensions (impulsivity, compulsion, craving, depressive, and anxiety symptoms).

The role of pharmacological treatments in behavioral addictions is more limited, and to date, there is no FDA-approved medication for any behavioral addiction. Current data indicate that pharmacotherapy is primarily used to target comorbid symptoms and underlying psychopathological dimensions rather than the addictive behavior itself.

In recent years, new approaches targeting neurocognitive processes have also been investigated. It has been suggested that neuromodulation techniques, such as transcranial magnetic stimulation, may be effective in influencing impulse control and reward processing \cite{gomis2019non}. Mindfulness-based interventions have been shown to reduce reactivity to cravings and trigger experiences associated with addictive behaviors \cite{wilson2017mindfulness}. However, the clinical efficacy of these methods is currently supported by a limited number of studies, and more evidence is needed for routine clinical use.

The level of evidence regarding treatment also varies among behavioral addictions. Gambling disorder stands out as the condition with the strongest evidence in this field \cite{goslar2020treatments}. Numerous studies have demonstrated that cognitive-behavioral therapy leads to a significant reduction in the frequency and severity of gambling behavior \cite{pfund2023effect}. Cognitive distortions that sustain gambling behavior, particularly the illusion of control and the desire to recoup losses, are among the primary treatment targets \cite{fortune2012cognitive, ledgerwood2020nancy}.

From a pharmacological perspective, opioid receptor antagonists—particularly naltrexone and nalmefene-stand out as the only class of medications to have consistently demonstrated positive results in double-blind, placebo-controlled trials for gambling disorder \cite{ioannidis2025pharmacological}. Chamberlain and Grant \cite{grant2016pharmacotherapy} highlight two separate studies demonstrating naltrexone’s superiority over placebo in reducing gambling-related behaviors and thoughts, as well as two multicenter studies showing that nalmefene produced significant improvements in gambling urges and behaviors. In addition, placebo-controlled studies have shown that glutamate modulators such as N-acetylcysteine also produce a significant reduction in gambling symptoms. In contrast, the results of studies using SSRI-class antidepressants (sertraline, fluvoxamine, paroxetine) have been inconsistent, and bupropion has also failed to demonstrate superiority over placebo \cite{lubman2022pharmacological}. While lithium has been found effective in patients with gambling disorder accompanied by bipolar spectrum disorders, two studies using olanzapine showed no difference from placebo. Therefore, pharmacological treatments are generally planned according to the accompanying psychiatric conditions and symptom profile.

Treatment approaches for internet gaming disorder are primarily based on psychosocial interventions \cite{zajac2017treatments}. Cognitive-behavioral therapy is the most commonly used method for reducing gaming time, restructuring usage patterns, and modifying dysfunctional thoughts \cite{han2020efficacy, reangsing2025effects}. It has also been shown that family-based interventions can increase treatment adherence, particularly among adolescents.

Among pharmacological treatment options, agents such as bupropion, methylphenidate, and atomoxetine have been studied, and it is believed that these medications may be beneficial for some patients, particularly due to their effects on attention and impulsivity. Dell'Osso et al. \cite{dell2006epidemiologic} reported that in open-label studies, escitalopram improved weekly online time and overall functioning in internet addiction, but failed to distinguish itself from placebo when the study transitioned to a double-blind phase; while bupropion significantly reduced online gaming time and depressive symptoms in a randomized controlled trial. Additionally, methylphenidate has been reported to provide significant improvements in both attention capacity and internet use in internet gaming addiction accompanied by ADHD. However, most current studies are limited to small sample sizes, and the generalizability of the results appears limited.

Meta-analyses indicate that psychological interventions can have significant effects on both the duration of use and symptoms of addiction \cite{winkler2013treatment}. Dell'Osso et al. \cite{dell2006epidemiologic} suggest that the psychopathological dimensions shared by internet addiction with other behavioral and substance addictions (impulsivity, compulsion, and reward system dysfunction) may guide treatment approaches, and that planning pharmacological interventions targeting these common dimensions could enhance clinical benefits.

Evidence regarding treatment for other behavioral addictions, such as compulsive shopping, hyper-sexual behavior, and exercise addiction, is more limited \cite{aguilar2024systematic}. Chamberlain and Grant \cite{grant2016pharmacotherapy} report that, of the three placebo-controlled studies conducted for compulsive shopping, only citalopram yielded positive results, while fluvoxamine studies failed to distinguish themselves from placebo. Regarding compulsive sexual behavior, a single double-blind study demonstrated that citalopram resulted in a significant reduction in sexual urges and behaviors. Regarding kleptomania, there is a single positive placebo-controlled study indicating that naltrexone is effective in reducing the urge and behavior to steal; however, escitalopram failed to demonstrate superiority over placebo in the double-blind phase. Data on the efficacy of psychological, pharmacological, and combined treatments in this area suggest they may provide meaningful improvement in the short term; however, the level of evidence remains quite limited \cite{goslar2020treatments}.

In conclusion, although the treatment of behavioral addictions is shaped by common mechanisms, there are significant differences in the level of evidence across these disorders \cite{brand2025current}. While cognitive-behavioral therapy stands out as the primary approach for many behavioral addictions, pharmacological treatments are used in more limited and specific cases. Chamberlain and Grant \cite{brand2016integrating} emphasize that most pharmacological treatment studies for behavioral addictions have been conducted with small sample sizes, over limited time periods, and likely with clinically underrepresented groups, highlighting the need for more comprehensive research to develop treatment algorithms. Dell'Osso et al. \cite{dell2006epidemiologic} also highlight the lack of large-sample randomized controlled trials and long-term follow-up data specifically regarding internet addiction. There is a clear need for studies conducted with larger samples and evaluating long-term outcomes, particularly in areas beyond gambling and internet gaming disorder \cite{boumparis2022internet}.

\newpage
\section{Digital Technologies in Addiction Management}\label{sec3}
Digital technologies are increasingly used in the management of behavioral (process) addictions, including gambling disorder, internet gaming disorder, and problematic digital media use. These disorders are characterized by impaired behavioral control, reinforcement-driven repetition, and high relapse risk. Clinical assessment largely relies on behavioral indicators rather than biological markers. Therefore, continuous monitoring of usage patterns, cognitive states, and contextual triggers is essential. Digital systems enable fine-grained behavioral data collection in real-life settings and support timely, context-aware psychological interventions. This approach aligns well with contemporary models that conceptualize behavioral addictions as dynamic and self-regulation–based disorders rather than episodic conditions \cite{brand2019interaction,berczik2012exercise}.

Recent advances in artificial intelligence and computational behavioral modeling have further enhanced digital interventions for behavioral addictions \cite{imrani2026interventions}. Machine learning methods can analyze interaction logs, screen-time metrics, and self-report data to identify maladaptive trajectories and predict escalation or relapse risk. These outputs support personalized intervention strategies, particularly within cognitive behavioral and inhibitory control frameworks \cite{park2025treatment}. Digitally delivered interventions also reduce treatment barriers related to stigma and accessibility, especially among younger populations. Evidence from controlled studies suggests that structured, theory-driven digital interventions can achieve clinically meaningful reductions in symptom severity when combined with appropriate ethical design and clinical supervision \cite{montag2021overcome, cho2025consort,svendsen2026reporting, imrani2026interventions}.

The following sub-sections review the main digital intervention modalities used in behavioral addiction management. Mobile applications and wearable or IoT-based sensors support continuous behavioral monitoring and self-regulation in daily life. Web-based platforms provide structured, evidence-based psychological interventions with high scalability. More advanced approaches, including AI-powered DTx, VR/AR-based exposure therapy, and conversational agents, enable adaptive, personalized, and context-aware support. Telehealth systems extend clinical services beyond traditional settings, while community platforms, peer-support networks, and gamification-based interventions enhance engagement, motivation, and treatment adherence \cite{ferreri2018addictology}.

\begin{figure*}[h!]
	\centering
	\includegraphics[width=\textwidth]{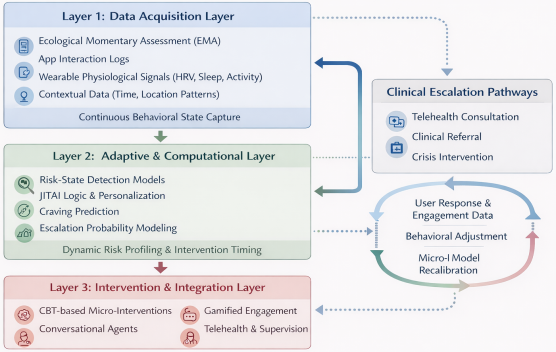}
	\caption{Multilayer digital intervention architecture for behavioral addiction management}
	\label{fig:digital_architecture}
\end{figure*}

Figure~\ref{fig:digital_architecture} shows the general digital intervention architecture discussed in this section. The model presents behavioral addiction management as a continuous and feedback-based process. Behavioral and contextual data are first collected through mobile apps and sensors. These data are processed to detect risk states and determine appropriate intervention timing. Brief interventions are then delivered through digital platforms, with clinical escalation options when needed. This model helps to understand how the different digital approaches discussed in the following sections relate to each other.

\subsection{Mobile apps}
Mobile applications are widely used in the management of behavioral addictions, including gambling disorder, IGD, and problematic digital media use. Their rapid expansion is mainly driven by high smartphone availability and low access barriers. Unlike clinic-based interventions, mobile apps operate within daily routines. This makes them suitable for conditions that require continuous self-regulation rather than short-term clinical treatment.

Most evidence-informed applications are grounded in CBT, motivational interviewing, and self-regulation models. Common features include craving monitoring, behavioral diaries, goal setting, feedback on risky behaviors, and brief coping prompts. These components target impaired inhibitory control and cue-reactivity processes described in contemporary addiction frameworks \cite{brand2019interaction}.

A key methodological strength of mobile interventions is ecological momentary assessment (EMA). EMA collects repeated real-time data on mood states, urges, and contextual triggers. Compared with retrospective clinical reporting, EMA improves ecological validity and temporal sensitivity \cite{torous2021growing}. When combined with usage logs, EMA can support time-sensitive micro-interventions and adaptive feedback.

\begin{figure*}[h!]
	\centering
	\includegraphics[width=\textwidth]{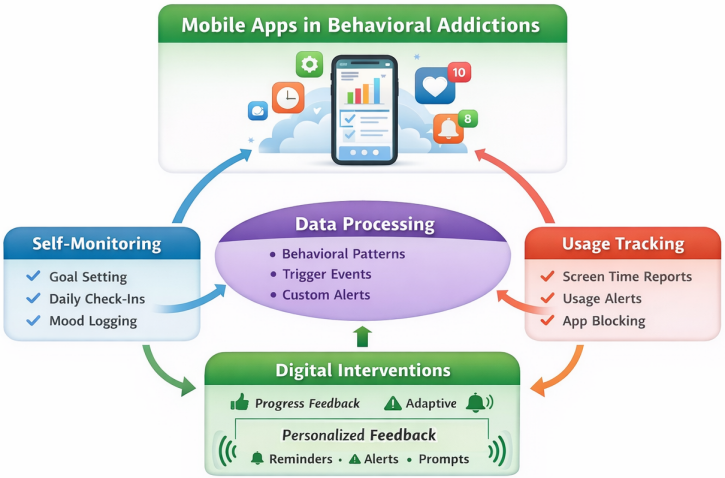}
	\caption{The core functional components of mobile applications used
		in the management of behavioral addictions}
	\label{fig:mobileapps}
\end{figure*}

Figure~\ref{fig:mobileapps} illustrates the core functional components of mobile applications used in the management of behavioral addictions. Mobile apps typically integrate self-monitoring tools (e.g., goal setting, mood logging), usage tracking features (e.g., screen-time reports, alerts), and digital intervention modules. Behavioral and contextual data are processed within the application to detect risk patterns and trigger adaptive responses. Personalized feedback mechanisms, including reminders, prompts, and progress reports, support self-regulation and sustained engagement. The framework reflects a closed-loop model in which monitoring, data processing, and intervention delivery operate continuously within daily life contexts.

Empirical evidence is strongest in prevention and early-risk populations. The ready4life cluster-randomized controlled trial (n=1351) reported significant reductions in at-risk drinking (OR $\approx$ 0.75) and PIU at 6-month follow-up \cite{haug2022efficacy}. Similarly, the MobileCoach Alcohol program (n=954) showed significant reductions in binge drinking, although follow-up retention was 28\% \cite{haug2023mobile}. These findings indicate moderate preventive efficacy but also highlight engagement challenges.

Mobile applications may also support treatment adherence. The DIDE eHealth tool demonstrated high usability (SUS $\approx$ 82) and frequent after-hours engagement in adults with substance use disorder \cite{stocker2024acceptability}. However, symptom-level outcomes were not evaluated.

Quality assessments of publicly available gambling apps reveal considerable heterogeneity. Applications incorporating structured CBT components scored higher on engagement and information quality metrics, yet most lack formal clinical validation \cite{mccurdy2023smartphone}.

A structured comparison of representative mobile app–based interventions is presented in Table~\ref{tab:mobile_apps}. Overall, mobile applications provide scalable prevention and relapse-support tools. Current evidence does not support their use as standalone treatment for severe gambling disorder. Long-term engagement, regulatory oversight, and integration with clinical care pathways remain essential \cite{torous2021growing}.

\begin{table*}[ht]
	\centering
	\caption{Evidence summary of mobile app-based interventions in behavioral addictions}
	\label{tab:mobile_apps}
	\small
	\begin{tabularx}{\textwidth}{p{2cm} p{3cm} p{2cm} p{2.5cm} p{2.5cm}}
		\toprule
		\textbf{Study} & \textbf{Population \& Design} & \textbf{Target behavior} & \textbf{Key findings} & \textbf{Interpretation} \\
		\midrule
		
		ready4life \cite{haug2022efficacy} &
		Apprentices (n=1351); cluster-RCT; 6-month follow-up &
		At-risk drinking; problematic internet use &
		OR $\approx$ 0.75 for at-risk drinking; significant reduction in PIU &
		Strong prevention-level evidence \\
		\\
		MobileCoach Alcohol \cite{haug2023mobile} &
		Adolescents (n=954); pre--post design &
		Binge drinking &
		Significant reduction (p $<$ .05); 28\% follow-up retention &
		Promising preventive effect; engagement decline over time \\
		\\
		DIDE App \cite{stocker2024acceptability} &
		Adults with SUD (n=24); feasibility study &
		Treatment adherence &
		SUS $\approx$ 82; frequent after-hours engagement &
		Adherence-support adjunct; no symptom-level efficacy \\
		\\
		Gambling Apps Review \cite{mccurdy2023smartphone} &
		14 commercial apps; quality review &
		Gambling behaviors &
		CBT-based apps scored higher on MARS engagement metrics &
		Quality heterogeneous; limited empirical validation \\
		\\
		Digital Psychiatry Review \cite{torous2021growing} &
		Multi-condition; narrative review &
		Cross-addiction &
		Scalable; high real-world dropout rates reported &
		Augmentation role recommended; not standalone care \\
		
		\bottomrule
	\end{tabularx}
\end{table*}

\subsection{Wearables \& IoT sensors}
Wearable devices and IoT-based sensors extend digital addiction management beyond self-report measures. Mobile applications primarily rely on active input, whereas wearables enable passive and continuous physiological monitoring. This is important because craving and loss-of-control states fluctuate dynamically and may not be captured reliably through retrospective reports.

Within the architecture shown in Figure~\ref{fig:digital_architecture}, wearable technologies mainly contribute to the data acquisition layer. They provide objective signals that may precede observable behavioral relapse. Common physiological markers include heart rate variability (HRV), electrodermal activity (EDA), sleep patterns, and physical activity.

\begin{figure*}[h!]
	\centering
	\includegraphics[width=0.9\textwidth]{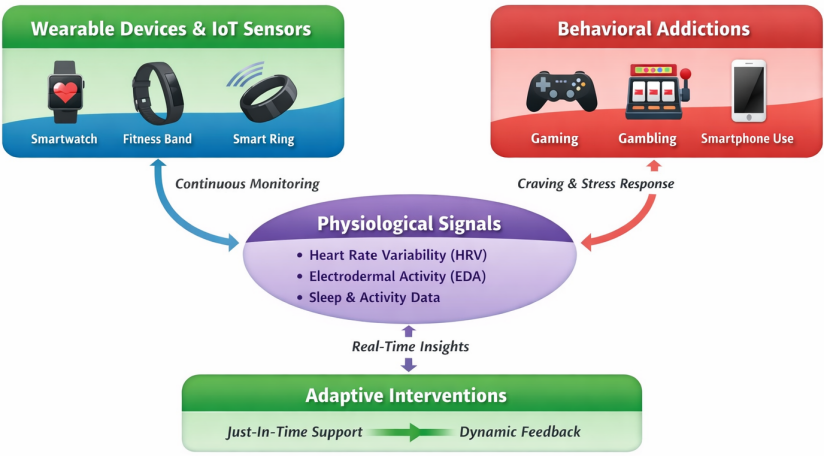}
	\caption{Wearable and IoT sensor integration in behavioral addiction monitoring}
	\label{fig:wearabledevices}
\end{figure*}

Figure~\ref{fig:wearabledevices}  illustrates the role of wearable devices and IoT sensors in behavioral addiction management. Continuous physiological monitoring through smartwatches, fitness bands, or similar devices enables the detection of autonomic signals such as HRV and electrodermal activity. These signals reflect stress and craving-related responses during engagement with addictive behaviors. When processed in real time, such data can inform adaptive, just-in-time interventions and dynamic feedback mechanisms. This approach supports timely risk detection and personalized regulation strategies within digital treatment frameworks.

HRV has been widely examined as an index of autonomic regulation in behavioral addictions. Studies in IGD report altered HRV responses during gaming exposure, indicating dysregulated physiological adaptation to addictive cues \cite{lee2018altered,long2023heart}. In addition, HRV reactivity features have been used in computational models to classify IGD severity with high accuracy, supporting the feasibility of integrating wearable-derived signals into digital monitoring frameworks \cite{hong2023severity}. In gambling research, electrodermal activity has been shown to differentiate win–loss outcomes and correlate with subjective excitement levels \cite{lole2012electrodermal,fukushima2025heart}. These findings suggest that reward-related arousal can be objectively captured and may inform time-sensitive intervention triggers.

Adaptive intervention research further supports this direction. The GamblingLess microrandomized trial protocol illustrates how repeated ecological assessments can guide just-in-time adaptive interventions in gambling disorder \cite{dowling2025gamblingless}. Similarly, studies in problematic smartphone use report associations between HRV changes and craving-related stress responses during behavioral reduction efforts \cite{dale2025influence}. Together, these findings indicate that wearable-derived autonomic markers may contribute to dynamic risk detection and intervention timing in behavioral addictions.

Despite promising mechanistic evidence, large-scale randomized controlled trials directly evaluating wearable-triggered interventions in behavioral addictions remain limited. Standardization of sensing protocols, cross-device validation, signal reliability, and ethical data governance frameworks are necessary for clinical implementation. Table~\ref{tab:wearables_iot} presents some findings from wearable and IoT-based studies.

\begin{table*}[ht]
	\centering
	\small
	\caption{Wearables and IoT-based monitoring in behavioral addictions}
	\label{tab:wearables_iot}
	\begin{tabular}{p{1cm} p{3.5cm} p{3.5cm} p{3.5cm} p{3cm}}
		\toprule
		\textbf{Study} & \textbf{Population \& Design} & \textbf{Target behavior} & \textbf{Key findings} & \textbf{Interpretation} \\
		\midrule
		
		\cite{lee2018altered} &
		IGD vs controls; real-time gaming; HRV monitoring &
		Internet Gaming Disorder &
		Altered HRV response during gaming; associated with symptom severity &
		HRV reflects cue-related autonomic imbalance \\
		\\
		\cite{hong2023severity} &
		N=70; HRV reactivity + deep learning &
		IGD severity classification &
		Accuracy = 95.10\% across four severity levels &
		Feasible physiological severity modeling \\
		\\
		\cite{lole2012electrodermal} &
		Lab electronic gambling; EDA \& HR &
		Gambling arousal &
		EDA differentiated win--loss outcomes; correlated with excitement ratings &
		Objective reward-reactivity signature \\
		\\
		\cite{dowling2025gamblingless} &
		Microrandomized protocol; 28-day EMA &
		Gambling behavior change &
		Repeated randomization; adaptive timing design &
		Supports JITAI integration \\
		\\
		\cite{dale2025influence} &
		Smartphone reduction RCT secondary analysis; HRV &
		Problematic smartphone use &
		HRV associated with craving-related stress responses &
		Autonomic sensitivity to withdrawal-like states \\
		
		\bottomrule
	\end{tabular}
\end{table*}

\subsection{Web-based intervention platforms}
Web-based intervention platforms are among the most established digital modalities for behavioral addiction management. They usually deliver structured, session-based content. This content is often based on CBT and motivational approaches. Compared to mobile apps, web platforms tend to provide longer modules and more complete therapy sequences. They are also easier to standardize and scale.

Web-based programs commonly include psychoeducation, cognitive restructuring, coping skills, relapse-prevention planning, and goal tracking. Many platforms add therapist guidance or brief support contacts. Evidence suggests that guided formats generally show better adherence than fully self-guided programs. This is important in behavioral addictions, where dropout is frequent.

\begin{figure*}[ht]
	\centering
	\includegraphics[width=\textwidth]{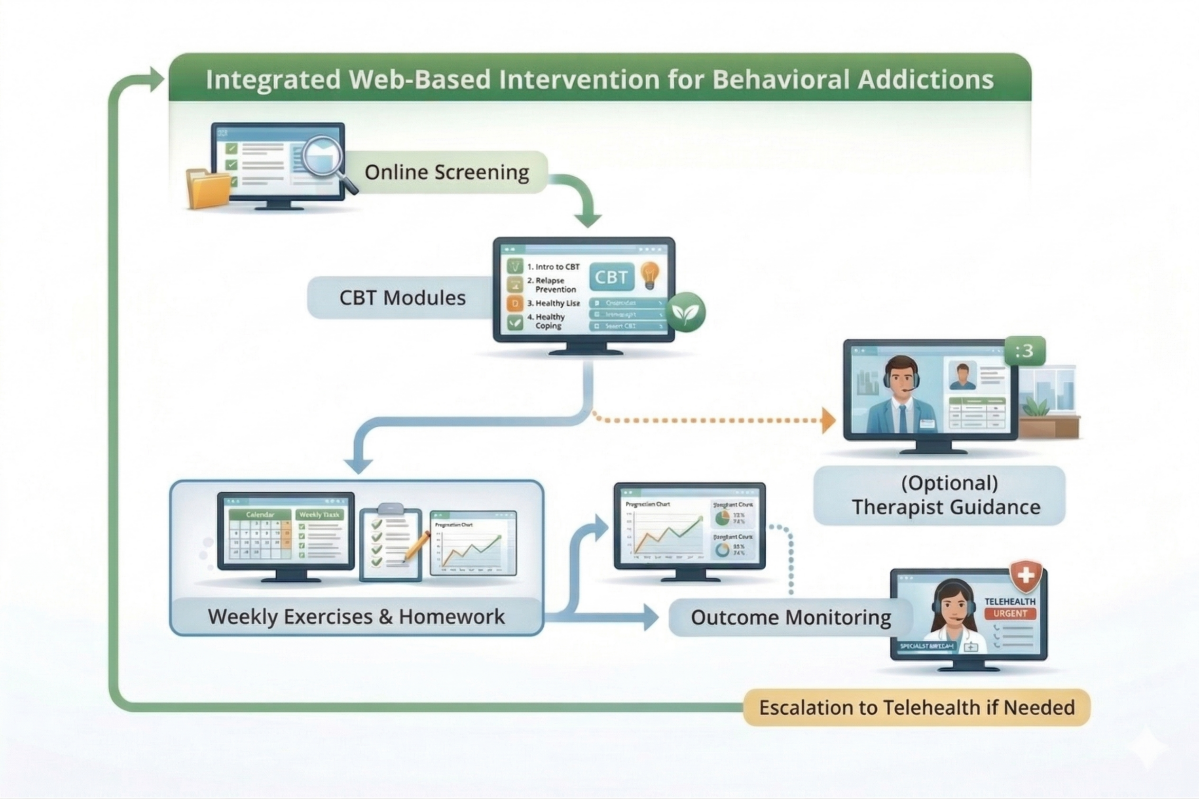}
	\caption{Workflow of web-based intervention platforms for behavioral addictions}
	\label{fig:webbased}
\end{figure*}

Figure~\ref{fig:webbased} presents a typical workflow of web-based intervention platforms used in behavioral addictions. The process often starts with online screening and then delivers structured CBT-based modules with exercises and relapse-prevention planning. Progress tracking and symptom monitoring are commonly integrated. Many platforms add brief therapist guidance to support adherence. Escalation pathways (e.g., telehealth referral) are used when risk increases or symptoms persist.

Evidence in gambling disorder indicates that web-based CBT can reduce symptom severity, but outcomes depend on intervention format and engagement. For example, a number of studies tested an internet-delivered CBT program with telephone guidance against an internet-delivered motivational enhancement intervention with motivational interviewing, in treatment-seeking individuals with gambling disorder \cite{mide2023internet,helvich2025gambling}. A separate self-guided internet intervention trial (N=150) examined feasibility and effectiveness, with improvements observed over the intervention period but no clear advantage over the control condition, highlighting adherence and study-design issues that often arise in unguided formats\cite{bucker2021randomized}. Implementation evidence also shows that therapist-supported iCBT can be embedded in routine addiction care (comparative cohort design), supporting real-world scalability\cite{molander2024introducing}. Meta-analytic evidence confirms overall beneficial effects of online psychological interventions for problem gambling/gambling disorder, while also emphasizing heterogeneity and engagement limitations\cite{augner2022psychological}.
For gaming disorder and related problematic internet use, structured web-based or digitally delivered CBT-based programs have shown sustained symptom reduction in prevention/early-risk contexts \cite{panova2021internet}. The PROTECT cluster randomized clinical trial in at-risk adolescents reported a greater reduction in symptom severity over 12 months in the intervention group (39.8\% vs 27.7\%; Cohen’s d=0.67), while incidence-rate differences were not statistically significant\cite{lindenberg2022effectiveness}. Some findings from web-based intervention studies are presented in Table~\ref{tab:web_platforms}.

\begin{table*}[ht]
	\centering
	\small
	\caption{Web-based intervention platforms in behavioral addictions}
	\label{tab:web_platforms}
	\begin{tabular}{p{1cm} p{3.5cm} p{3cm} p{3cm} p{3cm}}
		\toprule
		\textbf{Study} & \textbf{Population \& Design} & \textbf{Target behavior} & \textbf{Key findings} & \textbf{Interpretation} \\
		\midrule
		\cite{mide2023internet} & Single-blind RCT; n=71; iCBT+telephone guidance vs internet-delivered motivational enhancement + MI & Gambling disorder & Two active digital treatments compared with repeated symptom assessments & Guided iCBT is a strong clinical web-based example; supports structured delivery \\
		\\
		\cite{bucker2021randomized} & RCT; N=150; self-guided internet intervention vs control; baseline--post & Gambling problems & Improvements observed; between-group advantage not clearly demonstrated & Highlights adherence and control-condition challenges in unguided web formats \\
		\\
		\cite{molander2024introducing} & Routine addiction care; therapist-supported iCBT; comparative cohort & Gambling disorder & Implementation in routine care with outcome monitoring reported & Real-world scalability and clinical integration evidence \\
		\\
		\cite{augner2022psychological} & Systematic review + meta-analysis (2010--2021) & Problem gambling / GD & TCC: Hedges g=0.41; PCC: g=1.28 & Overall benefit; heterogeneity and engagement remain key issues \\
		\\
		\cite{lindenberg2022effectiveness} (PROTECT) & Cluster RCT; n=422 at-risk adolescents; 12-month follow-up & Gaming disorder / unspecified IUD & Symptom reduction 39.8\% vs 27.7\%; Cohen's d=0.67; incidence ns & Supports prevention/early-intervention role of structured digital CBT \\
		\bottomrule
	\end{tabular}
\end{table*}

\subsection{AI-powered DTx}
DTx aim to move digital care from “static content delivery” to “adaptive treatment”. In behavioral addictions, risk states change fast. Craving, cue exposure, loss-of-control, and relapse vulnerability are not stable. AI methods can model these changes using digital traces (e.g., app logs, screen-time), EMA responses, and sometimes sensor signals. This supports earlier risk detection and more precise timing of support.

\begin{figure}[!t]
	\centering
	\includegraphics[width=0.7\textwidth]{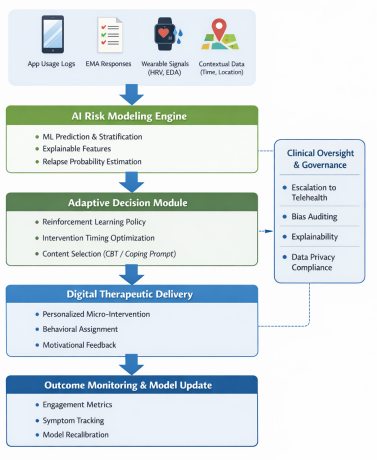}
	\caption{AI-driven closed-loop digital therapeutic architecture for behavioral addictions}
	\label{fig:AIbased}
\end{figure}

Figure~\ref{fig:AIbased} shows that AI-powered DTx operate across both the risk-detection and intervention-optimization layers. This figure illustrates a closed-loop AI-driven digital therapeutic architecture for behavioral addictions. Multimodal behavioral and contextual data feed a machine learning–based risk modeling engine that estimates vulnerability states. A decision module, potentially supported by RL, optimizes intervention timing and content selection. Personalized therapeutic elements are then delivered digitally, while engagement and outcome data are continuously monitored to update the model. Clinical oversight and governance mechanisms remain essential to ensure transparency, safety, and appropriate escalation pathways.

A key use-case is prediction and stratification. Explainable machine learning has been used to predict internet addiction status and to identify influential psychosocial features in large longitudinal youth datasets \cite{liu2025explainable}. In gambling, machine-learning models using real-world player tracking data can predict self-reported problem gambling and highlight behavioral predictors such as deposit patterns and session features \cite{auer2026using}. These studies support the role of AI as a risk-detection layer that can inform stepped-care and referral decisions.
The other use-case is adaptive intervention optimization, often framed as JITAI. Gambling JITAIs operationalize “right support, right time” by combining repeated EMA decision points with rule-based tailoring. For example, GamblingLess: In-The-Moment delivers tailored CBT/third-wave micro-interventions when users report cognitive vulnerability (high craving, low self-efficacy, or positive outcome expectancies) via EMAs administered three times per day in a 28-day microrandomized trial design \cite{dowling2022gambling}. Similar logic is used in Gambling Habit Hacker, another gambling-focused JITAI protocol evaluated via a 28-day microrandomized trial \cite{rodda2022smartphone}. Acceptability work in this line also suggests that dynamic “in-the-moment” delivery is feasible and can be evaluated using usage metrics and app quality scales \cite{rodda2025acceptability}.

AI can also be used as the decision engine for personalization (beyond static rules). RL has been proposed to learn intervention timing and content selection policies under changing contexts, which conceptually fits DTx “precision dosing” \cite{gonul2021reinforcement}. At the system level, a recent DTx framework proposes where AI can be embedded across therapeutic units, decision-making, and support components, rather than only in chat interfaces \cite{baumel2025more}. Still, addiction-focused AI research remains fragmented, and many studies are predictive rather than treatment-outcome trials. Reviews emphasize the need for transparent models, bias checks, and clinical validation before deployment at scale \cite{suva2024artificial}. Some findings from representative AI-supported DTx work are presented in Table~\ref{tab:ai_dtx}.

\begin{table*}[ht]
	\centering
	\small
	\caption{AI-powered DTx and AI-enabled personalization relevant to addiction-focused digital care.}
	\label{tab:ai_dtx}
	\begin{tabular}{p{1cm} p{3.5cm} p{3cm} p{3cm} p{3cm}}
		\toprule
		\textbf{Study} & \textbf{Population \& Design} & \textbf{Target
			behavior} & \textbf{Key findings} & \textbf{Interpretation} \\
		\midrule
		\cite{liu2025explainable} & Longitudinal youth data (2019--2022); n=8{,}824; explainable ML; IA via IAT$\ge$40 & Internet addiction & 33.3\% classified with IA; explainable ML used for prediction & Risk stratification and interpretability \\
		\\
		\cite{auer2026using} & Real-world online gamblers; N=1{,}611; player tracking and ML & Problem gambling & Logistic regression / random forest among best; predictors include deposits and session features & Digital traces for early risk detection \\
		\\
		\cite{dowling2022gambling} & Gambling JITAI protocol; 28-day MRT; EMA 3$\times$/day & Gambling reduction & User testing: MARS subscales $>$3/5; ease $>$8/10; helpfulness $>$7/10 & Monitoring, vulnerability, and micro-intervention \\
		\\
		\cite{rodda2022smartphone} & Gambling Habit Hacker JITAI protocol; 28-day MRT & Gambling limits & In-the-moment BCT delivery (planning, coping, self-monitoring) & Structured JITAI for adherence support \\
		\\
		\cite{gonul2021reinforcement} & RL-based personalization algorithm; timing/frequency/type optimization & Adaptive DTx logic & RL learns opportune moments and intervention selection policies & AI decision engine for precision dosing \\
		\bottomrule
	\end{tabular}
\end{table*}

An emerging direction in AI-powered DTx is their potential to approximate prescriptive treatment logic rather than just adaptive support. In substance use disorders, for example, FDA-authorized prescription DTx such as reSET-O (for opioid use disorder) demonstrate that rigorous clinical validation and regulatory oversight are possible for digital treatments. Although no analogous prescription DTx currently exists for behavioral addictions, these models establish a standard for clinical effectiveness and safety that AI-driven systems should aspire to meet. Embedding explainable AI, bias auditing, and outcome validation within DTx pipelines can support not only personalized intervention timing but also clinically robust treatment delivery, bridging the gap between predictive modeling and therapeutic efficacy.

\subsection{VR/AR exposure therapy}
Virtual Reality (VR) and Augmented Reality (AR)–based exposure therapies extend digital intervention modalities by immersing users in controlled, realistic simulations of addiction-related contexts. Unlike standard digital platforms, VR/AR provides ecological fidelity to environmental triggers, which can be used to elicit and modify cue reactivity and craving in safe settings. Immersive environments allow users to practice coping strategies and desensitization under quasi-real conditions that are difficult to recreate in traditional exposure paradigms.

Evidence from systematic reviews suggests that VR exposure can reduce craving and engage multiple sensory channels, enhancing habituation and emotional regulation compared with non-immersive approaches \cite{taubin2023systematic,glavak2025virtual}. Although most VR research has focused on substance use disorders, this work is relevant for behavioral addictions because cue reactivity and craving are core maintenance mechanisms in gambling and internet gaming disorder. Preliminary studies of VR cue reactivity indicate that immersive gambling environments can evoke real-world physiological and psychological responses, supporting the potential of VR for both assessment and treatment\cite{park2015effect}. A recent scoping review protocol highlights growing interest in VR/AR therapies for behavioral addictions, including gambling and gaming disorders, emphasizing the need for systematic clinical evaluation \cite{chiew2025virtual}.
VR interventions may combine cue exposure with CBT-based coping skills training, guided therapeutic feedback, and interactive desensitization tasks.

\begin{figure*}[htbp]
	\centering
	\includegraphics[width=0.5\textwidth]{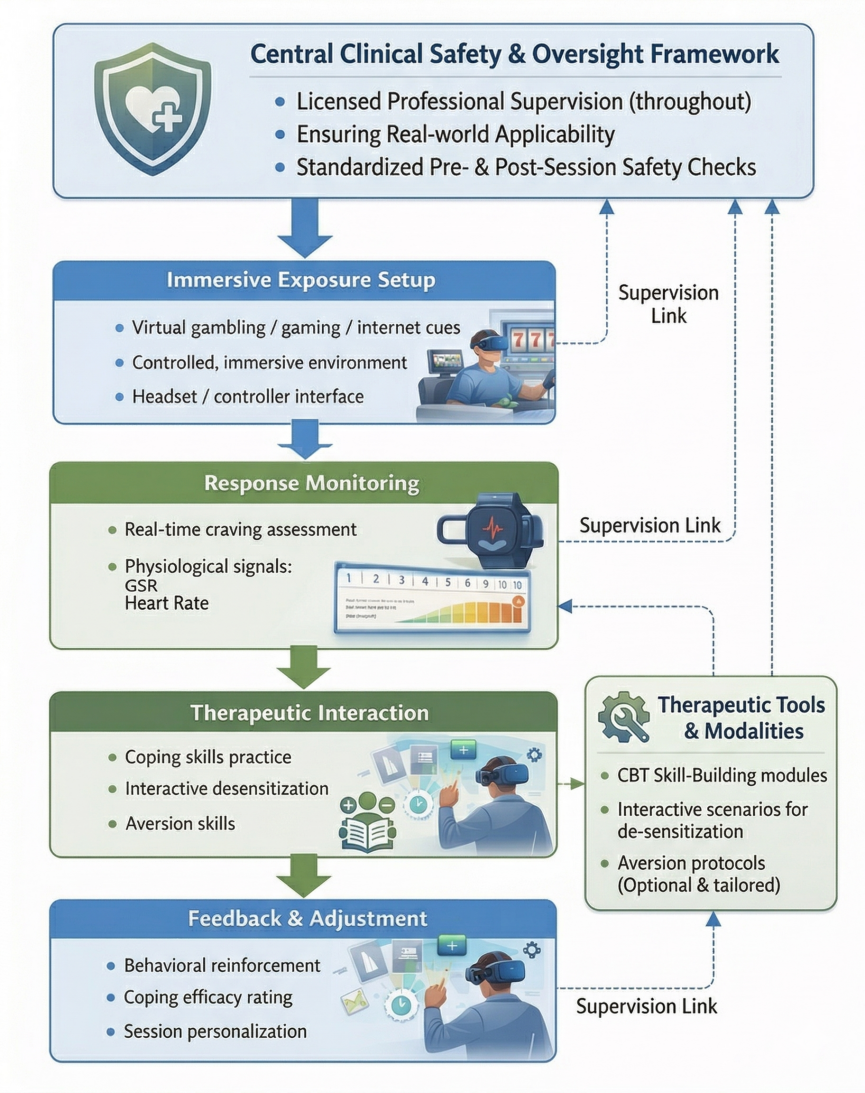}
	\caption{VR/AR exposure therapy framework for behavioral addictions}
	\label{fig:vr_ar}
\end{figure*}

The conceptual workflow of VR/AR-based exposure therapy is illustrated in Figure~\ref{fig:vr_ar}. This figure illustrates a structured VR/AR exposure therapy framework for behavioral addictions. Immersive environments present addiction-related cues (e.g., gambling or gaming stimuli) under controlled conditions. Real-time craving and physiological responses are monitored during exposure. Therapeutic interaction integrates CBT-based coping skills, desensitization tasks, and optional aversion components. Feedback mechanisms support behavioral reinforcement and session-level personalization. Clinical oversight ensures safety, supervision, and appropriate escalation when risk increases.
Early evidence also indicates that multi-session VR cue exposure, particularly when integrated with aversion or refusal skills components, can reduce craving intensity and improve coping self-efficacy in substance use disorders, a model that could be adapted for behavioral addiction cues \cite{huang2025virtual}. However, heterogeneity in intervention design, small sample sizes, and limited follow-up data constrain firm efficacy conclusions. Standardization of intervention parameters, systematic outcome measurement, and larger controlled trials are needed to translate VR/AR exposure therapy into clinical practice. A number of studies relevant to immersive exposure in addictive behaviors are summarized in Table~\ref{tab:vr_ar}.

\begin{table*}[ht]
	\centering
	\small
	\caption{VR/AR exposure therapy and immersive interventions relevant to addictive behaviors.}
	\label{tab:vr_ar}
	\begin{tabular}{p{1cm} p{3.5cm} p{3cm} p{3.5cm} p{3cm}}
		\toprule
		\textbf{Study} & \textbf{Population \& Design} & \textbf{Target behavior} & \textbf{Key findings} & \textbf{Interpretation} \\
		\midrule
		\cite{taubin2023systematic} & Systematic review; 7 VR studies & Substance use cue reactivity & Majority show reduced craving; mixed mood outcomes & VR exposure promising for craving modulation \\
		\\
		\cite{glavak2025virtual} & Broad review of VR programs & SUD & VR shows craving/use reduction; heterogeneity & Immersive exposure could translate to BA settings \\
		\\
		\cite{huang2025virtual} & RCT; VR cue exposure ± aversion; n=89 & MUD & Tonic craving, post-VR exposure; refusal self-efficacy & VR cue exposure can reduce craving and support self-regulation \\
		\\
		\cite{chiew2025virtual} (protocol) & Scoping review protocol & Behavioral addictions & Protocol to map VR/AR literature for gaming/gambling & Highlights need for systematic evaluation in BA \\
		
		\bottomrule
	\end{tabular}
\end{table*}

\subsection{Chatbots \& conversational agents}

Chatbots and conversational agents represent an increasingly visible modality within digital addiction management. These systems simulate structured dialogue using rule-based scripts or artificial intelligence models. In behavioral addictions, they are primarily used for psychoeducation, motivational enhancement, craving management, and brief cognitive restructuring.

From a theoretical perspective, conversational agents target impaired self-regulation, maladaptive beliefs, and ambivalence toward behavior change. By delivering structured prompts and reflective feedback, chatbots can operationalize motivational interviewing principles and CBT-based micro-interventions in a scalable format\cite{ccakmak2025effect}. Unlike static web modules, conversational interfaces provide interactive engagement, which may enhance perceived personalization and immediacy.

\begin{figure*}[htbp]
	\centering
	\includegraphics[width=\textwidth]{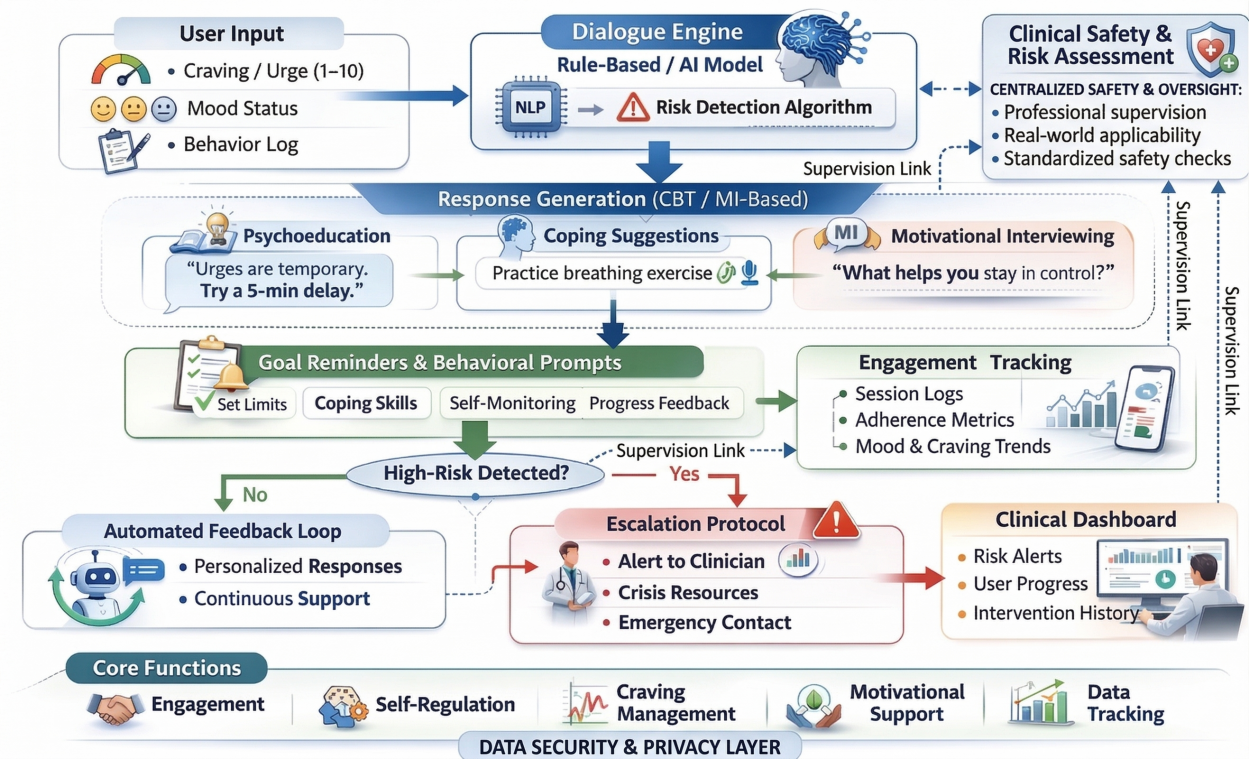}
	\caption{The conceptual architecture of an AI-driven chatbot system for behavioral addiction management}
	\label{fig:chatbot}
\end{figure*}

The conceptual workflow of chatbot-based support systems is illustrated in Figure~\ref{fig:chatbot}. This figure presents a structured conversational agent model for behavioral addiction support. User-reported inputs such as craving intensity, mood status, and behavioral logs are processed through a dialogue engine based on rule-based logic or artificial intelligence. Risk detection algorithms identify potential high-risk states. The system generates CBT- or motivational interviewing–based responses, including coping strategies, goal reminders, and motivational feedback. Engagement metrics and outcome indicators are continuously monitored. High-risk cases trigger escalation pathways involving clinician notification or crisis support. Data privacy, transparency, and clinical oversight remain integral components of the architecture.

Emerging evidence from mental health research suggests that AI chatbots can provide feasible, scalable support for emotional regulation and behavior change. A recent systematic review identified that chatbot-assisted interventions often aim to change cognitive and behavioral patterns, with intervention durations varying from short (e.g., 2–8 weeks) to longer formats (e.g., 6 months) \cite{lee2024systematic}. Another systematic work highlights that conversational agents can enhance diagnostic accuracy and support treatment planning in broader healthcare contexts \cite{huynh2025applications}. Specific work on digital conversational agents for youth mental health reported high acceptability and engagement, although clinical effects are still emerging and evidence remains limited \cite{hawke2025digital}.
The historical chatbot programs (e.g., GAMBOT) showed feasibility for reducing gambling severity in controlled cohorts, although replication and rigorous RCTs are sparse \cite{so2020unguided}. Research in substance use and smoking cessation also supports the potential of conversational agents to facilitate behavior change, suggesting a route for further adaptation to gambling and gaming disorders. When integrated with risk detection systems (e.g., usage patterns, EMA inputs), conversational agents may serve as the user-facing delivery channel for adaptive DTx, providing immediate, interactive responses based on predictive risk states. Selected findings from chatbot intervention research are summarized in Table~\ref{tab:chatbot}.

\begin{table*}[ht]
	\centering
	\small
	\caption{Chatbots and conversational agents in behavioral health and addiction contexts.}
	\label{tab:chatbot}
	\begin{tabular}{p{1cm} p{3.5cm} p{3cm} p{3.5cm} p{3cm}}
		\toprule
		\textbf{Study} & \textbf{Population \& Design} & \textbf{Target behavior} & \textbf{Key findings} & \textbf{Interpretation} \\
		\midrule
		
		\cite{lee2024systematic} &
		Systematic review; 28 chatbot programs &
		Substance use and mental health &
		Most chatbots aimed at cognitive/behavior change; varied duration (2 weeks–6 months) &
		Feasible and diverse applications in behavior change \\
		\\
		\cite{huynh2025applications} &
		Review of AI conversational agents in healthcare &
		Broad clinical settings &
		Improved diagnostic and treatment support capabilities reported &
		Potential support tool in addiction care pathways \\
		\\
		\cite{hawke2025digital} &
		Review of digital agents for youth; feasibility and acceptability studies &
		Mental health and substance use contexts &
		High acceptability; early symptom improvement trends &
		Engagement promising; long-term effects unclear \\
		\\
		GAMBOT \cite{so2020unguided} &
		Controlled messaging chatbot; n$\approx$197 &
		Problem gambling &
		Reduced self-reported gambling severity vs assessment-only &
		Early proof-of-concept in gambling support \\
		\\
		\cite{ogilvie2025overview} &
		Narrative review of chatbot use in substance use disorders &
		Substance use disorders &
		Limited high-quality RCT evidence; underutilized clinically &
		Need for rigorous addiction-specific trials \\
		
		\bottomrule
	\end{tabular}
\end{table*}

\subsection{Telehealth \& tele-psychiatry}

Telehealth and tele-psychiatry extend addiction treatment beyond physical clinical settings through synchronous (video-based) and asynchronous (messaging-based) communication systems. In behavioral addictions, telehealth provides structured psychotherapy, psychiatric evaluation, medication management when indicated, and relapse monitoring through remote platforms. This modality addresses accessibility barriers, stigma concerns, and geographic limitations that frequently delay treatment initiation.

\begin{figure*}[htbp]
	\centering
	\includegraphics[width=\textwidth]{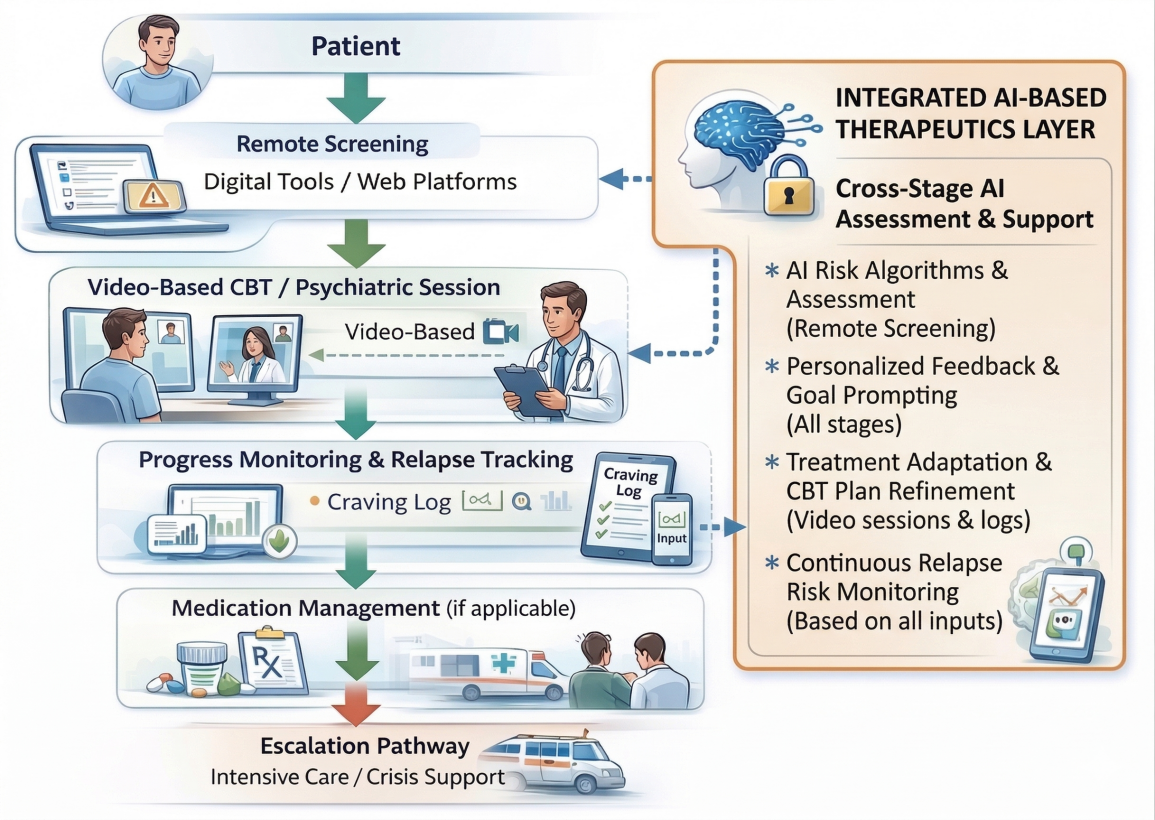}
	\caption{Telehealth and hybrid digital care model for behavioral addictions}
	\label{fig:telehealth}
\end{figure*}

From a mechanistic perspective, tele-psychiatry supports behavioral addiction management by maintaining therapeutic alliance and structured CBT delivery while enabling frequent follow-up. Regular remote contact may reduce relapse vulnerability by increasing accountability and reinforcing coping strategies during high-risk periods. Compared to fully automated digital tools, telehealth preserves clinician involvement, which is particularly relevant in moderate-to-severe gambling disorder. The telehealth workflow in behavioral addiction care is presented in Figure~\ref{fig:telehealth}. This figure shows a hybrid telehealth care model for behavioral addictions. Digital tools are used for remote screening and preliminary monitoring. Structured video-based CBT sessions and psychiatric consultations provide clinician-guided treatment. Ongoing progress tracking and relapse monitoring support continuity of care. Medication management may be integrated when clinically indicated. Escalation pathways enable referral to intensive or crisis services in high-risk cases. Integration with AI-based risk detection systems can enhance personalization while preserving clinical oversight.

Recent literature demonstrates that telehealth is a viable alternative to in-person treatment across a range of mental health and substance use outcomes. Systematic reviews indicate that telehealth does not significantly differ from face-to-face modalities in symptom reduction and quality of life measures for general mental health conditions, suggesting transferability to addiction contexts \cite{mark2022addiction}. In addiction treatment specifically, telemedicine use increased significantly during and after the COVID-19 pandemic, improving access to outpatient services without increasing adverse outcomes such as relapse or crisis events \cite{pusnik2024comparison}. Telehealth also improved treatment engagement and compliance, especially in rural or underserved populations \cite{qeadan2025impact}. Emerging work on telehealth for gambling disorder suggests it may expand specialized treatment for problem gambling nationwide, though rigorous controlled trials remain limited and require proper training and infrastructure for effective implementation \cite{smith2022telehealth}. Hybrid models that combine digital screening tools, AI risk detection, and tele-psychiatry consultations may further enhance personalization and clinical oversight. Some studies on the application of telehealth in addiction and mental health care are summarized in Table~\ref{tab:telehealth}.

\begin{table*}[ht]
	\centering
	\small
	\caption{Telehealth and tele-psychiatry in addiction and related behavioral health care}
	\label{tab:telehealth}
	\begin{tabular}{p{1cm} p{3.5cm} p{3cm} p{3.5cm} p{3cm}}
		\toprule
		\textbf{Study} & \textbf{Population \& Design} & \textbf{Target behavior} & \textbf{Key findings} & \textbf{Interpretation} \\
		\midrule
		
		\cite{mark2022addiction} &
		Systematic review; telehealth addiction services &
		Substance use disorders \& mental health &
		Telehealth comparable to in-person care in retention and symptom outcomes &
		Telehealth is a viable alternative to face-to-face addiction treatment \\
		\\
		\cite{pusnik2024comparison}&
		Review of telehealth utilization in addiction services &
		Addiction treatment delivery &
		Improved engagement and provider confidence; increased access to care &
		Supports continuity and scalability of remote addiction care \\
		\\
		\cite{qeadan2025impact} &
		Retrospective EHR analysis; multi-center outpatient data &
		Substance use disorders &
		Higher outpatient visit rates without increase in adverse clinical events &
		Telehealth sustainable with maintained safety outcomes \\
		\\
		\cite{smith2022telehealth} &
		Narrative review; COVID-era telehealth in gambling disorder &
		Gambling disorder &
		Expanded access to specialist treatment; feasibility noted &
		Promising for gambling care; further controlled trials needed \\

		\bottomrule
	\end{tabular}
\end{table*}

\subsection{Community platforms \& peer-support systems}
Community platforms and peer-support systems constitute an important social dimension of digital addiction management. These platforms include moderated online forums, structured peer-support networks, recovery communities, and social media–based support groups. In behavioral addictions, they primarily address social isolation, shame, and relapse vulnerability by providing shared experiences and accountability.

\begin{figure*}[htbp]
	\centering
	\includegraphics[width=0.5\textwidth]{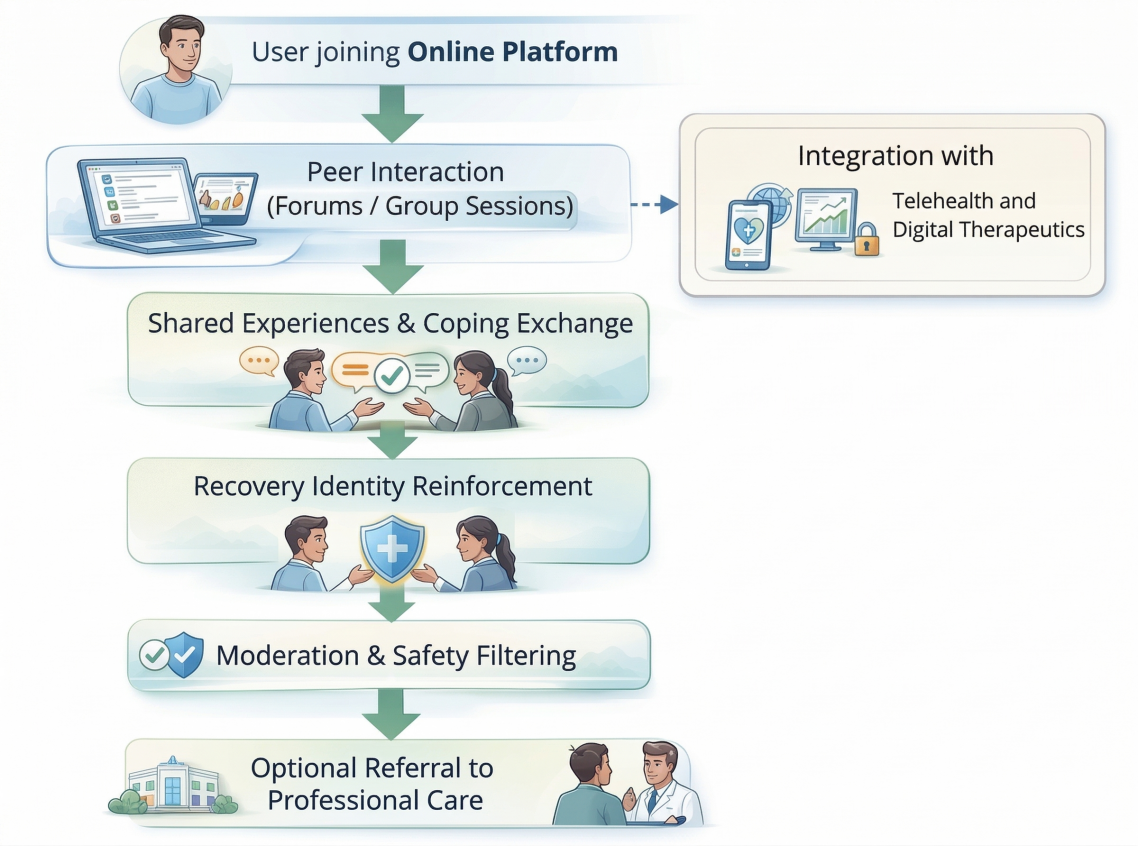}
	\caption{Digital peer-support and community-based recovery model for behavioral addictions}
	\label{fig:community}
\end{figure*}

From a theoretical perspective, peer-support systems operationalize social learning and recovery capital frameworks. Observational learning, normalization of relapse struggles, and reinforcement of recovery-oriented identity may strengthen motivation and self-efficacy. In gambling disorder and problematic digital media use, online communities often serve as first contact points before formal treatment engagement. The conceptual architecture of digital peer-support systems in behavioral addictions is presented in Figure~\ref{fig:community}. This model emphasizes the role of moderated interaction, recovery identity reinforcement, and integration with formal care pathways. Individuals enter moderated online platforms where structured peer interaction occurs through forums or group-based exchanges. Shared experiences and coping strategy discussions reinforce recovery-oriented identity and enhance perceived social support. Moderation and safety filtering mechanisms aim to prevent misinformation and triggering content. When risk levels increase, referral pathways enable connection to professional care. The model highlights the complementary role of community platforms within hybrid digital addiction care systems.

Online community platforms and peer-support systems represent an important social layer within digital behavioral addiction management. Earlier research has shown that peer-based online interventions can improve engagement and strengthen perceived coping capacity in addiction and mental health settings \cite{naslund2016future,prescott2020online,bergman2021online}, while structured digital group models such as the COPE.er framework demonstrate how moderated peer environments can be intentionally designed for problematic technology use \cite{alrobai2019cope}. More recent evidence further consolidates this view. A 2024 scoping review of web-based forums for substance use and gambling disorders indicates that such platforms often function as early access points for individuals who have not yet entered formal treatment \cite{peart2024web}. Qualitative analyses of gambling forums show that users openly discuss stigma, discrimination, and relapse-related struggles, highlighting the capacity of online communities to capture lived experiences that may remain underreported in clinical settings \cite{penfold2024gambling}. In addition, large-scale NLP analyses of online gambling discussions reveal heterogeneous thematic and emotional patterns \cite{van2025online}, suggesting both opportunities for scalable harm-reduction communication and the need for active moderation to prevent normalization of risky behaviors. Beyond informal forums, systematic reviews of peer recovery support services report consistent improvements in treatment engagement and retention, although evidence for direct behavioral outcome changes remains mixed \cite{eddie2025peer,kelly2025one}. Table~\ref{tab:community} presents selected findings from recent empirical studies addressing community platforms and peer-support systems in addiction contexts.

\begin{table*}[ht]
	\centering
	\footnotesize
	\caption{Some community platforms and peer-support systems}
	\label{tab:community}
	\begin{tabular}{p{1cm} p{3.5cm} p{3cm} p{3.5cm} p{3cm}}
		\toprule
		\textbf{Study } & \textbf{Population \& Design} & \textbf{Target behavior} & \textbf{Key findings} & \textbf{Interpretation} \\
		\midrule
		
		\cite{peart2024web} &
		Scoping review; 14 primary studies; database search and citation updates through Feb 2024 &
		Web-based peer forums (SUD/gambling) &
		Forums often alcohol-focused; where reported, users typically women in midlife; participation described quantitatively and qualitatively &
		Defines who engages and how; supports forum design and implementation \\
		\\
		\cite{penfold2024gambling} &
		Qualitative naturalistic forum analysis; data from three gambling-harm self-help forums &
		Gambling harm / stigma &
		Forum narratives used to characterize lived experiences of stigmatisation and discrimination &
		Community platforms capture sensitive barriers to care; informs stigma-aware support \\
		\\
		\cite{van2025online} &
		NLP analysis of r/onlinegambling &
		Online gambling communities &
		1,141 posts and 11,668 comments (2015--2023); sentiment shows positive, neutral, and negative content; varied harm-related themes &
		Online communities are potential harm-reduction channels; content governance is important \\
		\\
		\cite{eddie2025peer} &
		Systematic review (updated); quantitative multi-group PRSS studies &
		Peer recovery support services (SUD) &
		K=28; N=12,601; evidence supports improved engagement/retention; substance-use outcomes preliminary/inconclusive &
		Structured peer supports can strengthen care continuity (transferable logic to BA) \\
		\\
		\cite{kelly2025one} &
		Systematic review of recovery community centers (RCCs) &
		Recovery support infrastructure (SUD) &
		Evidence suggests promising associations with reduced use/improved well-being, but major methodological limitations; calls for comparative effectiveness research &
		Highlights potential value of community hubs; stronger study designs needed \\
		
		\bottomrule
	\end{tabular}
\end{table*}

Overall, the literature suggests that moderated community platforms and structured peer-support systems can play a supportive role in digital behavioral addiction care. They should be integrated into broader hybrid care models and used alongside professional treatment, not as a replacement for clinical services.

\subsection{Gamification-based interventions}
\begin{figure*}[htbp]
	\centering
	\includegraphics[width=0.7\textwidth]{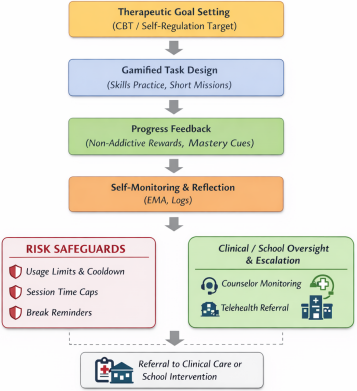}
	\caption{Therapeutic gamification model for behavioral addictions}
	\label{fig:gamification}
\end{figure*}

Gamification-based interventions have gained attention as supportive components in digital behavioral addiction management. Gamification refers to the integration of game design elements, such as points, levels, challenges, progress visualization, and feedback loops, into non-game therapeutic contexts. In behavioral addictions, these elements are typically embedded within self-monitoring systems, relapse-prevention modules, or digital CBT frameworks to increase engagement and adherence rather than to function as independent treatment modalities. Gamification mechanisms overlap with RL principles and self-determination theory. Structured feedback, incremental goal progression, and reward signals may enhance perceived competence and task persistence. However, in behavioral addictions, particularly gambling disorder and gaming disorder, reward sensitivity and reinforcement vulnerability are already central mechanisms. Therefore, gamification must be carefully designed to avoid replicating compulsive reward cycles.
The therapeutic logic of gamification-based interventions in behavioral addictions is presented in Figure~\ref{fig:gamification}. The process begins with therapeutic goal setting based on cognitive-behavioral and self-regulation targets. Gamified tasks are then introduced to support skills practice and short, structured missions. Progress feedback emphasizes mastery cues and non-addictive rewards rather than high-intensity reinforcement. Self-monitoring components, such as EMA and usage logs, promote reflection and behavioral awareness. Risk safeguards, including usage limits and session caps, aim to prevent overstimulation and compulsive engagement. Clinical or school-based oversight ensures monitoring and escalation when needed. The model highlights that gamification should reinforce therapeutic goals rather than replicate addictive reward mechanisms.

In gambling disorder, serious games have been proposed as adjunctive clinical tools. The e-Estesia serious game was developed to reduce arousal, improve emotion regulation, and increase well-being in individuals with gambling disorder \cite{mena2022estesia}. Gamification principles are also relevant for problematic digital media use and screen-time reduction. A randomized field experiment tested digital strategies for reducing screen time and provided evidence that structured digital prompts and behavior-change mechanisms can reduce use in real-world settings \cite{zimmermann2023digital}. In parallel, school-based prevention programs are becoming more evidence-driven. A recent systematic review and meta-analysis reported that school-based interventions can reduce problematic digital technology use and screen time, although heterogeneity across programs remains \cite{vzmavc2025effectiveness}. These findings support the role of gamified or game-like elements in prevention and early intervention, especially when integrated into structured educational settings. An updated systematic review summarized clinical treatments for internet gaming disorder and highlighted that intervention content and adherence remain central challenges, which strengthens the rationale for carefully designed engagement strategies, including gamified components, within clinical programs \cite{zajac2020treatments}. Table~\ref{tab:gamification} summarizes recent evidence on gamified approaches in behavioral addictions and problematic digital technology use \cite{mena2022estesia,zimmermann2023digital,vzmavc2025effectiveness,zajac2020treatments}.

\begin{table*}[ht]
	\centering
	\footnotesize
	\caption{Gamification and serious-game evidence relevant to behavioral addictions }
	\label{tab:gamification}
	\begin{tabular}{p{1cm} p{3.5cm} p{3cm} p{3.5cm} p{3cm}}
		\toprule
		\textbf{Study } & \textbf{Population \& Design} & \textbf{Target behavior} & \textbf{Key findings } & \textbf{Interpretation} \\
		\midrule
		
		\cite{mena2022estesia} &
		Clinical sample; CBT vs. CBT and e-Estesia (adjunct) &
		Gambling disorder &
		Between-group differences were limited; adjunct arm reported fewer relapses and better compliance indicators &
		Serious games may support adherence/relapse-related outcomes as adjunct modules \\
		\\
		\cite{zimmermann2023digital} &
		Randomized field experiment; objective screen-time tracking &
		Screen-time reduction &
		Design-friction (e.g., grayscale) reduced screen time; goal-setting effects were smaller; no clear short-term gains in well-being &
		Friction-based and prompt-based designs can reduce use, but benefits may not generalize \\
		\\
		\cite{vzmavc2025effectiveness} &
		Systematic review \& meta-analysis; school-based programs &
		Problematic digital technology use / screen time &
		Stronger effects for problematic use than for screen time; substantial heterogeneity across programs &
		Prevention can work in schools, but outcomes depend on program design and delivery  \\
		\\
		\cite{zajac2020treatments} &
		Systematic review of IGD treatment studies &
		Internet gaming disorder &
		Evidence remains limited and methods vary; stronger trials and consistent outcomes are needed &
		Engagement strategies are relevant, but should be tested in higher-quality clinical studies \\
		
		\bottomrule
	\end{tabular}
\end{table*}

\clearpage
\newpage
\section{Functions and Use-cases enabled by Digital Systems}\label{sec4}

Digital systems enable a wide range of functional capabilities in the management of behavioral addictions, transforming traditionally episodic and clinic-centered care into continuous, data-driven, and adaptive processes. These systems support end-to-end workflows spanning screening, diagnosis, risk prediction, intervention, and long-term recovery \cite{torous2021growing,torous2025evolving,balhara2019behavior}. 

\begin{figure*}[htbp]
	\centering
	\includegraphics[width=\textwidth]{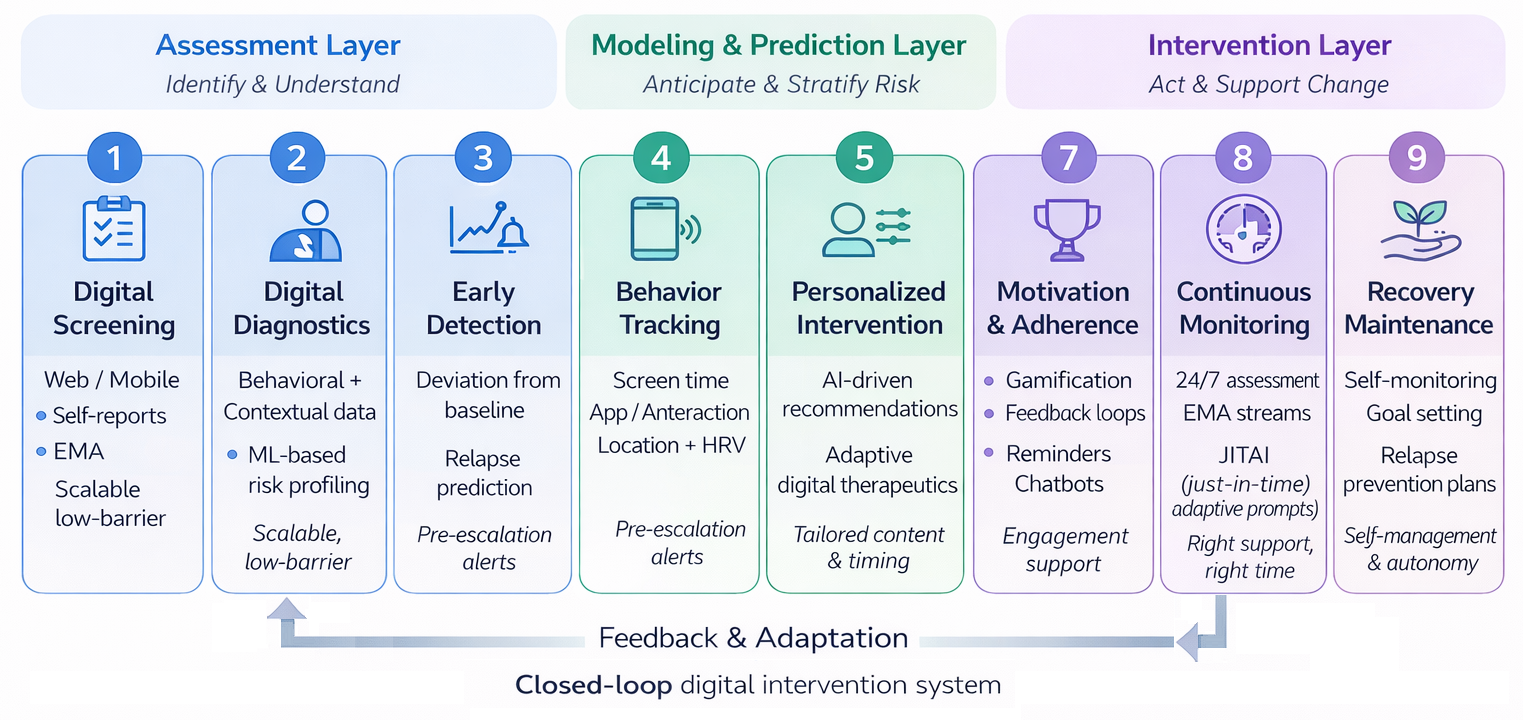}
	\caption{Functional framework of digital systems for behavioral addiction management.}
	\label{fig:functional_framework}
\end{figure*}

From a functional perspective, digital addiction management can be conceptualized as a set of interconnected layers that operate across assessment, modeling and prediction, and intervention domains. As illustrated in Fig.~\ref{fig:functional_framework}, these capabilities are integrated within a closed-loop framework, where multimodal data streams are continuously collected, analyzed, and utilized to dynamically adapt intervention strategies. This aligns with emerging paradigms in precision psychiatry and AI-driven healthcare systems \cite{topol2019high,baumel2025more,suva2024artificial}.

\subsection{Digital screening and assessment}

Digital platforms enable scalable screening through mobile and web-based systems, significantly improving accessibility and reducing barriers associated with traditional clinical assessment \cite{ferreri2018addictology,torous2025evolving}. 

A key advancement is EMA, which captures real-time data on craving, mood, and contextual triggers. EMA improves ecological validity and reduces recall bias compared to retrospective self-reports, enabling fine-grained temporal analysis of addictive behaviors \cite{toulami2025craving}. However, digital screening approaches remain dependent on user compliance and self-report accuracy. Engagement variability and missing data can limit reliability. These limitations motivate the integration of passive sensing and behavioral tracking. Overall, digital screening serves as a critical entry point in closed-loop systems, enabling early detection and continuous monitoring.

\subsection{Digital diagnostics \& risk stratification}

Digital diagnostics extend traditional assessment by integrating behavioral, contextual, and physiological data to estimate individual risk profiles. Machine learning models have demonstrated effectiveness in identifying addiction severity and predicting problematic usage patterns \cite{kato2023computational,liu2025explainable,auer2026using}. Risk stratification frameworks leverage multimodal features such as usage intensity, interaction patterns, and contextual signals to classify individuals into risk tiers. Unlike static diagnostic models, these systems continuously update risk estimates, enabling adaptive care pathways. Nevertheless, challenges remain regarding model generalizability, interpretability, and bias, particularly in heterogeneous populations. Addressing these limitations is essential for clinical deployment.

\subsection{Early detection and relapse prediction}

Relapse in behavioral addictions is a dynamic and context-dependent process influenced by cognitive, emotional, and environmental factors \cite{brand2019interaction,topbacs2025does}. 

Digital systems enable early detection by monitoring behavioral trajectories and identifying deviations from baseline patterns. Predictive models based on longitudinal data—including usage logs, physiological signals, and contextual features—have demonstrated strong potential in detecting relapse precursors \cite{wyant2024machine,yang2025relative,gezgin2025using}. However, predictive accuracy can be affected by data sparsity and variability across individuals. Additionally, false positives may lead to unnecessary interventions. Future work should focus on explainable and personalized prediction models. Early detection remains one of the most critical components for effective intervention and relapse prevention.

\subsection{Behavior tracking \& digital phenotyping}

Digital phenotyping enables continuous and passive monitoring of user behavior through smartphones and wearable devices. This approach provides objective, real-world behavioral data that complement self-reported measures \cite{montag2021overcome,kbah2025review}.

Behavior tracking systems collect multimodal signals such as screen time, app usage, mobility patterns, and physiological responses. Studies have shown that physiological markers such as HRV are associated with addiction-related processes and can support severity assessment \cite{hong2023severity,long2023heart}.

Despite these advantages, challenges related to data privacy, sensor reliability, and cross-device standardization persist. Ethical considerations and data governance frameworks are essential for large-scale implementation. Digital phenotyping forms the backbone of continuous monitoring and predictive modeling in addiction management systems.

\subsection{Personalized intervention \& adaptive care}

Digital systems support personalized interventions by adapting the content, timing, and mode of delivery to individual user profiles and behavioral patterns. In this context, AI-driven recommender systems and RL methods are increasingly used to optimize intervention strategies and improve the alignment between user needs and therapeutic responses \cite{isinkaye2015recommendation,chaudhari2024hybrid,gonul2021reinforcement}. In parallel, DTx integrate predictive modeling with intervention delivery within a unified framework, enabling adaptive treatment pathways that can respond to user behavior in near real time and continuously refine therapeutic decisions \cite{baumel2025more,shah2024cognitive}. Despite these advantages, personalization also raises important challenges related to transparency, fairness, explainability, and the risk of overfitting, making clinical validity and interpretability essential for safe and effective deployment. Overall, personalized intervention constitutes a critical step toward precision mental health.

\subsection{Harm reduction strategies}

Harm reduction approaches aim to minimize the negative consequences of addictive behaviors rather than enforce complete abstinence \cite{marlatt1996harm}. These strategies are particularly relevant in early-stage or high-risk populations.

Digital tools such as usage dashboards, real-time alerts, and application-blocking mechanisms support behavioral awareness and self-regulation \cite{giebel2023problems,torous2025evolving}. Nevertheless, the effectiveness of harm reduction strategies depends on user engagement and adherence. Passive interventions may be insufficient without behavioral reinforcement mechanisms. Harm reduction provides a pragmatic pathway toward gradual behavioral change.

\subsection{Motivation enhancement \& adherence support}

Sustained engagement is a major challenge in digital interventions. Gamification, feedback systems, and persuasive technologies are widely used to enhance motivation and adherence \cite{oyebode2024persuasive,torous2025evolving, ferreri2018addictology}. Conversational agents and chatbot-based systems have shown potential in supporting behavioral change and improving user engagement \cite{laranjo2018conversational,hawke2025digital,lee2024systematic}. However, long-term engagement remains difficult to sustain, with high dropout rates commonly reported. Designing adaptive and user-centered engagement strategies remains an open challenge.

\subsection{Continuous monitoring \& just-in-time adaptive interventions}

Continuous monitoring enables real-time assessment and supports JITAIs, which deliver interventions at moments of highest vulnerability \cite{nahum2025just,dowling2022gambling,rodda2022smartphone}. JITAIs leverage EMA data, behavioral tracking, and predictive models to dynamically determine intervention timing and content. This approach aligns intervention delivery with real-world user states, improving effectiveness. Despite promising results, challenges remain in optimizing decision policies and ensuring robustness in real-world settings.

\subsection{Recovery maintenance \& self-management}

Digital systems support long-term recovery through self-monitoring, goal setting, and relapse prevention strategies. Evidence from digital CBT and online interventions demonstrates their effectiveness in supporting sustained behavioral change \cite{molander2024introducing,augner2022psychological}. These systems promote self-efficacy and enable patient-centered care models. However, long-term adherence and integration with clinical workflows remain open challenges.

In summary, digital systems provide a unified, closed-loop framework for behavioral addiction management, integrating assessment, prediction, intervention, and recovery processes. This functional paradigm enables scalable, adaptive, and personalized mental health solutions aligned with emerging trends in digital psychiatry and artificial intelligence.

\newpage
\section{Data Sources}\label{sec5}
The rise of digital technologies has led to their rapid adoption in the medical field. This has brought significant changes in understanding diseases, developing treatment plans, and designing post-treatment follow-up plans. One of the major concepts that have been at the center of this transformation is digital phenotyping \cite{onnela2016harnessing}. Digital phenotyping refers to digital data generated from personal devices that can be used to assess and predict an individual’s health status \cite{jain2015digital}.

Data that constitute digital phenotypes can be generated in various ways. The information shared on social media platforms, posting times, and internet search history have the potential to provide insights into the mental state and behavioral patterns of individuals. In addition, information on an individual’s lifestyle, such as location tracking using mobile applications and smartphone usage history, is significant in developing digital phenotypes. Because it provides information on the behavioral patterns of individuals. With regard to this, smartphone applications used to monitor individuals’ health have the potential to provide information on individuals’ behavioral patterns. Moreover, wearable devices with sensors have the potential to monitor individuals’ behavioral and physiological characteristics. These digital platforms contribute significantly to the identification of digital phenotypes by providing multidimensional data on individuals’ daily lives \cite{loi2019digital}.

The process of measuring and analyzing the digital traces produced through the personal devices of an individual in real-time and naturalistic settings is referred to as digital phenotyping \cite{huckvale2019toward}. The digital phenotyping process begins with the acquisition of raw data from multiple digital sources. The next part is the processing and interpretation of the collected data. At this stage, the application of artificial intelligence technologies, particularly machine learning and deep learning, emerges as an important data analysis tool to extract relevant patterns from large and multidimensional data sets \cite{mendes2022sensing}. Machine learning helps in the prediction and learning process through statistical and mathematical modeling without explicit programming, using large volumes of data \cite{dos2024machine}. The application of these technologies helps in the prediction and, hence, plays an important role in the psychiatric practice \cite{dlima2022digital}. Artificial intelligence technologies help in the recognition of hidden patterns in the data using supervised, unsupervised, and deep learning techniques \cite{sarker2021deep}. Deep learning is an application of artificial intelligence and machine learning, which uses multiple layers of neural networks and data processing techniques to solve complex problems \cite{sarker2021machine}. Furthermore, NLP technologies help in the analysis of the textual data generated by the individual and provides relevant insights \cite{marengo2022mining}. These technological layers extend beyond facilitating administrative processes in healthcare, offering data-driven and proactive support for clinical decision-making. They enable the processing and analysis of multiple types of data, namely, time series data, textual data, and sensor data, and the feature extraction from the raw data, thus helping in the accurate modeling of the behavior of an individual. Especially the high-volume and continuous data generated through the application of the digital phenotyping process helps in the accurate prediction and classification performance of artificial intelligence algorithms \cite{dos2024machine}.

In the digital phenotyping process, collected data are generally categorized two main groups: active data and passive data \cite{bhugra2017wpa}. Active data involve data obtained from the direct involvement of participants in the research process, whereas passive data are collected automatically through background sensors and digital systems without requiring active user involvement \cite{oudin2023digital}.

Active data are typically collected through self-report-based assessments of mood, cognitive processes, and motor functioning. One major method of obtaining active data involves EMA \cite{onnela2016harnessing}. EMA involves allowing participants to report their feelings and behaviors in real-time during their daily activities \cite{shiffman2008ecological}. EMA has been considered a more reliable method of obtaining data, as opposed to self-report methods that rely on retrospective recall during clinical sessions \cite{nam2021ecological}.

EMA is typically administered via mobile devices, allowing broader accessibility \cite{battaglia2022use}. EMA involves three procedures: signal-contingent assessments (based on random prompts during the day), event-contingent assessments (triggered by specific events), and time-contingent assessments (conducted at predetermined intervals). EMA has been considered an effective method of obtaining data, including psychological states and addiction behaviors, in real-time during daily activities \cite{wray2014using, serre2015ecological}. The data obtained through EMA allow for the implementation of Ecological Momentary Interventions (EMIs) that offer support during times of need by tailoring the type and intensity of intervention, depending on the state of the participant \cite{heron2010ecological}.

The potential for passive data collection methods is substantial for psychiatric studies, as these allow for detailed monitoring of behavior and patterns in daily life \cite{jilka2024digital}. Passive data are collected through a multitude of sensor and measurement technologies that are available on digital technologies such as smartphones, tablets, smartwatches, sleep monitors, and fitness trackers \cite{heckler2025digital}. That data may include keyboard dynamics, accelerometer, gyroscope, GPS, time spent on the screen, temperature, light exposure, number of times a phone is checked, level of physical activity, number of messages and emails sent, and time and number of phone calls \cite{bufano2023digital}. Furthermore, health-related factors such as heart rate, level of exercise, number of stairs climbed, and sleep quality can also be monitored.

Such multidimensional digital data form the core data sources of digital phenotyping by enabling the examination of behavioral patterns and psychological states within real-life contexts \cite{mendes2022sensing}. It is important to note that the sensor technology does not measure mental health directly; rather, it measures behaviors that may indicate mental health. Accordingly, studies emphasize that digital phenotyping is not to be used as a tool for diagnosis or treatment; rather, a complementary approach that supports traditional assessment methods in clinical decision-making and treatment processes \cite{liang2019survey, jilka2024digital}.

Passive data collection not only enables the monitoring of behavioral patterns, but it also allows for the detection and intervention of risk situations. In this context, passive data collection can support the implementation of JITAIs, which are often implemented through smartphones and wearable devices, and use sensor and user information to detect risk states based on the individual’s context, providing them with personalized intervention or therapeutic content when they are in a critical situation, such as when they are at risk of relapse \cite{klasnja2015microrandomized}.

In today’s fast-paced digital world, several researchers have emphasized the potential of digital trace analysis to improve the assessment, treatment, and monitoring of mental disorders \cite{markowetz2014psycho}. The accessibility of smartphones among the global population offers an opportunity to use smartphones as an important source of data to be used in psychiatric research \cite{montag2019new}. Indeed, the effectiveness of the data collected using smartphone sensors, such as location, accelerometer, Bluetooth, and microphone sensors, has been established in the identification of behavioral patterns associated with mental disorders such as anxiety and depression \cite{choi2024digital}. Similarly, digital traces, such as touchscreen and typing patterns, have been identified as potential markers for the early prediction of various mental health disorders. Additionally, digital phenotyping has been found to have potential use in schizophrenia research, especially in the early detection of schizophrenia symptoms and monitoring the progression of the symptoms over time \cite{jung2025key}.

Digital phenotyping approaches have been found to demonstrate high sensitivity and specificity in detecting early signs of mental disorders. Also, the use of digital phenotyping provides important information on the prediction of mental disorders while at the same time supporting gold-standard clinical assessments \cite{bufano2023digital}. These findings have emphasized the need to use digital phenotypes in the assessment of behavioral addictions. Due to the limitations of traditional psychiatric assessment tools, especially recall bias and the constraints of self-report data, interest in digital phenotyping has been steadily increasing over the years, with an increasing number of studies on the use of digital phenotyping in recent years.

The types of data collected for the purpose of digital phenotyping are mostly behavioral, physiological, psychological, environmental, and social in nature \cite{zhang2025comprehensive}. Therefore, the types of data used in digital phenotyping will be examined under five categories in the sections that follow. Table \ref{tab:digital_phenotyping} shows the used data digital phenotyping.

\begin{table*}[h!]
	\caption{Used data in digital phenotyping}
	\label{tab:digital_phenotyping}
	\centering
	\resizebox{\textwidth}{!}{%
		\footnotesize
		\begin{tabular}{p{2cm} p{3cm} p{4.5cm} p{5cm}}
			\toprule
			\textbf{Data category} & \textbf{Sensor type} & \textbf{Data type} & \textbf{Example features} \\
			\midrule
			
			\textbf{Physiological data} 
			& PPG & Cardiovascular data & Heart rate, HRV \\
			& PPG & Oxygen saturation & SpO$_2$ \\
			& EDA & Autonomic nervous system activity & Electrodermal activity \\
			& Thermal sensor & Physiological temperature changes & Skin temperature \\
			
			\midrule
			
			\textbf{Behavioral / activity data} 
			& Accelerometer, gyroscope & Movement and activity data & Step count, physical activity level, activity intensity, sedentary time \\
			& GPS / Wi-Fi & Location data & Location changes, movement routes \\
			& Accelerometer, ambient light sensor, phone data, microphone & Sleep patterns & Sleep duration, sleep onset, number of awakenings, sleep quality, circadian rhythm \\
			
			\midrule
			
			\textbf{Environmental data} 
			& Ambient light sensor & Environmental data & Light exposure \\
			& Microphone & Environmental data & Ambient noise \\
			
			\midrule
			
			\textbf{Social and communication data} 
			& Microphone & Speech features & Voice tone, speech rate and amount, speaking duration, pause durations, call logs \\
			& Phone usage logs (SMS/email/call logs) & Digital behavior & Text message and word analysis, online social activity, keyboard use (typing speed, error rates), screen usage (number of screen unlocks) \\
			
			\bottomrule
	\end{tabular}}
\end{table*}

\subsection{Self-reported data}
One of the primary data sources used in digital phenotyping studies is self-reported data. This type of data is based on individuals directly reporting their own behaviors, emotional states, or experiences related to addictive behaviors through smartphone applications or web-based platforms \cite{ferreri2018addictology}. In this approach, the data collection process requires the active participation of the user. Data are typically obtained through short questionnaires, electronic diary entries, and assessments of experiences such as craving or withdrawal related to addiction. Among these methods, EMA approaches hold a particularly important place, as they allow individuals to record their momentary experiences within their daily life context, thereby enabling the collection of data that are naturalistic, context-sensitive, and ecologically valid \cite{serre2015ecological}.

EMA methods have been widely used in the development of digital health interventions for alcohol use \cite{gustafson2011explicating, wilson2017steering} and substance use disorders \cite{baldacchino2016decision}. In addition to data collection, EMA approaches can also facilitate the delivery of treatment interventions. Randomized controlled trials have demonstrated that such interventions are effective in identifying high-risk situations, preventing relapse, and enhancing real-time coping skills \cite{gustafson2014smartphone, glass2017treatment, humphris2017engagement, mckay2022efficacy}. Studies utilizing EMA have revealed significant associations between momentary variables (e.g., craving levels, mood states) and addictive behaviors \cite{mcclernon2013your, wray2014using, serre2015ecological, liang2019survey}.

An example of an EMA-based application is Meine Zeit Ohne, which prompts users to select a specific addictive behavior (e.g., social media use, gambling) and set a goal to reduce or quit this behavior over a two-week period \cite{grahlher2024mobile}. Throughout the process, users receive daily notifications requesting self-efficacy assessments, thereby aiming to facilitate behavior change.

In EMA studies examining gambling behavior, data collection typically relies on brief self-report questionnaires administered via smartphones \cite{russell2018direct, hing2019hey, hawker2025exploring}. Participants report their momentary emotional states, gambling motivations, and gambling behaviors through prompts delivered at specific times during the day. These studies predominantly rely on active data collection methods.

In the field of gambling addiction, several mobile-based applications enable the monitoring of gambling behavior by allowing users to report momentary craving levels and behaviors \cite{mccurdy2023smartphone}. However, only a limited number of these applications are evidence-based \cite{dowling2025gamblingless}. Among the evidence-based applications are Gambling Habit Hacker \cite{dowling2024applying}, GamblingLess: Curb Your Urge \cite{merkouris2020gamblingless}, and GamblingLess: In-The-Moment \cite{dowling2022gambling}. The interventions offered by these applications are based on a “push” approach, meaning that decisions regarding when and how interventions are delivered are determined by algorithms rather than the individual \cite{klasnja2015microrandomized}.

Gambling Habit Hacker is a mobile intervention that analyzes self-reported data and delivers behavior change techniques when low self-efficacy or high-risk situations are detected, supporting individuals in maintaining financial limits related to gambling behavior \cite{rodda2025acceptability}. 

GamblingLess: Curb Your Urge is an EMI-based mobile application designed to reduce gambling urges. Through EMA prompts delivered at random times during the day, the application collects self-reported data on gambling episodes, craving levels, the ability to limit gambling, and self-efficacy in resisting urges  \cite{merkouris2020gamblingless}. Based on these data, the application provides intervention content such as brief cognitive-behavioral techniques and mindfulness exercises \cite{hawker2021smartphone}.

An enhanced version of this approach, GamblingLess: In-The-Moment, is based on a JITAI framework, delivering personalized interventions using self-reported EMA data \cite{dowling2022gambling}. In this application, participants complete EMA assessments three times per day, reporting cognitive variables such as craving intensity, perceived control over gambling, and positive outcome expectancies. These data are analyzed to deliver targeted cognitive-behavioral interventions at moments when the user is cognitively at risk.

Another approach used in gambling interventions is chatbot-based “pull” interventions. In these systems, users initiate engagement by accessing the application when needed, allowing them to use self-monitoring tools, receive feedback, and access therapeutic content \cite{klasnja2015microrandomized}. An example of this approach is GAMBOT, the only evidence-based chatbot intervention currently available, which enables users to monitor their gambling behavior, receive feedback, and access therapeutic resources \cite{so2024guided}. Randomized controlled trials have demonstrated that such applications show promising results in reducing problem gambling severity and supporting behavior change. A general overview of digital intervention-based applications for gambling is presented in Table \ref{tab:interventions}.

\begin{table*}[h!]
	\centering
	\footnotesize
	\caption{Examples of AI-driven digital interventions for behavioral addictions}
	\label{tab:interventions}
		\begin{tabular}{p{2cm} p{2.5cm} p{1.5cm} p{3cm} p{5cm}}
			\toprule
			\textbf{Application} & \textbf{Intervention Type} & \textbf{Data Type} & \textbf{Approach} & \textbf{Description} \\
			\midrule
			
			Gambling Habit Hacker	
			& Push (JITAI)	
			& EMA	
			& CBT-based behavior change	
			& Financial control support following risk detection via EMA \\
			
			GamblingLess: Curb Your Urge	
			& Push (EMA-based EMI)	
			& EMA	
			& CBT, Mindfulness, and Relapse prevention	
			& Real-time intervention based on craving level \\
			
			GamblingLess: In-The-Moment	
			& Push (JITAI)	
			& EMA	
			& CBT and Mindfulness	
			& Personalized intervention at moments of risk \\
			
			GAMBOT	
			& Pull	
			& User interaction	
			& Chatbot	
			& Support on demand \\
			
			SPGeTTI	
			& Push (JITAI)	
			& GPS	
			& CBT-based real-time warning messages	
			& Repeated alerts when approaching gambling venues \\
			
			Don’t Go There	
			& Push (JITAI)	
			& GPS and accelerometer / gyroscope	
			& Warning messages	
			& Preventive notifications when entering high-risk locations, even when GPS is off \\
			
			Meine Zeit Ohne	
			& Push	
			& EMA	
			& Behavior change	
			& Supports behavior change through daily tracking of a selected goal \\
			
			mGaming Wellness	
			& AI-based self-monitoring	
			& Passive behavioral data and EMA 	
			& Behavior change	
			& Analyzes user data to predict addiction risk and provides personalized feedback \\
			
			\bottomrule
		\end{tabular}%
\end{table*}

In the broader field of behavioral addictions, studies utilizing EMA and EMI methods remain relatively limited and are still in the developmental stage \cite{meyer2026identifying}. Nevertheless, findings from the studies discussed above provide an important conceptual framework for behavioral addictions, particularly in understanding the dynamics of addictive behaviors in daily life, monitoring relapse processes, and developing digital intervention strategies \cite{hawke2025digital}.

\subsection{Physiological data}
The advancement of smartphones and wearable technologies such as smartwatches and fitness bands has enabled the integration of various sensors into these devices \cite{onnela2016harnessing}. Today, these digital devices allow for the continuous monitoring of a wide range of physiological indicators, including heart rate, skin temperature, physical activity levels, sleep patterns, and energy expenditure \cite{lee2024smartphone}. In addition, sensors embedded in smartphones and wearable devices can capture data related to individuals’ movement and mobility patterns. Such sensor-based measurements enable the real-time monitoring of behavioral and physiological changes in daily life, thereby constituting a significant data source in digital phenotyping studies \cite{mena2022estesia}.

In digital phenotyping research, passively collected data offer the potential to evaluate individuals’ behaviors more objectively within real-life contexts. It has been demonstrated that individuals’ self-estimations of smartphone usage duration or online behaviors often differ substantially from their actual usage patterns \cite{lin2015time}, indicating that self-reported data may be influenced by recall bias and subjective evaluation. In contrast, sensor-derived passive data provide a more objective means of monitoring digital behaviors and physiological changes \cite{jung2025key}.

Physiological measures such as heart rate, HRV, and EDA are considered critical for understanding an individual’s contextual state. These indicators of autonomic nervous system functioning can provide valuable information about an individual’s stress levels and psychological state at a given moment \cite{hickey2021smart}. Currently, such physiological measurements are increasingly being used as potential clinical assessment tools for monitoring disease progression and detecting early warning signs across various domains of addiction \cite{carreiro2015imstrong, imtiaz2017development, nandakumar2019opioid}.

Research in the field of behavioral addictions has also highlighted the relevance of these measures. Changes in HRV have been observed in association with smartphone use and problematic internet use \cite{cheng2023can, dale2025influence}. Similarly, a relationship has been identified between gaming addiction severity and HRV levels \cite{hong2023severity}. These physiological indicators complement self-report-based mental health assessments by providing objective biological markers, thereby enabling a more comprehensive examination of psychological processes within digital phenotyping frameworks \cite{an2024relationship}.

Sleep patterns can be assessed using data derived from smartphone sensors. In this context, variables such as phone charging times, cessation of prolonged phone use, and environmental conditions (e.g., quietness and darkness) can be evaluated collectively to monitor sleep patterns \cite{chen2013unobtrusive, borger2019capturing}. Smartphone-based sleep measurements have been shown to be useful in assessing symptoms of depression and anxiety \cite{knights2023associations}. Therefore, sleep-related data can be considered an important source for evaluating addiction severity and predicting potential relapse risk.

The integration of digital data with self-reported measures allows for a more comprehensive understanding of both individuals’ actual behavioral patterns and their subjective perceptions of these behaviors, thereby facilitating a more holistic examination of psychopathology. However, when studies in the field of gambling addiction are examined, it appears that there are currently no published studies adopting this integrated approach, although an ongoing study exists \cite{meyer2026identifying}. This study aims to develop predictive models of gambling episodes by analyzing physiological data such as heart rate, sleep patterns, and physical activity obtained from smartwatch sensors together with EMA data. It is anticipated that such an approach, particularly when combined with machine learning methods, may contribute to the identification of early warning indicators of gambling behavior.

\subsection{Behavioral \& usage data}
Behavioral data refer to a type of data that enables the assessment of individuals’ health status through the monitoring and analysis of their daily behavioral patterns. These data are typically obtained from digital technologies such as smartphones and wearable devices and include information on daily behaviors such as step count, physical activity levels, sleep patterns, phone usage duration, and social media activity. Such data can provide important indicators of individuals’ daily activity levels, sleep–wake cycles, and social behaviors \cite{zhang2025comprehensive}.

Smartphones and digital platforms, which represent important data sources for the assessment of internet use disorders, enable the objective monitoring of individuals’ online behavioral patterns \cite{montag2021potential}. Digital traces left on social media platforms (e.g., likes, posts, and textual content) allow for meaningful inferences regarding individuals’ personality traits and behavioral patterns \cite{montag2019new}. Data such as application usage frequency, social media interactions, time spent online, and switching behavior between applications can contribute to identifying addiction tendencies, including problematic internet use, online gambling, and gaming behaviors \cite{markowetz2014psycho, hitcham2023relationship}. Clinicians have also emphasized that usage patterns related to social media, gaming, or gambling platforms may serve as important indicators of addictive behaviors and can support treatment processes \cite{schmidt2023clinician}. The analysis of such data allows for the examination of individuals’ online behaviors within the context of daily life and facilitates the early detection of risky usage patterns. For example, it has been demonstrated that platform-specific behavioral data such as the number and frequency of Facebook posts, textual content, language use, and page likes can be analyzed using machine learning methods to predict tendencies toward social network use disorders \cite{marengo2022mining}.

To prevent problematic smartphone and social media use, smartphone-based applications have been developed that automatically monitor individuals’ digital behaviors and provide feedback based on these data \cite{van2018problematic}. These applications passively track behavioral data such as screen time and application usage frequency and, based on the collected data, provide users with feedback to increase awareness of their usage habits, restrict access to certain applications, or deliver real-time alerts and coping suggestions in risky situations \cite{rahmillah2023evaluating}. An example is mGaming Wellness, an artificial intelligence–based application developed for mobile gaming addiction in adolescents \cite{khoziasheva2025applying}. The application analyzes usage data collected from smartphones to estimate an individual’s risk level for gaming addiction and provides tailored feedback. It also offers various tools to enhance behavioral awareness, reduce gaming time, and promote healthier digital habits.

In digital phenotyping studies, behavioral data derived from individuals’ interactions with digital devices are also considered a crucial data source \cite{zhang2025comprehensive}. Interactions with smartphones are largely mediated through software-based keyboards, and the touch interactions generated during this process produce valuable behavioral data. Features such as typing speed, inter-keystroke intervals, touch pressure, error rates, and swipe typing behaviors have been associated with individuals’ mood, stress levels, and cognitive load \cite{gao2012does, ciman2015isensestress, lee2015influence}. Therefore, metadata derived from keyboard interactions can provide insights into users’ psychological states without accessing content information \cite{ghosh2019emotion}. These behavioral digital traces are also considered potential data sources for understanding and early detection of behavioral addictions such as social media use, online gaming, and problematic internet use. Indeed, a study conducted with adolescents demonstrated that keyboard interaction data obtained from tablet devices such as typing duration, deletion behaviors, inter-word intervals, and changes in typing speed may serve as meaningful indicators for identifying digital phenotypes associated with internet gaming disorder \cite{cho2024digital}.

Transactional data related to betting behaviors recorded on online gambling platforms are also used as data sources in digital phenotyping studies \cite{murch2024establishing}. Variables such as betting frequency, wager amounts, total time spent on online platforms, smartphone usage, and gaming behavior logs are analyzed using machine learning methods to identify users at risk of problematic gambling behavior \cite{auer2026using}.

\subsection{Contextual \& environmental signals}
In digital phenotyping studies, the identification of environmental context is considered an important data source for understanding the conditions under which behaviors occur. Individuals’ geographic location, travel patterns, and frequently visited places can be determined using GPS sensors embedded in smartphones. Furthermore, other factors such as noise levels and temperature can also be measured \cite{zhang2025comprehensive}.

The integration of EMA data with passive data collected through smartphones can also help to make a multidimensional evaluation of mental states. These data can help to develop methods that can detect risky situations and deliver timely interventions through technology \cite{nahum2016just}. EMA studies have shown that individuals’ environmental and emotional factors are highly related to substance use. In this regard, location-based studies have shown that GPS data can be used to detect substance craving. Similarly, a study on a mobile application called Q Sense, designed for smoking cessation, found that individuals’ smoking locations were detected through GPS data, and timely support was provided to individuals when approaching these locations through geofencing technology \cite{naughton2016context}.

Findings from studies utilizing GPS sensors indicate that location-based interventions are effective in identifying high-risk environments and delivering interventions \cite{skinner2017digital}. In a study that aimed to examine problematic smartphone use, application-based EMA data was analyzed, and passive sensor data collected through smartphones, such as GPS, accelerometer, and step count, was also used \cite{ahn2025app}. The results showed that behavioral patterns detected through GPS can be potential digital markers for problematic smartphone use prediction.

These findings suggest that location-based digital interventions could also be applicable in the case of behavioral addictions. However, the application of location-based digital interventions in the case of gambling is still limited. There are mobile applications developed in the domain of gambling, and they are intended to send alerts and messages using GPS sensors. However, the majority of the mobile applications are still in the early development phase \cite{dowling2024applying}. An example is the SPGeTTI application, which provides cognitive-behavioral messages and coping strategies, as well as supportive messages, while simultaneously detecting the entry into gambling venues through GPS-based location tracking. In this case, the application sends real-time alerts and messages using the JITAI method to avoid the continuation of gambling behavior \cite{humphrey2019spgetti}. Another application, Don’t Go There, is intended to reduce gambling behavior using the JITAI method. The application sends real-time alerts when the user is close to gambling venues and attempts to identify the location using other smartphone sensors, such as the accelerometer and gyroscope, even when the GPS is turned off \cite{coral2020don}.

\subsection{Social and communication traces}
The analysis of the social interactions and communication patterns of individuals in digital platforms is an important part of digital phenotyping. These data can be obtained from various digital sources, including social media platforms, messaging applications, and smartphone usage logs. Indicators such as the call and messaging frequencies, the number of contacts, the intensity of interactions in social networks, and social media patterns are important and objective information related to social connectivity and communication patterns \cite{chen2025current}.

The content and linguistic characteristics of social media posts can be examined using NLP techniques, allowing researchers to draw significant inferences about the mood, psychological state, and behavioral patterns of individuals \cite{marengo2022mining, garg2023mental}. Researchers have found that shifts in content, linguistic, and communicational behaviors in social media posts may be associated with shifts in mental state \cite{diniz2022boamente}.

In addition, communication-based indicators derived from phone calls, such as speech rate and speech quantity, can provide important markers associated with individuals’ emotional states \cite{cummins2023multilingual}. Similarly, the analysis of individuals’ communication data collected through digital communication tools enables researchers to monitor individuals’ levels of social interaction objectively \cite{wang2014studentlife}.

User-generated content posted on online forums has also been recognized as a vital source of data in digital phenotyping studies. In particular, messages posted on gambling forums provide valuable data for investigating gambling behavior and associated problems \cite{james2021use}. Analyzing such text-based data through NLP and machine learning approaches helps researchers detect risky conditions associated with behavioral addictions \cite{smith2024automatic}. Further, through such studies, researchers have also shown that the emotional, cognitive, and social aspects of gambling addiction can be revealed through linguistic markers \cite{van2025lived}.

\subsection{Multimodal data fusion challenges}
Multimodal data fusion constitutes a central yet technically demanding component of digital addiction management systems. Behavioral addictions are complex, dynamic, and context-dependent phenomena that cannot be fully characterized through a single data stream \cite{zhao2024deep}. Effective digital phenotyping often requires integrating heterogeneous sources such as self-reports, physiological measurements, behavioral usage logs, contextual signals, and linguistic traces. While multimodal integration promises improved predictive accuracy and personalization, it introduces significant methodological, computational, and practical challenges \cite{sui2023data}.

One of the primary difficulties lies in heterogeneity across data modalities \cite{yilmaz2021multimodal}. Self-reported questionnaires are typically low-frequency, structured, and subjective, whereas behavioral logs are high-frequency and event-driven. Physiological signals such as HRV or electrodermal activity are continuous time-series data with sensor noise, while textual data from chat interactions are unstructured and semantically rich. These modalities differ not only in format and dimensionality but also in temporal resolution, reliability, and semantic meaning. Designing architectures that can meaningfully align and integrate such disparate representations remains a non-trivial task \cite{li2024multimodal}.

Temporal synchronization presents an additional challenge \cite{piatkowski2025lightweight}. Behavioral addictions often involve rapidly fluctuating states, such as momentary cravings or stress spikes. However, different modalities may be sampled at different rates and with varying latency. Aligning daily self-reports with second-level physiological signals or event-based app usage logs requires careful temporal modeling. Early fusion strategies (combining raw features) may suffer from misalignment, while late fusion strategies (combining model outputs) may fail to capture cross-modal interactions \cite{gadzicki2020early}. Hybrid and attention-based fusion approaches attempt to balance these trade-offs but increase model complexity \cite{nemati2019hybrid}.

Missing and incomplete data further complicate multimodal systems \cite{zhang2024multimodal}. Wearable sensors may be removed, self-reports may be skipped, and contextual data such as GPS signals may be intermittently unavailable. Missingness itself can carry behavioral meaning (e.g., disengagement during high-risk periods), yet most machine learning models are not inherently robust to irregular or sparse inputs. Imputation strategies \cite{adhikari2022comprehensive}, probabilistic modeling, and modality dropout training techniques are often required to maintain stability.

Another critical issue concerns modality dominance and bias \cite{zheng2025reducing}. High-dimensional modalities such as behavioral logs may overshadow lower-frequency but clinically meaningful signals like self-reported craving intensity. Without careful normalization and weighting strategies, models may overfit to the most abundant data source, reducing interpretability and potentially overlooking subtle but important predictors \cite{gao2025asymmetric}.

From a computational perspective, multimodal deep learning models are resource-intensive and require substantial training data to generalize effectively. This raises concerns about scalability, especially in real-world healthcare deployments where computational resources and labeled data may be limited \cite{zhou2026model}. Moreover, integrating sensitive multimodal data streams increases privacy and security risks, necessitating robust encryption, access control, and possibly federated architectures \cite{li2025truth}.

Finally, interpretability becomes more complex as modalities are fused \cite{li2022integrating}. Explaining how a combination of linguistic sentiment shifts, increased nighttime screen use, and physiological stress patterns jointly contributed to a relapse prediction requires sophisticated explanation frameworks that remain understandable to clinicians.

\newpage
\section{AI and Machine Learning Approaches}\label{sec6}
\begin{figure*}[htbp]
	\centering
	\includegraphics[width=0.8\textwidth]{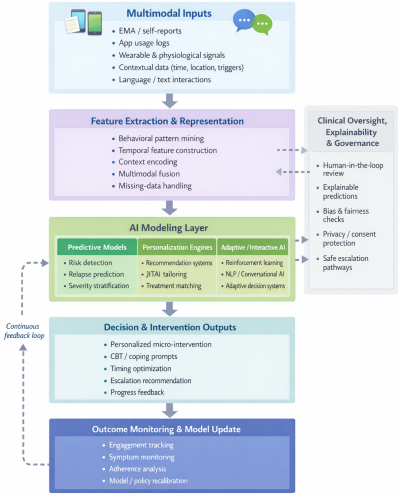}
	\caption{Conceptual AI pipeline for prediction, personalization, and intervention in behavioral addictions}
	\label{fig:section6}
\end{figure*}

Recent advances in artificial intelligence have opened new possibilities for the assessment and management of behavioral addictions. Machine learning methods enable the analysis of complex and heterogeneous data generated through digital platforms, wearable devices, and self-reported behavioral records. As illustrated in Figure~\ref{fig:section6}, these systems can be conceptualized as an AI pipeline that transforms multimodal behavioral data into predictive insights and personalized intervention strategies. Within this framework, AI models support tasks such as risk prediction, early relapse detection, and adaptive intervention planning. Various computational approaches have been explored in the literature, including classical machine learning models, deep learning architectures, RL frameworks, and NLP techniques. These methods facilitate the integration of multimodal data and enable the identification of behavioral patterns that are difficult to capture using traditional analytical approaches. At the same time, issues such as model transparency, reliability, and clinical interpretability remain critical for real-world deployment. Therefore, recent research increasingly emphasizes not only predictive performance but also explainability and the safe integration of AI systems into digital therapeutic environments.

\subsection{Risk prediction models}
\begin{figure*}[htbp]
	\centering
	\includegraphics[width=0.8\textwidth]{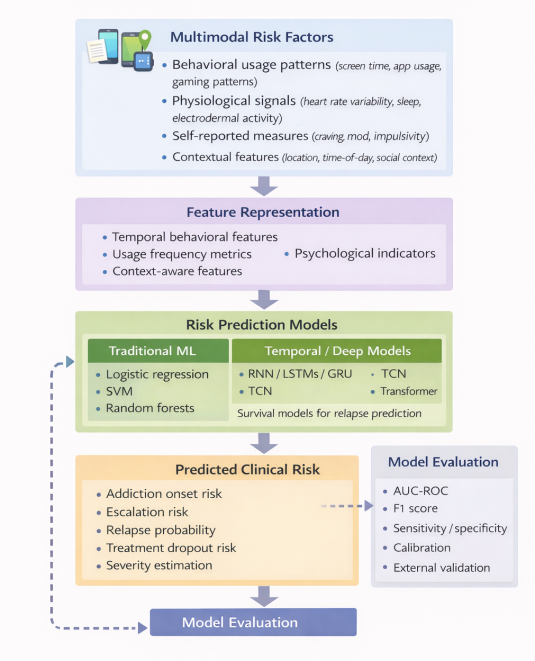}
	\caption{AI-based risk prediction models for behavioral addictions}
	\label{fig:section6.1}
\end{figure*}

Risk prediction models constitute one of the most critical AI-driven components in the digital management of behavioral addictions \cite{balhara2019behavior, park2022content}. These models aim to estimate the likelihood of addiction onset \cite{kleinman1978onset}, escalation \cite{uppal2019prediction}, relapse \cite{topbacs2025does, aytop2025correlation}, or treatment dropout \cite{sirvent2025prediction} by leveraging heterogeneous data sources such as self-reports, behavioral logs, physiological signals, and contextual information. Accurate risk stratification enables early intervention, personalized care planning, and optimized allocation of clinical and digital resources \cite{zilberman2018personality}. As illustrated in Figure~\ref{fig:section6.1}, AI-based risk prediction systems typically integrate multimodal behavioral and contextual signals to estimate addiction-related risks, transforming heterogeneous digital data into clinically meaningful risk estimates.

Risk prediction in behavioral addictions is typically framed as a supervised learning problem. Depending on the clinical objective, tasks may include:

\begin{itemize}
	\item Binary classification (e.g., high-risk vs. low-risk individuals) \cite{wang2022hybrid}
	\item Multi-class classification (e.g., mild, moderate, severe addiction risk) \cite{thity2025illustration}
	\item Regression models (e.g., predicting addiction severity scores) \cite{shah2015addiction}
	\item Time-to-event (survival) analysis (e.g., predicting time until relapse) \cite{devi2015application}
	\item Sequential prediction (e.g., forecasting near-term craving episodes) \cite{wyant2024machine}
\end{itemize}

In addition, short-term relapse prediction and long-term vulnerability estimation \cite{redish2008addiction} represent two complementary modeling paradigms in digital systems. Risk models integrate diverse multimodal signals, including:

\begin{itemize}
	\item Behavioral usage data: screen time, app switching patterns, gaming duration, transaction logs (e.g., gambling frequency) \cite{zeng2024concurrent}.
	\item Physiological signals: HRV, sleep quality, electrodermal activity \cite{kbah2025review}. 
	\item Self-reported measures: craving intensity, mood scales, impulsivity indices \cite{toulami2025craving}. 
	\item Contextual features: geolocation, time-of-day patterns, social context indicators \cite{yang2025relative}.
	\item Linguistic traces: sentiment, cognitive distortions, and emotional markers extracted from chat or journaling data \cite{siregar2024effective}.
\end{itemize}

The fusion of these modalities enables richer digital phenotyping but introduces challenges related to missing data, synchronization, and temporal alignment.

Traditional machine learning methods such as logistic regression, support vector machines (SVM), random forests, and gradient boosting \cite{kato2023computational} have been widely used for structured addiction datasets due to their robustness and interpretability \cite{gezgin2025using}. With the increasing availability of longitudinal and high-frequency data, deep learning architectures-including recurrent neural networks (RNNs), LSTMs, GRUs, temporal convolutional networks (TCNs), and transformers-are employed to capture temporal dependencies and behavioral trajectories \cite{jing2023ta, shah2024extraction}.

For relapse prediction and dynamic risk estimation, survival models (e.g., Cox proportional hazards models \cite{moeeni2016predictors}, deep survival networks \cite{wang2024resdeepsurv}) and state-space models are particularly relevant. Graph-based learning approaches are emerging for modeling social influence effects \cite{xu2025sigrl}, especially in online gaming \cite{gu2024let} or social media addiction contexts.

Unlike static clinical assessments, digital systems enable continuous risk updating. Real-time prediction pipelines can support JITAI \cite{nahum2025just} by detecting early warning signals (e.g., increased nighttime usage, negative sentiment shifts) and triggering timely micro-interventions. 

Key performance metrics include AUC-ROC, F1-score, sensitivity/specificity, calibration curves, and Brier scores. However, in clinical settings, calibration and clinical utility (e.g., decision curve analysis) are as important as discrimination performance. External validation across populations, cultural contexts, and age groups remains limited. Longitudinal validation and prospective clinical trials are necessary to assess real-world reliability and impact.

\subsection{Personalized recommendation systems for interventions}
\begin{figure*}[htbp]
	\centering
	\includegraphics[width=0.8\textwidth]{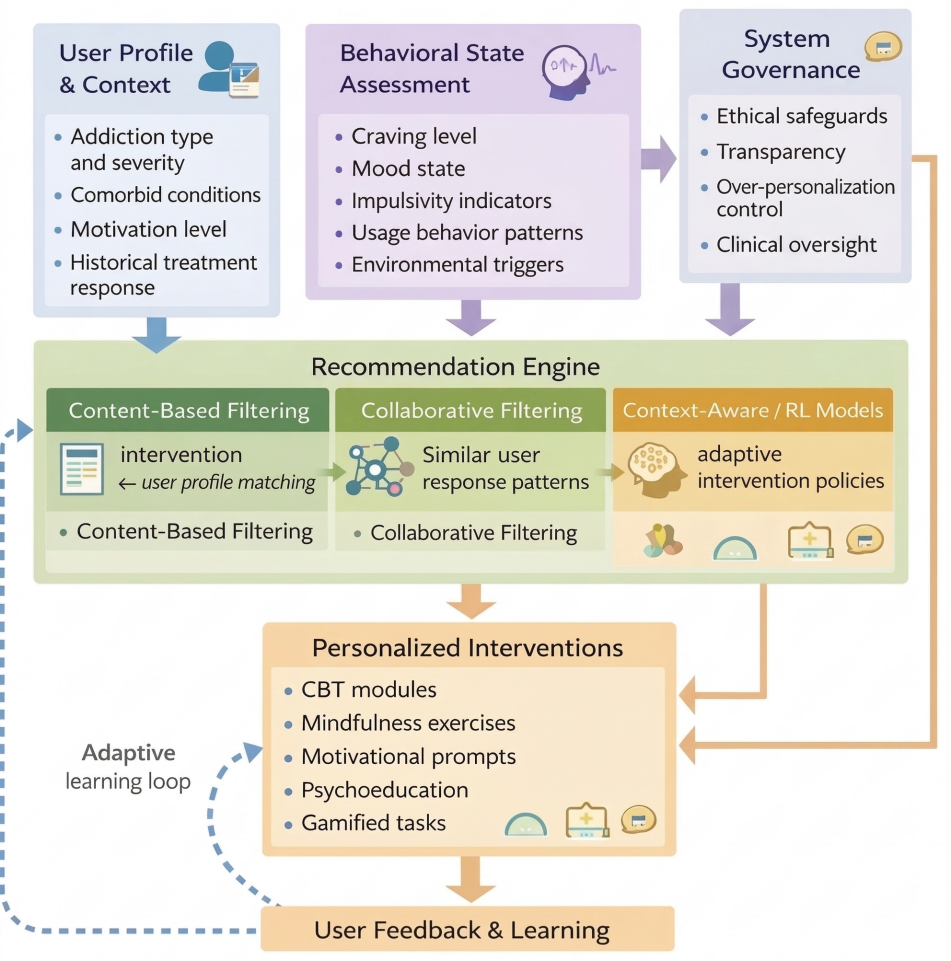}
	\caption{AI-based personalized recommendation framework for addiction interventions}
	\label{fig:section6.2}
\end{figure*}

Personalized recommendation systems constitute a core component of AI-enabled digital addiction management platforms, aiming to tailor interventions to the unique needs, behaviors, and risk profiles of individuals \cite{uludag2023personalised}. In contrast to traditional standardized treatment protocols, these systems dynamically determine which therapeutic content, coping strategy, motivational message, or behavioral exercise should be delivered, at what time, and under which contextual conditions \cite{klein2014intelligent}. Such personalization enhances engagement, perceived relevance, and ultimately therapeutic effectiveness by aligning intervention delivery with the individual’s evolving psychological and behavioral state \cite{hors2018tailoring}. Figure ~\ref{fig:section6.2} presents a conceptual overview of personalized recommendation systems used in digital addiction interventions. Personalized recommendation systems integrate user profiles, behavioral states, and contextual signals to determine appropriate interventions for individuals with behavioral addictions. Recommendation engines employ different strategies such as content-based filtering, collaborative filtering, and context-aware learning models to match therapeutic content with user needs. The system continuously adapts through user feedback and behavioral responses, enabling dynamic optimization of intervention strategies in digital therapeutic environments.

At a conceptual level, personalized recommendation in addiction management integrates multiple dimensions, including addiction type and severity, comorbid conditions such as anxiety or depression, motivational readiness for change, historical treatment responses, and real-time contextual factors \cite{oyebode2024persuasive}. By leveraging these multidimensional profiles, digital systems can move beyond static rule-based approaches and instead generate adaptive care pathways \cite{alexandrou2008adaptive} that evolve over time. This shift reflects the broader paradigm of precision mental health, where interventions are continuously calibrated based on individual-level data rather than population averages \cite{wang2025harnessing}.

From a technical perspective, recommendation systems in this domain typically rely on content-based, collaborative, or hybrid filtering approaches \cite{isinkaye2015recommendation}. Content-based models match intervention modules to user characteristics such as symptom patterns, impulsivity levels, or cognitive distortions \cite{lops2010content}. Collaborative filtering leverages patterns observed among similar users to identify interventions that have been effective in comparable profiles \cite{chen2018survey}. Hybrid systems combine these strategies to mitigate sparsity and cold-start problems \cite{chaudhari2024hybrid}. Increasingly, context-aware models incorporate real-time signals such as mood fluctuations, geolocation, time-of-day risk patterns, or physiological stress indicators to enhance ecological validity and responsiveness.

Given the longitudinal and dynamic nature of addiction recovery, sequential modeling approaches are particularly relevant. Context-Aware / RL Models \cite{intayoad2020reinforcement} frameworks enable systems to learn optimal intervention policies by maximizing long-term outcomes such as relapse reduction or sustained engagement \cite{ban2024neural}. These models treat intervention selection as a sequential decision-making problem, continuously updating strategies based on user feedback and behavioral responses \cite{chen2021user}. Such adaptive optimization allows DTx to refine treatment pathways over time rather than relying on static predefined sequences.

The scope of recommended interventions may include CBT modules \cite{shah2024cognitive}, mindfulness exercises, motivational interviewing prompts \cite{levensky2007motivational}, psychoeducational materials \cite{buder2012learning}, gamified tasks \cite{tondello2017recommender}, peer-support suggestions, and digital nudges \cite{jesse2021digital}. In higher-risk situations, systems may escalate care by recommending telehealth consultations or clinician contact \cite{ignatowicz2019internet}. The ability to personalize both content and intensity of intervention helps reduce cognitive overload and supports sustained adherence.

Despite their promise, personalized recommendation systems face significant challenges. Data sparsity, evolving behavioral patterns, ethical concerns around persuasive design, and the risk of over-personalization require careful system design and governance. Moreover, transparency and explainability are essential to ensure trust among clinicians and users. Rigorous clinical validation through longitudinal studies and randomized controlled trials remains necessary to establish efficacy and safety.

\subsection{RL for habit formation \& habit suppression}
RL provides a powerful computational framework for modeling, supporting, and optimizing behavior change processes in the context of behavioral addictions \cite{kato2023computational}. Because addiction is fundamentally linked to maladaptive habit formation and dysregulated reward learning, RL-based systems align naturally with the neurobehavioral mechanisms underlying craving, reinforcement, and relapse. In digital addiction management, RL is primarily used to dynamically optimize interventions that either promote healthy habit formation \cite{lisowska2021personalized} or suppress harmful habitual behaviors \cite{heinicke2020reducing}.

\begin{figure*}[htbp]
	\centering
	\includegraphics[width=0.5\textwidth]{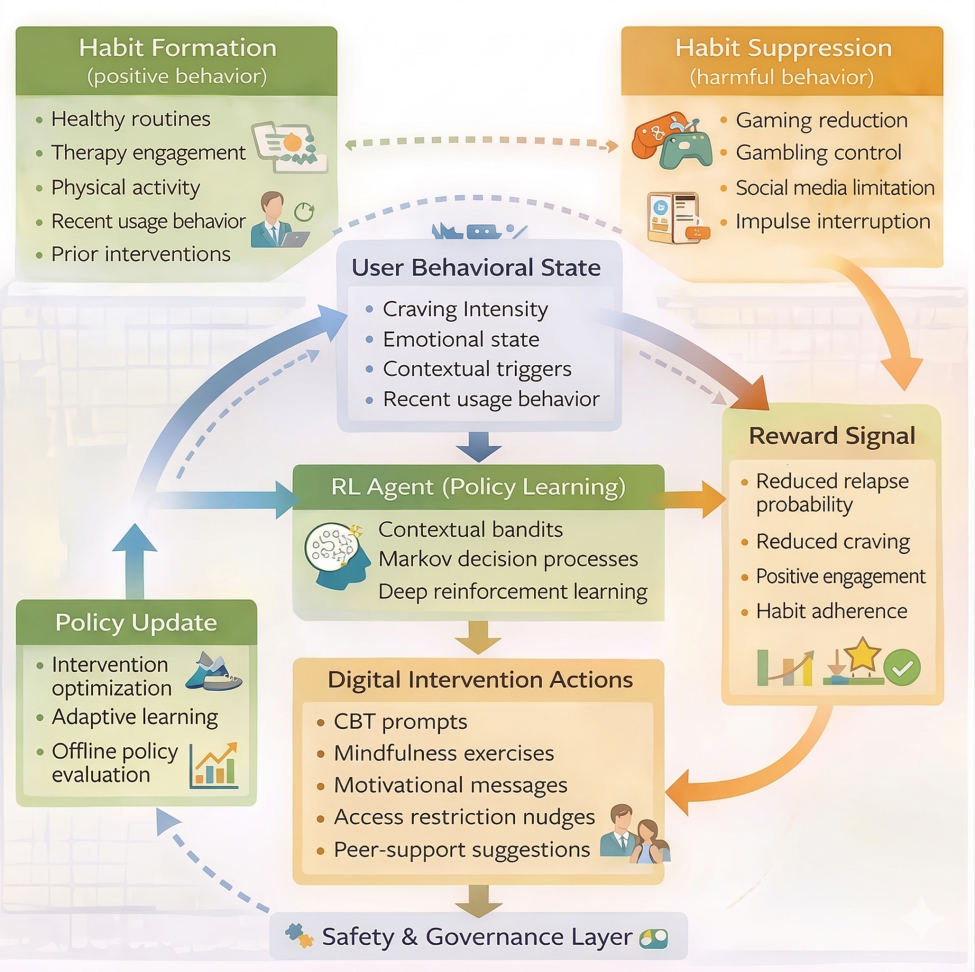}
	\caption{RL framework for habit formation and habit suppression in behavioral addictions}
	\label{fig:section6.3}
\end{figure*}

From a conceptual standpoint, RL formalizes behavior change as a sequential decision-making problem. An individual’s behavioral state-characterized by factors such as craving intensity, emotional status, contextual triggers, and prior actions-can be modeled as the “state” in an RL framework. Digital interventions (e.g., mindfulness prompts, CBT exercises, motivational messages, access restrictions, or peer-support suggestions) represent the “actions.” The system receives feedback in the form of “rewards,” which may reflect reduced craving, lower relapse probability, improved engagement, or sustained abstinence. Over time, the algorithm learns a policy that maximizes cumulative long-term reward, effectively identifying which interventions work best under which circumstances \cite{vlaev2015action}. Figure ~\ref{fig:section6.3} presents a conceptual view of how RL models behavior change as a sequential decision-making process in which the user’s behavioral state informs the selection of digital interventions. Within this framework, contextual factors such as craving intensity, emotional status, recent usage patterns, and environmental triggers define the system state, while adaptive interventions, including CBT prompts, mindfulness exercises, motivational messages, or access restriction nudges, represent the actions. The RL agent continuously updates its policy based on reward signals derived from behavioral outcomes such as reduced relapse probability, sustained abstinence, or improved engagement. Through this iterative learning process, the system can both reinforce positive behavioral routines (habit formation) and suppress maladaptive behaviors associated with addiction (habit suppression).

In habit formation scenarios, RL can be used to reinforce adaptive behaviors \cite{cai2017adaptive} such as regular sleep patterns, physical activity, structured daily routines, or engagement in therapeutic exercises. By continuously analyzing user responses and adherence patterns, the system can adjust intervention timing, intensity, and modality to strengthen positive behavioral loops. For example, contextual bandit models may determine the optimal timing for delivering a motivational message to maximize the likelihood of completing a coping exercise \cite{foster2020instance}. In this way, healthy routines gradually become automated and self-sustaining.

Conversely, in habit suppression contexts-such as reducing compulsive gaming, gambling, or excessive social media use-RL can help identify high-risk states and apply targeted countermeasures \cite{brands2025extended}. The system may learn that specific temporal patterns, emotional states, or environmental cues precede harmful behavior and proactively deploy micro-interventions. These may include friction-inducing mechanisms (e.g., delay prompts), cognitive reframing exercises, or escalation to social or professional support. By optimizing long-term outcomes  \cite{jarraya2025analyzing} rather than immediate short-term responses, RL frameworks are particularly suited to addressing the delayed reward structures characteristic of addiction.

Technically, several RL paradigms are applicable. Contextual bandits are often used for short-term adaptive message selection \cite{rastogi2025multiscale}. More complex Markov Decision Process (MDP) models capture transitions between behavioral states over time \cite{perrusquia2022human}. Deep RL extends these approaches to high-dimensional and multimodal data environments, enabling scalable personalization \cite{petrenko2025shapley}. However, the deployment of RL in clinical contexts requires careful reward design, safety constraints, and human oversight to prevent unintended behavioral reinforcement or ethically problematic nudging \cite{islam2024reinforcement}.

Despite its promise, applying RL to addiction management poses challenges. Sparse and delayed rewards, non-stationary behavioral dynamics, limited training data, and safety-critical decision-making constraints complicate model design. Offline policy evaluation, simulation-based validation, and hybrid human-AI control mechanisms are therefore essential before real-world implementation.

\subsection{NLP for interaction \& sentiment}
NLP plays a pivotal role in digital addiction management systems by enabling meaningful human–AI interaction and extracting clinically relevant insights from textual and conversational data \cite{preiss2022using}. Since behavioral addictions are closely intertwined with emotional states, cognitive distortions \cite{sage2025survey, utti2025conversational}, motivational fluctuations, and social context, language becomes a rich and sensitive signal for assessing risk, monitoring progress, and delivering personalized support \cite{agarwal2025exploratory}. NLP technologies therefore function both as an interaction interface and as an analytical engine within AI-powered DTx.

At the interaction level, NLP enables conversational agents \cite{laranjo2018conversational, tudor2020conversational}, chatbots \cite{lamprou2025customizable}, and virtual therapists \cite{kamene2025physical} to engage users in structured or semi-structured dialogue. These systems may deliver psychoeducation, conduct cognitive restructuring exercises, guide mindfulness practices, or implement motivational interviewing techniques. Advances in LLMs have significantly improved contextual understanding, coherence, and empathetic response generation, allowing digital systems to simulate supportive therapeutic communication. In addiction contexts, such conversational agents can provide immediate, low-barrier access to assistance, particularly during high-risk moments when human clinicians may not be available. 

\begin{figure*}[htbp]
	\centering
	\includegraphics[width=0.5\textwidth]{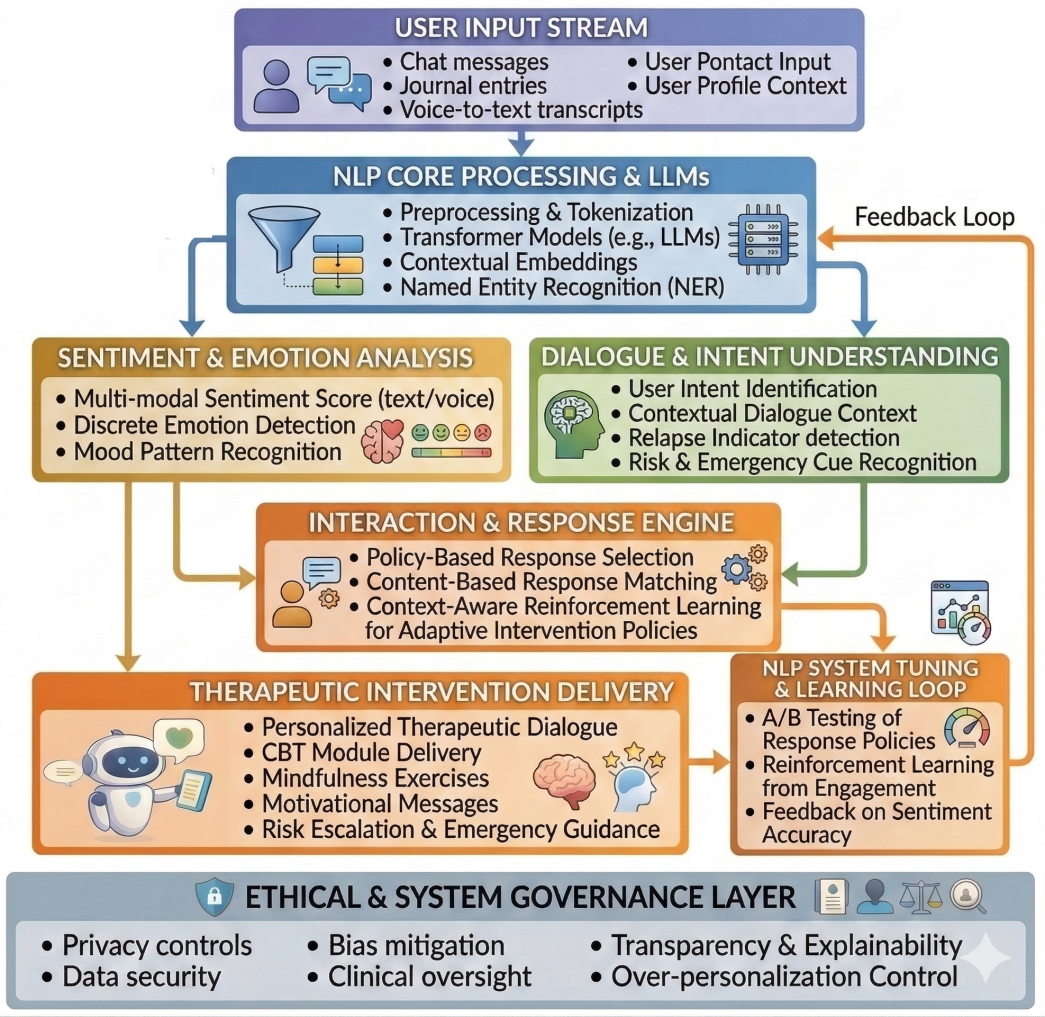}
	\caption{NLP framework for interaction and sentiment analysis in digital addiction management systems}
	\label{fig:section6.4}
\end{figure*}

Figure  ~\ref{fig:section6.4} presents the conceptual framework of NLP-based interaction and sentiment analysis in digital addiction management systems. In this framework, user-generated language inputs such as chat messages, journaling entries, and conversational transcripts are processed by NLP models to extract emotional signals, cognitive patterns, and contextual indicators associated with addiction risk. At the same time, conversational agents utilize language generation capabilities to deliver psychoeducation, motivational support, and therapeutic dialogue. By integrating interactive communication with linguistic analysis, NLP systems enable real-time user engagement, continuous monitoring of emotional states, and early detection of relapse-related linguistic markers while supporting personalized digital interventions under appropriate safety and ethical oversight.

Beyond dialogue generation, NLP methods are widely used for sentiment analysis and affect detection. By analyzing user-generated text-such as chat logs, journaling entries, forum posts, or transcribed speech-systems can detect emotional polarity (positive, negative, neutral), intensity of distress, hopelessness indicators, impulsivity-related language, or craving expressions \cite{li2022word}. More advanced models incorporate emotion classification, cognitive distortion detection, and stress-level estimation. Temporal tracking of linguistic patterns \cite{zhang2017time} enables early identification of deteriorating mental states or relapse precursors, supporting proactive intervention strategies.

Recent transformer-based architectures and contextual embeddings have enhanced the ability to capture nuanced semantic and pragmatic signals in addiction-related discourse \cite{lokala2024detecting, wang2025neuroclip}. Topic modeling and semantic clustering can reveal recurring themes such as triggers, social isolation, financial stress (in gambling addiction), or performance anxiety (in gaming addiction) \cite{gauthier2022will}. Additionally, dialogue modeling techniques \cite{bouzoubaa2024words} can assess engagement quality and conversational alignment, which are critical for sustaining adherence in digital therapeutic programs.

However, NLP-driven systems face important challenges. Language is highly context-dependent and culturally sensitive, making cross-population generalization difficult \cite{hayat2025context}. Data scarcity \cite{sufi2024addressing} in clinically labeled addiction datasets can limit supervised model performance. Privacy concerns are particularly acute, as textual data may contain deeply personal information \cite{corwin2025comprehensive}. Moreover, ensuring that AI-generated responses remain clinically appropriate, non-judgmental, and safe requires robust guardrails, alignment strategies, and human oversight.

\subsection{Explainability \& interpretability concerns}
Explainability and interpretability are critical considerations in the deployment of AI-driven systems for behavioral addiction management \cite{stanimirovic2024applying, gezgin2025using}. Unlike many commercial recommendation environments, digital addiction interventions operate within clinically sensitive and ethically regulated contexts. Decisions such as identifying a user as high risk for relapse, recommending specific therapeutic modules, or escalating care to a clinician can have significant psychological and practical consequences. Therefore, AI models must not only achieve high predictive performance but also provide transparent, understandable, and clinically meaningful explanations for their outputs \cite{linardatos2020explainable}.

In addiction management, interpretability serves multiple stakeholders. Clinicians require insight into which features-such as increased nighttime usage, negative sentiment shifts, or elevated physiological stress indicators-contributed to a risk prediction \cite{xu2025risk}. Such transparency facilitates clinical validation, supports shared decision-making, and allows professionals to override or contextualize algorithmic outputs when necessary. For users, understandable explanations enhance trust, perceived fairness, and engagement with digital systems. If individuals understand why a particular coping strategy or intervention is recommended, they are more likely to adhere to it.

From a technical standpoint, explainability approaches can be categorized into intrinsically interpretable models and post-hoc explanation techniques \cite{slack2021reliable}. Traditional models such as logistic regression, decision trees, and rule-based systems provide relatively transparent decision boundaries but may lack the expressive power required for complex multimodal and temporal data. Conversely, deep learning architectures, such as recurrent networks or transformers-offer superior modeling capacity but function as “black boxes.” Post-hoc techniques, including feature attribution methods (e.g., SHAP, LIME), attention visualization, counterfactual explanations, and saliency mapping, are increasingly used to bridge this gap \cite{marcinkevivcs2020interpretability}. In temporal addiction modeling, explaining how risk evolves over time presents an additional layer of complexity \cite{sood2022feature, tang2023explainable}.

However, explainability in digital addiction systems extends beyond feature importance. It also involves ensuring that models behave reliably across demographic groups, addiction types, and cultural contexts \cite{smart2025beyond}. Hidden biases or spurious correlations \cite{ye2024spurious} may lead to unfair risk stratification or inappropriate intervention recommendations. Therefore, interpretability must be coupled with fairness auditing, calibration analysis, and robustness testing. Moreover, explanations must be clinically meaningful rather than merely statistically descriptive; identifying that “screen time increased” is less useful than contextualizing it within recognized behavioral relapse patterns.

Another key concern involves the tension between personalization and transparency \cite{schelenz2024transparency, schelenz2020applying}. Highly adaptive RL systems may continuously update policies in ways that are difficult to summarize or justify in simple terms. Developing explainable RL frameworks and incorporating human-in-the-loop oversight mechanisms are promising directions to ensure accountability \cite{chiodo2025formalising}.

Ultimately, explainability and interpretability are not optional add-ons but foundational requirements for safe and trustworthy AI deployment in addiction management. Transparent systems strengthen clinical integration, regulatory compliance, and user confidence, thereby facilitating responsible innovation in DTx. Ensuring that AI-driven decisions can be understood, audited, and ethically justified remains central to the sustainable advancement of personalized digital addiction care.

\subsection{Data scarcity, bias, and generalization issues}
Data scarcity, bias, and limited generalization represent some of the most significant methodological and practical challenges in applying AI and machine learning to behavioral addiction management \cite{bansal2022systematic}. Although digital systems generate large volumes of behavioral traces, clinically validated, high-quality, and longitudinal addiction datasets remain relatively scarce \cite{schmidt2019generalization}. This imbalance creates a paradox: while raw data may be abundant, labeled, structured, and outcome-verified data suitable for robust model training are often limited.

Data scarcity is particularly pronounced in clinically annotated datasets, where reliable labels such as confirmed relapse events, diagnostic categories, or treatment outcomes require expert evaluation \cite{magnini2025cost}. Behavioral addictions such as gaming, social media use, or online shopping-also lack universally standardized diagnostic criteria compared to substance use disorders, further complicating dataset harmonization \cite{loscalzo2025methodological}. Moreover, relapse events are relatively rare compared to non-relapse periods, leading to severe class imbalance in predictive modeling. Small sample sizes and short follow-up durations reduce statistical power and increase the risk of overfitting, particularly for deep learning architectures.

Bias constitutes another critical concern. Digital addiction datasets are often collected from specific platforms, demographic groups, or cultural contexts, resulting in sampling bias \cite{wang2021user}. For example, app-based data may overrepresent younger, digitally literate populations, while underrepresenting older adults or individuals with limited technological access. Behavioral patterns may differ substantially across cultures, socioeconomic backgrounds, and gender groups. If not carefully audited, predictive models may encode and amplify these biases, leading to unequal risk stratification or inappropriate intervention recommendations. Measurement bias \cite{tay2022conceptual} can also arise when self-reported data are influenced by social desirability, stigma, or recall inaccuracies.

Generalization challenges emerge when models trained on one addiction type, platform, or population fail to perform adequately in different contexts \cite{hayat2025context}. A risk prediction model developed using gaming data may not transfer effectively to gambling addiction, even if behavioral dynamics appear superficially similar. Likewise, models trained in one cultural setting may not generalize across countries due to differences in social norms, coping mechanisms, or technology usage patterns. Concept drift—where behavioral patterns evolve over time—further complicates long-term deployment, as user habits and digital environments continuously change \cite{brand2019interaction}.

Several methodological strategies are being explored to address these issues. Data augmentation, transfer learning, and self-supervised pretraining can mitigate small-sample limitations. Federated learning frameworks enable collaborative model training across institutions without direct data sharing, enhancing diversity while preserving privacy. Domain adaptation techniques improve cross-population transferability, while fairness-aware learning algorithms attempt to reduce demographic disparities. Additionally, continuous model monitoring and recalibration are essential to maintain performance under evolving behavioral conditions.

\newpage
\section{Ethical, Legal, and Societal Considerations}\label{sec7}
Digital therapeutics (DTx) represent a distinct category of digital health interventions, defined by explicit therapeutic intent, software-based mechanisms of action, and increasing regulatory scrutiny. Positioned between pharmaceuticals and consumer-facing digital applications, DTx operate within fragmented and still-developing governance frameworks that often lag behind rapid software innovation, creating an inherently ambiguous regulatory space \cite{refolo2022ethics, rassi2022digital}. These tensions are particularly visible in behavioral addictions -such as gambling, gaming, and problematic social media or smartphone use- where digital tools rely on continuous data collection, algorithmic personalization, and real-time behavioral feedback, situating them at the intersection of healthcare systems, consumer technology markets, and data-driven economies. As a result, their governance requires ethical and legal approaches extending beyond traditional clinical ethics and individual patient–provider relationships, demanding a broader systemic perspective \cite{torous2021growing}.

Unlike face-to-face psychosocial treatments, digital interventions for behavioral addictions embed therapeutic reasoning directly into software infrastructures, rendering ethical risk a persistent feature of the technological system rather than a discrete clinical event. Continuous monitoring, behavioral tracking, automated feedback, and algorithmic inference reshape power relations among users, developers, clinicians, and institutions, often diffusing responsibility and blurring accountability for adverse outcomes \cite{torous2025evolving}. Behavioral addictions share core mechanisms with substance addictions -particularly compulsion and reinforcement- which are amplified in digital environments that simultaneously facilitate both harm and intervention through habit-forming design and pervasive connectivity \cite{griffiths2017behavioural, king2020problematic}. While digital interventions offer significant benefits in terms of accessibility, anonymity, and scalability, these advantages are accompanied by heightened risks related to privacy, autonomy, justice, and social inequality.

The growing integration of AI, mHealth, and DTx into addiction care necessitates a paradigm shift where ethical, legal, and societal considerations are treated as foundational to responsible intervention design rather than secondary safeguards. The vulnerability of users struggling with impulse control and compulsive behaviors makes the “safety-by-design” principle an ethical prerequisite to prevent iatrogenic harm, such as technology-induced dependency or algorithmic discrimination \cite{topol2019high}. Furthermore, as digital tools increasingly reach marginalized groups, the potential for digital health technologies to reproduce or exacerbate existing social disparities must be critically addressed within governance models \cite{park2022content}. Consequently, the transition toward digitally mediated addiction care requires robust legal frameworks that can accommodate the dynamic nature of software while protecting the fundamental rights and psychological well-being of users in an increasingly data-driven therapeutic landscape.

\subsection{Privacy \& data protection}
Privacy and data protection constitute one of the most structurally embedded ethical challenges in AI-driven digital interventions for behavioral addictions, such as gambling, gaming, and compulsive social media use. Tools developed for these conditions routinely collect highly granular, longitudinal behavioral, psychological, and contextual data -including time-stamped logs, emotional states, impulse control markers, and relapse indicators. Compared to conventional health records, these data are inferential and predictive, substantially increasing their ethical sensitivity and potential for harm if misused \cite{torous2021growing, mittelstadt2016ethics}. Intensive collection methods, such as EMA and passive monitoring, allow for the precise recording of user habits like “infinite scrolling” or “loot-box” spending, but uncertainties surrounding data access and storage -highlighted by incidents like the BetterHelp controversy- undermine trust and negatively affect sustained therapeutic engagement \cite{alhammad2024patients, babu2025digital}. Although regulatory frameworks like GDPR and HIPAA establish baseline legal protections, compliance alone does not guarantee ethically robust data governance or meaningful user control in the specialized context of addiction care \cite{lustgarten2020digital, hamdoun2023ai}.

The technical architecture of DTx amplifies these risks through continuous monitoring and cloud-based infrastructures that often involve third-party analytics, diffusing accountability and increasing the threat of cyberattacks \cite{ajayi2025ai}. Many applications operate outside strict health data regulations, creating an “inaccurate expectation of privacy” where users assume clinical-level confidentiality that does not exist. This is particularly concerning for free applications that rely on data-driven business models where sensitive information becomes the “currency of care,” being on-sold for commercial purposes or used for behavioral targeting within the global “infosphere” \cite{landers2023ethical, herzog2022digital}. Consequently, behavioral addiction data intersect with commercial platform economies in Industry 4.0 environments, where algorithmic profiling can detect user vulnerabilities to tailor content that exploits compulsive patterns \cite{helm2024co, topol2019high}.

To mitigate these threats, the literature emphasizes the integration of advanced technical and design safeguards. AI-driven systems must move toward Responsible UX design and Value Sensitive Design that prioritize user diversity and digital well-being during development \cite{zannat2023mindful}. Key strategies include:

\begin{itemize}
	\item Privacy-preserving AI: Utilizing techniques such as federated learning, which trains models on decentralized devices without raw data transfer, and differential privacy, which adds statistical noise to prevent individual identification \cite{ali2025artificial}.
	\item Transparent governance: Establishing clear protocols for data stewardship, strong encryption, and secure sharing mechanisms to prevent unauthorized secondary data use \cite{ajayi2025ai, khakpaki2025ai, suva2024artificial}.
	\item Ethical validation: Implementing robust frameworks for consent and clinical validation, especially as systems process detailed doctor-patient interactions \cite{orhan2025benefits}.
\end{itemize}

Ultimately, privacy in digital addiction interventions must be understood as a foundational ethical condition rather than a purely technical requirement. Secure, private, and confidential environments are paramount for building the trust necessary for users to disclose habitual impulses -such as those found in internet addiction and gaming disorder- and engage in long-term recovery \cite{humphrey2022adapting, thomas2025exploring, ni2025scoping}. Ensuring that AI technologies do not expose individuals to profiling or unauthorized data use remains essential for the integrity of digital health systems and the protection of service users' informational autonomy \cite{khakpaki2025ai, van2018problematic}.

\subsection{Security \& safety concerns}
Digital interventions for behavioral addictions -including compulsive gambling, gaming, and problematic social media or smartphone use- introduce security and safety risks that extend beyond conventional data protection, encompassing cybersecurity, clinical safety, algorithmic reliability, and the prevention of technology-induced harm. Audits of digital mental health applications across the spectrum of addictive behaviors reveal widespread governance deficiencies, including opaque privacy practices and extensive third-party data tracking \cite{huckvale2019assessment}. These vulnerabilities are especially consequential in addiction contexts, where security breaches involving behavioral or financial data may intensify stigma, facilitate economic exploitation, or expose users to targeted marketing that triggers addictive impulses, directly undermining therapeutic goals. Furthermore, the safety of addiction-related data is uniquely compromised by the behavioral patterns of the target population, such as the increased frequency of device loss, sale, or theft, which can expose sensitive health records to unauthorized acces. The absence of consistent legal frameworks across jurisdictions further complicates the protection of these digital therapeutic infrastructures, often leaving users without a clear understanding of how their data is handled or protected \cite{landers2023ethical}.

Beyond data breaches, DTx generate novel iatrogenic risks through algorithmic errors, inadequate crisis escalation, and the creation of a “false sense of security” \cite{babu2025digital}. Safety assessments identify significant clinical vulnerabilities, including misleading therapeutic feedback, incorrect risk calculations, and “AI hallucinations” -fabricated or incorrect outputs from generative models that increase cognitive burden and emotional distress during crisis situations \cite{song2025ai}. In behavioral addictions, where interventions rely on automated relapse prediction and craving regulation, misclassification or system failure may normalize excessive engagement or delay urgent clinical intervention. Furthermore, many generic chatbots lack clinical validation and may provide unsafe guidance, leading users to rely on self-guided tools rather than seeking necessary professional care \cite{singh2024human}. This reliance is particularly dangerous when applications have not undergone formal accreditation by evaluation bodies like Organisation for the Review of Care and Health Apps (ORCHA) or the Mental Health Apps Library (MIND database), leaving their therapeutic effectiveness and usability uncertain.

Safety concerns are further amplified by design features that may unintentionally replicate the very addictive mechanisms they aim to treat. Gamification, reward-based nudging, and ``techno-social closed-loop systems" risk mirroring problematic gaming or social media dynamics when insufficiently regulated \cite{king2020problematic, herzog2022digital}. While intended to support behavior change, constant monitoring and performance metrics can generate pressure, guilt, or shame, potentially reinforcing compulsive self-monitoring patterns. In environments designed for internet or smartphone use, persuasive and habit-forming strategies -such as infinite content feeds, autoplay, and personalized notifications- are often crafted to maximize engagement and user retention, which can contribute to addiction-like behavioral loops \cite{helm2024co, montag2019addictive}. Consequently, gamified interventions can function as a "double-edged sword," where motivational mechanisms intended for recovery instead encourage compulsive overuse or lead users into physically or psychologically risky engagement patterns \cite{spahl2024gamified, sahni2025healing}.

A critical safety gap exists in the management of high-risk situations, such as suicidal ideation, which frequently co-occur with severe behavioral addictions and other conditions like ADHD or depression \cite{gamble2020artificial}. Autonomous AI systems often struggle with the complexity of psychiatric crises; some may provide inadequate responses or struggle to demonstrate professional judgment, while others abruptly terminate conversations upon detecting risk, highlighting the limitations of fully autonomous digital formats \cite{ni2025scoping}. To ensure safety, digital addiction tools must incorporate immediate access to emergency services, robust moderation systems to prevent the normalization of harmful behaviors, and rigorous testing before being deemed clinically effective. These findings underscore that in behavioral addiction care, safety-by-design, human-in-the-loop oversight, and rigorous post-market monitoring are ethical prerequisites rather than optional enhancements to prevent unintended outputs and adversarial manipulation \cite{klouvckova2025psychological}.

\subsection{Equity, access, and digital divide}
Digital interventions are frequently presented as scalable solutions to longstanding access gaps in behavioral addiction treatment, offering a potential response to the global shortage of mental health professionals and bridging geographical barriers through telemedicine and mobile platforms \cite{panahi2025future, efthymiou2023transforming, landers2023ethical}. By increasing anonymity and providing psychoeducation or early support for those on waiting lists, these technologies can reduce the stigma that often prevents addicted individuals from seeking face-to-face care. Accordingly, digital interventions act as a ‘buffer zone’ or ‘intermediary space’, allowing for clinical support while maintaining anonymity and physical distancing. However, growing evidence indicates that these tools often reproduce existing inequalities shaped by digital literacy, device ownership, and reliable internet access, systematically excluding socioeconomically marginalized groups \cite{watson2023fda}. Approximately 40\% of the world’s population lacks reliable internet access, setting a structural ceiling for digital-first strategies and effectively transforming the digital divide into a new social determinant of health \cite{lodha2025ethical}. In the context of behavioral addictions, these barriers intersect with financial precarity and unstable housing, raising significant doubts about whether digital-first strategies align with public health equity goals, especially since gambling-related harm disproportionately affects lower-income groups who are least likely to have stable internet or private digital spaces.

The benefits of DTx are further stratified by income, education, age, gender, and language, mirroring broader health inequalities \cite{western2025bridging}. While remote participation can enhance privacy, it simultaneously creates new structural barriers linked to technological infrastructure and professional capacity \cite{doraiswamy2019empowering}. Many commercial platforms operate on subscription-based or tiered pricing models, placing essential therapeutic features behind paywalls and risking the transformation of DTx into ``boutique services" for privileged users rather than population-level public health interventions \cite{babu2025digital}. Furthermore, digital health applications often assume that users possess sufficient time, autonomy, and resources to implement recommended behavioral changes “assumptions that fail to account for the social conditions influencing addictive behaviors and treatment accessibility. This socio-technical context is critical, as problematic internet or smartphone use emerges through interactions between users and social environments; thus, social inequalities in the capacity to regulate digital behavior may influence both the development and the recovery from addiction \cite{van2018problematic}.

Equity claims are also undermined by algorithmic bias and the lack of representative data. AI models often perform poorly for underrepresented groups -including racial minorities, non-native speakers, and low-income communities- because they are frequently trained on limited or non-representative datasets \cite{ali2025artificial}. In behavioral addiction care, biased risk profiling and inaccurate risk assessments can exacerbate existing inequities in service distribution, particularly in low-resource settings where the addiction burden and treatment gaps are substantial. Industry 4.0 technologies therefore have the potential to widen social and economic inequalities if access remains uneven, reinforcing the digital divide \cite{sahni2025healing, zannat2023mindful}.

To ensure that digital innovation does not disproportionately benefit already advantaged populations, ethical technology design must prioritize inclusivity and accessibility. Socially responsible frameworks must explicitly address the digital divide by developing low-storage applications, offline functionality, and interfaces compatible with older devices to avoid excluding economically disadvantaged users \cite{spahl2024gamified}. Accessibility also requires adapting interventions for different literacy levels and communication preferences, including voice-over support and cultural adaptation. Ultimately, achieving health equity in digital addiction care requires public funding, free access initiatives, and collaborative frameworks involving policymakers, healthcare professionals, and technologists to ensure that AI tools bridge service gaps rather than widening the chasm of social inequality.

\subsection{Stigmatization \& discrimination risks}
Behavioral addictions remain highly stigmatized and are frequently framed through moralizing narratives that emphasize individual failure and a lack of self-control rather than structural or commercial determinants. While digital health tools are often promoted for their ability to provide a judgment-free, anonymous environment -potentially encouraging help-seeking among those who fear face-to-face exposure- they also risk reproducing and even amplifying stigma through surveillance-oriented design and algorithmic labeling \cite{jefferies2024telehealth}. In interventions for gambling, gaming, and internet addiction, continuous monitoring and risk-scoring systems may position users as objects of surveillance rather than active agents in their own recovery, potentially deterring individuals already shaped by shame (whakama) and concealment. Furthermore, if digital tools use explicit labels like "addicted," they may negatively affect a user's self-image and resistance to intervention, particularly among youth who frequently avoid treatment due to embarrassment or doubts about their ability to quit \cite{efthymiou2023transforming}.

The integration of AI into this landscape introduces significant risks of algorithmic discrimination and the reinforcement of social biases. NLP models used in psychiatry and behavioral prediction systems have demonstrated biases related to race, gender, religion, nationality, and age, which can lead to misclassification or inaccurate risk assessments for marginalized populations. When algorithms are trained on non-representative or industry-influenced datasets, they risk embedding cultural and socioeconomic assumptions into digital addiction care, effectively "moralizing" health behaviors by transforming everyday digital practices into measurable indicators of personal responsibility \cite{herzog2022digital}. These "techno-social closed-loop systems" can induce feelings of "bad conscience, guilt, and shame" when users fail to meet digital benchmarks, while gamified elements like leaderboards and performance metrics may intensify unrealistic social comparisons and feelings of inadequacy \cite{sahni2025healing}. Moreover, focusing exclusively on individual pathology through algorithmic labeling risks overlooking the role of platform design -such as habit-forming interfaces- in shaping compulsive behaviors, thereby framing problematic use solely as a personal failure.

Stigmatization and discrimination risks also extend beyond the clinical interaction into broader institutional settings. Unauthorized disclosure or secondary use of sensitive behavioral addiction data by insurers, employers, or financial institutions can lead to severe social exclusion and employment discrimination \cite{mittelstadt2016ethics}. Labels such as “high-risk user” derived from predictive analytics can bias decision-making in these sensitive contexts, making privacy protections a central component of ethical addiction treatment \cite{ferrara2024fairness}. Proprietary and opaque algorithms further complicate this issue, as it remains difficult to evaluate whether nudges and behavioral predictions are operating fairly or if they are prioritizing engagement metrics over therapeutic outcomes. For marginalized or historically mistrustful communities, these technological and institutional exclusions can reinforce a cycle of inequality and disengagement from healthcare services.

To ensure that digital interventions support recovery without reproducing shame or inequality, ethical digital addiction care must move toward culturally sensitive and inclusive design. This requires the use of diverse training datasets, the implementation of safeguards against stigmatizing language, and a clear separation from commercial gambling and gaming interests. Value-sensitive design frameworks and stronger oversight mechanisms are essential to prevent discriminatory practices and ensure that AI systems do not reinforce societal biases. Ultimately, addressing both interpersonal stigma and structural sources of algorithmic discrimination is necessary to build the trust required for long-term engagement and to protect the informational autonomy of vulnerable users \cite{ajayi2025ai}.

\subsection{Global policy recognition and international regulatory frameworks }
Behavioral addictions have become increasingly prominent in global public health, digital governance, and regulatory policy debates. The ubiquity of digital technologies has facilitated continuous engagement in potentially addictive activities, including gambling, gaming, social media use, online shopping, and the consumption of online sexual content. While research has expanded significantly, policy interventions remain fragmented. Gambling and gaming are the most comprehensively regulated sectors, whereas other behaviors are typically addressed under general digital platform regulations and consumer protection laws  \cite{griffiths2017behavioural}.

Global political recognition has been solidified through international diagnostic systems. The World Health Organization (WHO) officially recognized Gaming Disorder as a formal mental health condition by including it in the ICD-11\footnote{https://www.who.int/standards/classifications/classification-of-diseases}. Similarly, the American Psychiatric Association (APA) added Internet Gaming Disorder to the DSM-5 as a category requiring further clinical study \cite{sharma2015fifth}. At the European Union level, the Digital Services Act (DSA) mandates that platforms mitigate risks associated with addictive design\footnote{https://eur-lex.europa.eu/eli/reg/2022/2065/oj/eng}. Furthermore, the Pompidou Group has published reports emphasizing the necessity of harm-reduction policies specifically targeting digital gambling and gaming environments\footnote{https://www.coe.int/en/web/pompidou/online-addictions}.

Gambling remains the most strictly regulated behavioral addiction worldwide. National policies generally encompass licensing, advertising restrictions, and harm-reduction mechanisms \cite{gainsbury2014review}. The United Kingdom initiated a "digital age reform" through its 2005 Gambling Act and subsequent 2023 White Paper\footnote{ https://www.gov.uk/government/publications/high-stakes-gambling-reform-for-the-digital-age/high-stakes-gambling-reform-for-the-digital-age}. Sweden implemented Spelpaus, a national self-exclusion system, under its 2018 Gambling Act (Spelinspektionen, 2018) \cite{haakansson2020gambling}, while Australia oversees online betting through the 2001 Interactive Gambling Act\footnote{https://www.infrastructure.gov.au/department/media/publications/interactive-gambling-act-2001}.

Governments are taking increasingly stringent measures, particularly to protect minors. In 2021, China restricted youth gaming time to three hours per week and mandated a real-name verification system\footnote{https://www.loc.gov/item/global-legal-monitor/2021-09-13/china-new-rule-further-restricts-minors-from-playing-online-videogames/}. The Belgian Gaming Commission effectively banned loot boxes by classifying them under gambling laws\footnote{https://www.bbc.com/news/technology-43906306}. In the United States, the proposed Kids Online Safety Act (KOSA) aims to regulate algorithmic recommendation systems and addictive design features\footnote{https://www.congress.gov/bill/118th-congress/senate-bill/1409}. Across Europe, regulatory momentum has also intensified. For example, France enacted legislation in 2023 requiring parental consent for social media use by children under the age of 15, reflecting a broader shift toward age-based digital protection frameworks\footnote{https://www.vie-publique.fr/loi/289592-majorite-numerique-reseaux-sociaux}. At the supranational level, the EU’s Digital Services Act further strengthens safeguards for minors by imposing risk-assessment and mitigation duties on large platforms, including provisions addressing algorithmic amplification and targeted advertising practices\footnote{https://eur-lex.europa.eu/eli/reg/2022/2065/oj/eng}. Similar policy debates and inquiries on minimum age thresholds and platform accountability are ongoing in countries such as Australia, indicating a widening global trend toward stricter governance of youth engagement in digitally mediated environments. These measures represent a shift from individual responsibility to systemic age-verification mandates, aiming to protect minors from the dopamine-driven design architectures of modern digital platforms. Other behaviors, such as compulsive buying and online sexual content consumption, are generally addressed through consumer rights directives or safety laws requiring age verification\footnote{https://eur-lex.europa.eu/eli/dir/2011/83/oj/eng}.

In Türkiye, the legal framework for combating behavioral addictions is primarily executed through deterrent measures under Law No. 7258 and the Turkish Penal Code, particularly in areas with high economic impact such as illegal betting and gambling\footnote{https://www.mevzuat.gov.tr/mevzuat?MevzuatNo=5237\&MevzuatTur=1\&MevzuatTertip=5}. Individuals or institutions providing financial mediation or advertising for these illicit activities face prosecution with prison sentences. Digital oversight is conducted by the Information and Communication Technologies Authority (BTK) within the scope of Law No\footnote{https://www.mevzuat.gov.tr/MevzuatMetin/1.5.2813.pdf}. 5651; addictive gambling and gaming content classified under "catalog crimes" can be blocked ex officio as an administrative measure without awaiting a prior judicial decree.

Conversely, the legal process adopts a social law perspective focused on the ``protection" of addicted individuals and children at risk. Pursuant to the Child Protection Law\footnote{https://www.mevzuat.gov.tr/MevzuatMetin/1.5.5395.pdf}, if gaming or internet addiction threatens a child’s educational or health development, courts may impose mandatory counseling or health measures, monitored by the Ministry of Family and Social Services. At the clinical level, behavioral addictions have been officially defined as medical disorders under the 2024-2028 National Strategy Document, granting individuals the right to free treatment under the Social Security Institution (SGK) and access to AMATEM/ÇEMATEM (Alcohol and Substance Addiction Treatment Centers) units. Furthermore, YEDAM (Green Crescent Counseling Centers)\footnote{https://www.yedam.org.tr/} units within the Green Crescent (Yeşilay)\footnote{https://www.yesilay.org.tr/} complement this legal and medical infrastructure by providing free legal and psychological counseling to ensure social rehabilitation and reintegration.

As part of these regulations for treatment and rehabilitation, the regulatory oversight of DTx and telemedicine remains fragmented and poorly aligned with the rapid evolution of software-based addiction interventions. Functionally similar tools may be regulated as prescription DTx, Software as a Medical Device (SaMD), or unregulated wellness applications based on stated marketing claims rather than real-world clinical impact, which weakens accountability and enables market entry without proportional evidence. Existing frameworks, largely designed for static medical hardware, inadequately address adaptive algorithms, frequent software updates, and the real-time behavioral risks inherent in addiction care \cite{rassi2022digital}. This has created a “large grey market” of digital health applications offering behavior-change interventions without comprehensive evaluation of their risks or effectiveness \cite{herzog2022digital}. While some products, such as reSET®, have undergone randomized controlled trials, many digital addiction tools enter the market with minimal scrutiny, as technological innovation consistently outpaces the development of regulatory and governance frameworks -a phenomenon often associated with the technology sector’s “move fast and break things” ethos \cite{chesterman2021move}.

In the United States, the Food and Drug Administration (FDA) regulates some DTx as SaMD, yet many AI-driven chatbots and wellness apps remain outside this oversight, leading to a lack of transparency and harm monitoring for adaptive systems \cite{gamble2020artificial}. Similarly, in Europe and the United Kingdom, baseline regulation through the EU Medical Device Regulation (MDR), UKCA marking, and the NHS Digital Technology Assessment Criteria (DTAC) often fails to address addiction-specific ethical and clinical challenges\footnote{https://www.england.nhs.uk/digitaltechnology-assessment-criteria-dtac/}. While newer initiatives like the EU Artificial Intelligence Act and the European Health Data Space introduce risk-based oversight, they remain weakly integrated with addiction care standards, leaving many gambling and gaming interventions categorized as wellness tools despite targeting vulnerable populations \cite{saeidnia2024ethical}. This regulatory gap shifts the burden of risk onto users and clinicians, creating uncertainty for healthcare providers who must recommend treatments without established medical certification or formal ethical standards from professional organizations like the APA or ABPP \cite{landers2023ethical}.

The necessity for robust, adaptive regulatory frameworks is particularly acute for AI-enabled addiction care, where the boundary between therapeutic technology and commercial digital engagement systems is frequently blurred \cite{ni2025scoping}. Policymakers must balance patient safety with technological innovation by implementing mechanisms such as regulatory sandboxes, mandatory certification processes, and independent audits for algorithmic transparency. Effective governance should not only address diagnostic recognition within systems like the DSM and ICD but also consider the technological infrastructures -such as habit-forming platform designs- that contribute to problematic engagement patterns. Furthermore, maintaining fidelity to evidence-based treatment models, such as CBT, is essential during the digital adaptation process to ensure conceptual integrity \cite{helm2024co}.

As DTx become more autonomous in clinical decision-making, governance must ensure that AI supports rather than replaces human judgment, emphasizing the need for transparent communication regarding the role of AI in healthcare decisions \cite{el2024ai}. In behavioral addiction contexts, this includes integrating harm-reduction strategies like age-verification, pre-commitment tools, and self-exclusion registries into regulatory requirements. Ultimately, determine whether gamified health applications should be regulated as medical devices remains a major societal challenge, as this classification dictates the requirements for clinical validation and safety monitoring \cite{sahni2025healing}. Without standardized evaluation frameworks and enforcement of the same regulations that apply to in-person services, digital addiction interventions risk expanding commercially without meeting critical clinical and ethical standards.

\subsection{Insurance \& reimbursement models}
Insurance coverage and reimbursement represent one of the most critical structural bottlenecks for the ethical and sustainable implementation of DTx in behavioral addiction care. Although a limited number of prescription DTx have entered payer systems, reimbursement remains inconsistent and weakly aligned with long-term outcomes such as relapse prevention and psychosocial recovery \cite{jefferies2024telehealth}. This regulatory-reimbursement feedback loop often prevents digital tools from being recognized as legitimate clinical interventions, leaving them instead as peripheral consumer products. Payers frequently hesitate to fund interventions that lack standardized validation and post-market evaluation, which in turn limits the real-world evidence generation needed to justify further scale. In the absence of formal reimbursement, many digital addiction interventions are pushed into consumer or employer-based markets -an acute concern for individuals with gambling or gaming addictions who often face significant financial strain, debt, and social marginalization.

The integration of digital DBTs into institutional health strategies, including employer wellness initiatives and insurance-based prevention programs, raises significant ethical concerns related to fairness and accountability. When individuals are incentivized or pressured to adopt behavioral monitoring tools, responsibility for health risks being reconfigured from structural and commercial determinants toward individual compliance with digitally quantified performance metrics. In the context of behavioral addictions, this shift may further individualize both risk and recovery, framing them as matters of personal discipline rather than outcomes shaped by dynamic interactions between users, technologies, and their broader social environments \cite{herzog2022digital}. Furthermore, many commercial platforms utilize subscription-based or tiered pricing models, effectively placing high-quality therapeutic support behind paywalls and exacerbating disparities for the approximately 10\% of addiction patients who may lack any insurance coverage \cite{babu2025digital}.

Despite these challenges, hybrid and stepped-care models -where digital CBT serves as an entry point before in-person therapy- show promise for sustainable integration. From a systems perspective, AI-driven tools such as automated screening systems and virtual assistants can improve healthcare efficiency by streamlining clinical workflows, reducing waiting lists, and providing 24/7 support. In countries like the UK, digital tools are increasingly viewed as supplementary resources that extend therapy beyond the time-restricted sessions typical of publicly funded systems \cite{thomas2025exploring}. However, the long-term financial sustainability of these innovations is often hampered by a reliance on short-term research grants rather than stable public funding, with serious health games receiving significantly less development capital than commercial entertainment products \cite{spahl2024gamified}.

Ultimately, the economic recognition of behavioral addiction interventions remains underdeveloped, particularly for emerging areas like problematic smartphone use. While AI technologies offer the potential to lower long-term expenditures by reducing relapse rates and caseloads, their impact is limited by a lack of institutional recognition. Without dedicated reimbursement pathways that recognize digital interventions as an integral part of the behavioral addiction care continuum -rather than market-driven accessories- these tools will remain unevenly accessible and weakly accountable. Aligning technological innovation with institutional governance and stable funding is essential to ensure that DTx serve broader public health goals rather than purely commercial interests \cite{zannat2023mindful}.

\newpage
\section{Economic, Market, and Ecosystem Dimensions}\label{sec8}
Digital tools for behavioral addictions -including gambling, gaming, and compulsive digital engagement- are increasingly shaped by market logics, platform economies, and private-sector business models rather than public health priorities. Without ethical governance, these dynamics risk privileging scalability, data extraction, and user engagement over prevention, equity, and structural intervention \cite{saeidnia2024ethical}. Venture-capital-driven ecosystems often prioritize rapid growth and engagement metrics, creating a fundamental tension with clinical robustness and the long-term therapeutic outcomes required for sustainable addiction recovery \cite{rassi2022digital}. This market-driven approach is particularly concerning in the context of behavioral addictions, where the economic incentives of digital platforms often align with the very engagement patterns the interventions aim to mitigate.
From a macro-economic perspective, behavioral addictions are linked to broader socio-economic consequences, including reduced workplace productivity, increased healthcare utilization, and significant social service burdens. AI-driven scalable interventions -such as automated monitoring, chatbot-based support, and personalized relapse prevention- offer the potential to optimize resource utilization and support population-level management strategies, particularly in regions facing acute shortages of mental health professionals \cite{suva2024artificial, khakpaki2025ai}. By streamlining clinical workflows and providing low-cost screening, these technologies could theoretically lower long-term healthcare expenditures. However, the market readiness and ecosystem integration of such solutions remain severely constrained by a lack of empirical evidence and a shortage of real-world implementation studies. Consequently, without deliberate policy-level integration and a shift away from purely commercial metrics, digital addiction tools risk remaining fragmented accessories rather than integral components of a socially responsible healthcare ecosystem.

\subsection{Business models for digital health \& addiction solutions and the integration of artificial intelligence}
Digital addiction interventions currently operate within an economic framework driven by subscription-based services, employer or provider contracts, and data-driven monetization strategies. While these models facilitate market expansion, they frequently prioritize user engagement over harm reduction. In the context of behavioral addictions -such as gambling, gaming, and social media use- this tension is particularly acute, as reward-based and persuasive design elements often replicate the very mechanisms associated with problematic use, thereby blurring the ethical boundaries between treatment and harm \cite{billieux2015we}. This commercial trajectory mirrors the rapid growth of the AI healthcare industry, which is projected to expand from five billion dollars in 2020 to forty-five billion dollars by 2026 \cite{Mahajan2020AIHealthcare}. Within the “Healthcare 5.0” ecosystem, the integration of cloud computing, deep learning, blockchain, and the Internet of Things (IoT) supports innovation in service delivery \cite{klouvckova2025psychological, karri2025healthcare}.  AI-based chatbots and virtual assistants provide scalable support for mental health and addiction-related issues, yet they often conflict with the engagement-driven models of Industry 4.0 that prioritize user attention. Design features intended to maximize time spent on platforms are directly linked to technology overuse and addictive behavioral patterns \cite{zannat2023mindful}.

As digital addiction solutions proliferate, many applications -particularly free versions- have adopted data-driven business models where sensitive health data becomes the primary economic resource. This raises profound ethical concerns regarding the commercialization of behavioral health data, leading to a “currency of care” based on data extraction \cite{hamdoun2023ai}. The involvement of private companies often centers on sharing confidential personal information, creating a potential misalignment between profit incentives and patient protection. Furthermore, there are few regulatory barriers to prevent AI chatbot companies from directing vulnerable users toward paid advertising. Despite these risks, AI offers critical tools for addiction management; machine learning and real-time monitoring enable enhanced prevention, diagnosis, and recovery processes \cite{khakpaki2025ai}. For instance, by analyzing “infinite scrolling” patterns in social media or the frequency of “loot-box” purchases in gaming, AI systems can now construct personalized risk profiles that identify compulsive behavior more accurately than traditional self-reports \cite{dehbozorgi2025application, andrade2025mapping}.

Systems like SoberGrid combine social support with AI analytics to provide tailored feedback, while other platforms utilize educational modules and behavioral interventions for withdrawal management and relapse prevention \cite{ajayi2025ai}. However, as many popular chatbots (e.g., Shim, Woebot, Wysa) are developed by technology companies without clinical backgrounds, these tools are increasingly viewed as "complementary tools" within a "blended care model" rather than total substitutes for professional therapy \cite{gamble2020artificial}.

The paradox of digital addiction interventions lies in the fact that they often operate within the same technological ecosystems that contributed to the development of addictive environments. Engagement-oriented strategies such as gamification, streaks, and push notifications are used to maximize retention, mirroring persuasive technologies that contribute to compulsive usage patterns and digital dependency. Consequently, game-based mental health technologies function as a ``double-edged sword," promoting therapeutic engagement while simultaneously raising concerns about behavioral dependency \cite{babu2025digital}. While ecosystem-based harm-reduction strategies -such as real-time AI-driven pop-up interventions in gambling environments- show promise, many smartphone regulation apps for productivity and family monitoring are being commercialized with limited empirical evidence of effectiveness. Ultimately, while AI-enabled solutions can address workforce shortages by providing scalable, low-cost support to large populations, the structural mismatch between market logic and the ethical demands of addiction care necessitates greater accountability and transparency \cite{ramesh2016gaming, jefferies2024telehealth}. When supported by ethical regulation and transparent models, these technologies may redefine service delivery by combining automated care with professional oversight \cite{sahni2025healing, helm2024co}.

\subsection{Integration into healthcare \& insurance systems}
The integration of digital addiction tools into healthcare and insurance systems remains uneven, with many interventions operating parallel to, rather than within, established addiction services. This fragmentation limits continuity of care, clinical accountability, and coordinated follow-up, particularly in cases of gambling and gaming addiction which are complicated by financial, familial, and psychosocial factors \cite{watson2023fda}. Institutional barriers, such as limited clinician training, unclear liability, and misaligned reimbursement models, remain more significant than technological hurdles \cite{torous2025evolving}. Research consistently shows that clinician involvement is the cornerstone of trust and adherence in DTx. Clinician-mediated digital formularies can serve as vital ethical filters, ensuring safer adoption and contextual interpretation of digital data. While web-based interventions may reduce stigma and increase access, their long-term effectiveness depends on being embedded within formal care pathways rather than functioning as standalone solutions \cite{bernstein2023case, park2022content}.

Healthcare professionals (HCPs) are increasingly prompted to recommend or prescribe digital offerings as patients seek guidance on tech-based treatment options. The integration of AI technologies is closely linked to the global expansion of EHR \cite{evans2016electronic}. These systems combine EHR data with smartphone-based electronic phenotyping and social media data to create unprecedented resources for addiction research. AI-driven decision-support systems analyze speech patterns, physiological, and behavioral data to identify mental health conditions and addiction risks more effectively than traditional methods \cite{shatte2019machine, zainab2025innovative}. Telepsychiatry platforms further enhance care by using AI algorithms to detect indicators of emotional distress during consultations. However, a clear consensus exists that AI should augment -not replace- human-led clinical provisions.

Sustainable impact requires collaboration between developers, policymakers, and service providers to embed these tools within broader systems. AI-driven pop-up interventions and personalized behavioral feedback, initially proposed for gambling harm reduction, offer a scalable "preventive ecosystem" model applicable to various digital addictions, including compulsive smartphone and social media use \cite{efthymiou2023transforming}. AI and mobile health tools, such as the Manaaki app, address the gap for individuals who do not seek traditional services. These applications support the treatment pathway by providing psychoeducation before therapy, reinforcing tasks during sessions, and serving as ``step-down" tools for relapse prevention. Nonetheless, an over-reliance on individual behavior change may obscure structural determinants of health, such as working conditions and social inequality. Ultimately, successful integration requires a systemic transformation that addresses ethical uncertainties and incorporates AI as a decision-support tool within evidence-based clinical workflows.

\subsection{Interoperability with EHR/EMR platforms}
Interoperability with EHR and EMR systems remains underdeveloped in digital addiction care, a gap that affects not only gambling interventions but the entire spectrum of behavioral addictions. Most digital treatments for gaming disorder, social media addiction, and compulsive smartphone use operate in isolation, using non-standardized data architectures. This fragmentation prevents the integration of patient-generated data -such as app usage duration, social media engagement triggers, or late-night gaming sessions- into clinical monitoring and decision-making. This lack of synergy limits clinicians’ ability to interpret digital metrics within broader longitudinal and comorbidity contexts, such as the relationship between social media-induced anxiety and depressive episodes \cite{topol2019high}.

Regulatory frameworks are increasingly addressing this by framing interoperability as a safety requirement. The NHS Digital Technology Assessment Criteria emphasize that responsible adoption of addiction tools requires seamless data stewardship and clinical relevance. Without interoperable infrastructures, the high-frequency data generated by smartphone regulation apps or gaming monitors risk becoming a burden to healthcare providers rather than a clinically actionable asset, ultimately undermining the preventive potential of AI \cite{ramesh2016gaming}.

AI systems empower clinicians by analyzing multimodal data sources, which is essential for identifying the diverse triggers of different addictions. Research has shown that combining EHR data with linguistic patterns from social media or neuroscience data can effectively simulate the course of mental health conditions \cite{schmaal2015predicting}. In the broader field of ICT (Information and Communication Technology) addictions, machine learning can merge behavioral patterns with digital activity to identify subtle markers of internet and smartphone addiction that traditional self-reports -often subject to recall bias- fail to capture \cite{andrade2025mapping, klouvckova2025psychological}.

By analyzing diverse datasets -including wearables that monitor physiological stress during gaming or social media interactions- AI can predict addiction susceptibility and facilitate targeted prevention programs. While some specific platforms like Therachat or Recovery Record have begun linking patient logs with clinical providers, a system-wide ``eHealth ecosystem" is required to ensure that a user’s digital footprint leads to proactive care rather than just passive data collection \cite{gamble2020artificial}.

Ultimately, the effectiveness of these tools -whether for smartphone regulation or gambling harm reduction- depends on their seamless integration into clinical workflows. This integration must move beyond simple interface design and focus on a systemic transformation where AI acts as a decision-support tool, bridging the gap between a user's digital environment and the healthcare system's formal oversight.

\subsection{Public health implications \& policy integration}
From a public health perspective, AI-driven digital tools for behavioral addictions -including gambling, gaming, social media, and problematic smartphone use- offer transformative potential for early intervention, relapse prevention, and population-level harm reduction. However, without structured legal, regulatory, and governance frameworks, their implementation remains inconsistent and patient protections uneven \cite{jefferies2024telehealth}. Evaluation of these DTx must therefore extend beyond individual effectiveness to include distributive impact, governance quality, and their contribution to broader population-level prevention strategies \cite{refolo2022ethics}.

Digital therapeutics and AI technologies are frequently framed as a response to the global mental health crisis and the acute shortage of clinicians \cite{landers2023ethical, singh2021prevalence}. Approximately one in eight people globally live with a mental health condition, and AI promises to alleviate system pressures by efficiently allocating human resources and providing low-threshold support for individuals who may not seek traditional services. Yet, researchers caution that AI is a  ``tool, not a panacea”; technological investment must not supersede investment in people and infrastructure. A significant public health concern remains that up to 90\% of mental health apps lack evidence-based clinical value or suitable security policies, risking the introduction of unvalidated interventions into vulnerable populations. However, clinicians are only able to assess digital therapies if developers offer clear evidence of the product’s development and efficacy testing \cite{bauer2017ethical}.

The shift toward preventive mental health systems is a primary promise of AI, moving from reactive treatment to proactive monitoring \cite{ni2025scoping}. By analyzing large datasets -including digital activity, gaming behavior, and financial spending- AI systems can identify early risk indicators and predict crises or hospitalizations \cite{el2024ai}. This is particularly relevant for youth gambling and problematic smartphone use, where early warning systems can detect ``co-addictive human-machine configurations" -patterns where the addiction arises from the dynamic interaction between the user and the platform’s structural characteristics \cite{helm2024co, efthymiou2023transforming}.

Effective policy integration is the ``urgent need of the hour” to bridge the gap between technological availability and public health impact \cite{singh2021prevalence}. Digital interventions achieve meaningful effects only when embedded within coordinated regulatory, service, and financing infrastructures \cite{graham2006lost}. Coordinated action requires integrating prevention and harm-reduction strategies within national frameworks \cite{mcmahon2019effects, blank2021interventions}. Furthermore, ethical governance must ensure technologies do not amplify social inequalities or promote unsupervised self-treatment. Ultimately, while AI tools can reduce economic burdens and promote equitable healthcare, success depends on collaborative efforts to ensure that the rapid growth of digital media use remains aligned with societal well-being.

\newpage
\section{Challenges and Future Directions }\label{sec9}

\begin{figure*}[h!]
	\centering
	\includegraphics[width=0.5\textwidth]{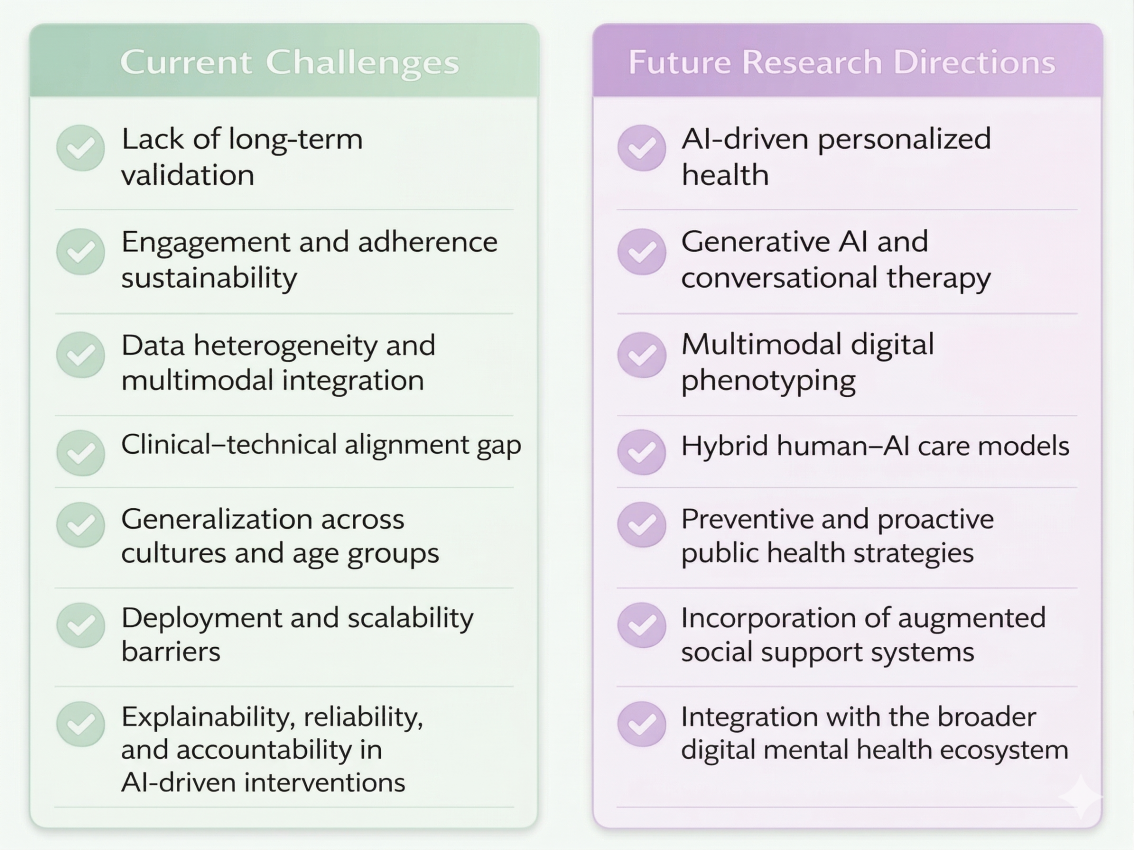}
	\caption{Current challenges and future research directions in digital behavioral addiction interventions}
	\label{fig:challenges}
\end{figure*}

Digital technologies have introduced new opportunities for the prevention and management of behavioral addictions. Mobile platforms, AI-driven tools, immersive technologies, and online intervention systems have expanded access to care and enabled continuous behavioral monitoring. However, despite these advances, several methodological, clinical, and implementation challenges remain. Addressing these limitations is necessary for translating experimental digital tools into reliable and scalable addiction care solutions. Figure~\ref{fig:challenges} provides a conceptual overview of this landscape by organizing the discussion around two interrelated dimensions: current challenges and future research directions. Specifically, the figure highlights that progress in the field depends not only on overcoming persistent limitations in validation, engagement, multimodal integration, clinical alignment, scalability, and AI accountability, but also on advancing more personalized, adaptive, and clinically integrated intervention models. Accordingly, this section summarizes the major challenges identified in the current literature and outlines key directions for future research.

\subsection{Challenges}
Although the literature on digital interventions for behavioral addictions has grown rapidly, the current evidence base remains uneven in terms of quality, consistency, and real-world applicability. A recent umbrella review showed that intervention effects for digital addiction are promising but still heterogeneous across addiction subtypes and treatment modalities, with several evidence classes remaining weak or inconclusive \cite{lu2025interventions}. Similarly, a recent systematic review of technology-based interventions for internet addictive behaviors emphasized the limited number of primary studies, substantial variability in intervention design, and insufficient reporting of software quality and interaction features \cite{penafiel2025technology}. Broader reviews in digital mental health also stress that implementation barriers, safety concerns, and limited service integration continue to restrict clinical translation \cite{lochner2025digital}. In addition, recent work on cultural and contextual adaptation has shown that digital interventions may fail to engage underserved populations when local social, linguistic, and economic conditions are not adequately addressed \cite{naderbagi2024cultural}. Overall, the field now faces a challenge that extends beyond demonstrating effectiveness: digital interventions must also be validated, sustained, and implemented reliably in real-world settings.

\subsubsection{Lack of long-term validation}

A major limitation of the current literature is the lack of long-term validation. Most digital interventions for behavioral addictions are evaluated over short follow-up periods, which limits the assessment of relapse, maintenance, and sustained behavior change. Recent evidence syntheses still emphasize that the field is characterized by heterogeneous designs, uneven evidence quality, and a limited number of robust primary trials with extended follow-up \cite{lu2025interventions,penafiel2025technology}. A notable exception is the OMPRIS line of research on internet use disorders. The original randomized clinical trial showed short-term efficacy of a therapist-guided webcam-based intervention, and a later 2-year follow-up study reported sustained symptom reduction among reassessed participants \cite{dieris2023efficacy,dieris20252}. However, such long-term outcome studies remain rare. More longitudinal trials, implementation studies, and real-world follow-up data are needed before digital interventions can be considered durable treatment options.

\subsubsection{Engagement \& adherence sustainability}

Sustaining engagement over time remains one of the main challenges in digital interventions for behavioral addictions. Recent work in digital mental health shows that engagement and adherence are still defined and measured inconsistently, and this makes comparison across studies difficult \cite{boucher2024engagement,lipschitz2023engagement}. A recent systematic review of technology-based interventions for internet addictive behaviors also emphasized substantial variation in user interaction design, software quality, and intervention structure, suggesting that adherence is shaped not only by clinical content but also by technical design \cite{penafiel2025technology}. Similar concerns appear in gambling-focused digital interventions. Recent evaluations of gambling JITAIs reported good acceptability, but they also highlighted user burden, fatigue, and the need for lower-intensity and more flexible designs to support sustained use \cite{dowling2025gamblingless,rodda2025acceptability}. More broadly, a recent umbrella review concluded that digital addiction interventions are effective overall, but their benefit varies across addiction subtype and modality, which further suggests that adherence cannot be treated as a uniform process across behavioral addictions \cite{lu2025interventions}. Taken together, these findings indicate that long-term adherence depends on a combination of clinical relevance, usability, burden, personalization, and contextual fit.

\subsubsection{Data heterogeneity \& multimodal integration}

Data heterogeneity and multimodal integration remain major challenges in digital interventions for behavioral addictions. Current systems increasingly combine EMA, smartphone usage logs, wearable sensor signals, and self-report measures. However, these data sources differ in format, sampling frequency, and reliability. A recent systematic review of technology-based interventions for internet addictive behaviors also showed substantial variation in intervention structure, software quality, and interaction design, which complicates comparison across studies \cite{penafiel2025technology}. Similar problems are reported in mobile sensing research more broadly. A recent scoping review of active and passive data collection identified persistent barriers related to user compliance, passive data consistency, and authorization across multimodal systems \cite{slade2025current}. In addition, a large real-world study based on data from 3{,}000 participants showed considerable variation in sensor data quality across Android and iOS devices, indicating that technical differences alone may confound behavioral prediction models \cite{halabi2024comparative}. In internet gaming disorder, recent biomarker reviews also suggest that multiple signal types are relevant, but integration across behavioral, physiological, and cognitive domains is still methodologically limited \cite{skok2024biomarkers}. Overall, these findings suggest that the main challenge lies not merely in collecting more data, but in integrating heterogeneous data sources into stable and clinically interpretable models.

\subsubsection{Clinical-technical alignment gap}

Many systems are designed primarily around technical capability, such as prediction accuracy, automation, or interface novelty, while their alignment with therapeutic workflows and clinical decision-making often remains limited. Reviews in digital mental health have emphasized that real-world adoption depends not only on technical performance, but also on usability, trust, regulation, and integration into routine care \cite{lochner2025digital,torous2025evolving}. A recent development guide for digital mental health interventions similarly noted that intervention content and technological features should be aligned from the outset, rather than treated as separate stages of development \cite{mertens2025did}. Similar concerns are also evident in participatory design research. A systematic review found that co-design with young people is still applied inconsistently, with varying levels of involvement and limited reporting quality \cite{mcgovern2025use}. In addiction care specifically, a recent qualitative study showed that even when digital tools are perceived positively, their implementation is still shaped by clinician concerns related to workflow burden, service fit, and operational feasibility \cite{getty2025addiction}. Overall, these findings suggest that the impact of digital addiction interventions will remain constrained unless technical design, clinical workflows, and service delivery are developed in closer alignment.

\subsubsection{Generalization across cultures and age groups}

The generalizability of current digital interventions for behavioral addictions remains limited, as much of the available evidence is derived from adolescents and young adults, particularly in studies focusing on internet gaming disorder and problematic internet use. A recent meta-analysis on psychological interventions for internet gaming disorder showed that treatment effects differed across domestic and international studies, suggesting that cultural context may influence intervention response \cite{wang2025comparative}. More broadly, a recent scoping review of digital mental health interventions for people aged 16--25 years found that most studies relied on convenience samples from universities and underrepresented marginalized groups, which restricts age-related and social generalizability \cite{potts2025digital}. Similar concerns are reported in the cultural adaptation literature. A recent narrative review emphasized that digital interventions are often transferred across settings without sufficient adaptation to local language, values, and service structures \cite{naderbagi2024cultural}. This concern is especially relevant in low- and middle-income settings. A 2025 scoping review showed that most digital mental health interventions were still developed in high-income countries and that fidelity, usability, and cultural adaptation were inconsistently addressed in LMIC contexts \cite{chau2025digital}. Overall, these findings suggest that digital addiction interventions cannot yet be considered widely generalizable without broader validation across cultures, age groups, and service settings.

\subsubsection{Deployment \& scalability barriers}
A major challenge lies in the gap between research success and real-world deployment, since promising outcomes in controlled studies do not automatically ensure scalable and sustainable implementation in routine practice. Recent reviews in digital mental health show that reimbursement, regulatory clarity, clinician adoption, workflow fit, and interoperability with existing services remain major barriers to scale-up \cite{berardi2024barriers,lochner2025digital,mohr2025digital}. This problem is also relevant for behavioral addictions. For example, therapist-supported online CBT for gambling disorder has been introduced in routine addiction care, but its implementation still depends on organizational capacity, structured support, and service-level integration \cite{molander2024introducing}. In practice, digital tools for gambling, gaming, and problematic digital media use will not scale sustainably unless they can be embedded into routine care pathways, supported by clear governance and feasible delivery models.

\subsubsection{Explainability, reliability, and accountability in AI-driven interventions}

Explainability, reliability, and accountability have become central concerns in AI-driven mental health interventions. In behavioral addictions, AI systems may influence risk detection, symptom monitoring, and intervention timing. For this reason, clinical decisions cannot rely only on predictive performance. Recent reviews in digital mental health emphasize that interpretability, fairness, and clinical relevance must be treated as core requirements rather than optional technical features \cite{torous2025evolving,ali2025artificial}. The explainability debate is still unresolved. A recent systematic review in healthcare showed no clear consensus on whether all AI-based decision-support systems should be fully explainable, but it also concluded that explainability requirements must be considered in relation to context, risk level, and clinical use \cite{freyer2024ethical}. Reliability is another major issue. Models that perform well in development datasets may not remain stable across settings, populations, or workflows. In a recent qualitative study from the UK, clinicians and healthcare staff highlighted accountability, transparency, and bias as major barriers to trust in AI-assisted decision-making, especially when responsibility for adverse outcomes remains unclear \cite{nouis2025evaluating}. Recent ethics-oriented work in computational psychiatry also argues that generic responsible-AI principles are not sufficient for mental health settings, where fluctuating autonomy, vulnerability, and clinician--patient relationships require more specific governance frameworks \cite{putica2025ethical,tavory2024regulating}. Taken together, these findings suggest that AI-driven interventions in behavioral addictions will require not only accurate models, but also transparent reasoning, stable real-world performance, and clearly defined accountability structures before they can be integrated into routine care.

\subsection{Future research directions}
Future work is likely to move toward more personalized, multimodal, and hybrid care systems that combine artificial intelligence with clinician oversight and social support structures. In this context, progress will depend on whether digital tools can achieve not only technical sophistication, but also clinical validity, implementation readiness, and long-term sustainability. The following research directions outline the main priorities for advancing the field.

\subsubsection{AI-driven personalized health}
AI-driven personalized health represents a transformative approach in behavioral addiction management, shifting care from generalized protocols toward individually tailored interventions \cite{mihai2025ai}. By leveraging AI and machine learning on rich multimodal data, digital systems can understand the unique behavioral patterns, risk factors, physiological responses, and psychological profiles of each user. This individualized understanding enables interventions that are context-sensitive, adaptive over time, and optimized for long-term recovery outcomes \cite{bhalsing2025ai}.

Central to AI-driven personalized health is the integration of continuous monitoring, predictive modeling, and adaptive decision-making. Data from self-reports, wearable sensors, app usage, social interactions, and natural language input are analyzed to generate dynamic health profiles. These profiles inform risk prediction models, personalized recommendation systems, and RL-based interventions, allowing the system to proactively suggest preventive or corrective actions. For example, an individual identified as at high risk for relapse during late-night social media use may receive a timely motivational prompt, access to a coping exercise, or a recommendation to engage with peer-support networks.

Moreover, personalization extends beyond behavioral recommendations to the optimization of intervention modalities, intensity, and timing. AI models can learn from historical responses to identify which strategies—such as cognitive-behavioral exercises, mindfulness practices, gamified tasks, or chatbot interactions—are most effective for a particular individual. Over time, these models continuously adapt, refining intervention policies based on observed outcomes, engagement patterns, and evolving behavioral states. This feedback-driven adaptation ensures that care remains relevant and effective as the user progresses through different stages of recovery \cite{vallabhaneni2024power}.

AI-driven personalized health also enables population-level insights without compromising individual specificity \cite{wang2022personalized}. Aggregated and anonymized data can reveal emerging risk patterns, common triggers, or effective intervention sequences, informing both digital system design and public health strategies \cite{edvardsson2025precision}. By combining individual and collective intelligence, AI facilitates scalable and precision-oriented addiction management that bridges clinical rigor with real-world applicability.

However, realizing AI-driven personalized health requires careful attention to data quality, model interpretability, ethical safeguards, and privacy protection. Clinicians must be able to understand and trust AI recommendations, while users need transparent explanations of how their data informs interventions. Federated learning, differential privacy, and explainable AI methods are increasingly employed to balance personalization, security, and accountability \cite{hu2020personalized}.

\subsubsection{Generative AI and conversational therapy}
Generative AI has emerged as a groundbreaking tool in the field of digital behavioral addiction management, particularly through its application in conversational therapy \cite{muralidhar2026human}. By leveraging large language models and advanced generative techniques, digital systems can simulate human-like dialogue, providing immediate, personalized, and context-aware therapeutic interactions \cite{tasnim2025context}. These capabilities allow for scalable, low-barrier access to psychological support, bridging gaps in traditional care and extending intervention reach beyond clinical settings.

Conversational therapy powered by generative AI goes beyond simple scripted chatbots. Modern generative models can interpret user input in real time, understand nuanced emotional cues, detect cognitive distortions, and tailor responses to the individual’s motivational state, risk level, and behavioral patterns \cite{babushkina2026disrupted}. For example, when a user expresses craving or stress, the system can generate empathetic responses, suggest coping strategies, or guide structured exercises such as cognitive restructuring, mindfulness, or behavioral activation. By continuously adapting dialogue content, tone, and pacing, generative AI can maintain engagement, reinforce adherence, and provide personalized behavioral guidance that aligns with each user’s recovery journey.

A key advantage of generative AI lies in its ability to integrate multimodal data into conversational therapy \cite{chu2025towards}. Textual input from journaling, voice transcriptions, or social media activity can inform real-time responses, while behavioral and physiological signals may modulate the timing and type of interventions. This enables a dynamic, context-sensitive therapeutic experience that mirrors human clinical reasoning but at scale. Additionally, generative AI systems can deliver proactive interventions, such as preemptively addressing relapse risks or motivational dips, by anticipating user needs based on historical patterns and predicted behavioral states \cite{tran2023multimodal}.

Despite these opportunities, deploying generative AI in addiction therapy presents critical challenges. Ensuring the accuracy, safety, and appropriateness of AI-generated guidance is paramount, as erroneous or misleading advice could have serious psychological consequences. Ethical considerations include maintaining user privacy, preventing over-reliance on AI, and mitigating biases encoded in training data that may affect different demographic groups unevenly. Regulatory compliance and alignment with evidence-based therapeutic frameworks are also essential to ensure clinical credibility and user trust \cite{bahadori2025evolving}.

Ongoing research is exploring hybrid models that combine generative AI with human oversight, allowing clinicians to supervise or intervene in high-risk situations \cite{holzinger2025human}. Such models can balance scalability with safety and maintain the therapeutic alliance while leveraging AI’s capacity for personalization and immediacy \cite{mesko2023imperative}. Moreover, explainable AI techniques are increasingly incorporated to provide users and clinicians with insights into why certain recommendations or conversational strategies are suggested, enhancing transparency and accountability.

\subsubsection{Multimodal digital phenotyping}
Multimodal digital phenotyping represents a cutting-edge approach for understanding and managing behavioral addictions by capturing a comprehensive, high-resolution picture of an individual’s behavioral, physiological, and contextual states \cite{moura2023digital}. Unlike traditional assessments that rely primarily on episodic clinical interviews or self-reports, digital phenotyping leverages continuous, real-world data streams collected from smartphones, wearable devices, social media, and other digital interactions \cite{aledavood2025multimodal}. Integrating multiple modalities allows for a more nuanced understanding of addiction-related behaviors, triggers, and risk factors.

In practice, multimodal digital phenotyping combines diverse data types including behavioral logs (e.g., app usage, gaming or browsing patterns), physiological signals (e.g., HRV, sleep quality, electrodermal activity), contextual information (e.g., geolocation, social environment, time of day), and linguistic traces from texts, chats, or voice inputs \cite{liu2025digital}. By fusing these heterogeneous streams, AI systems can identify latent behavioral patterns, detect early warning signs of relapse, and track subtle changes in mood, impulsivity, or engagement that may not be evident through self-report alone. This enables highly personalized risk assessment and intervention tailoring.

The analytical backbone of multimodal digital phenotyping often relies on machine learning and deep learning models capable of capturing complex temporal and cross-modal relationships \cite{barkat2025latent}. Sequential models, such as RNNs or transformers, can track behavioral trajectories over time, while attention-based mechanisms highlight which modalities or features are most informative at different moments \cite{ludwig2026trajectories}. Graph-based approaches may also be employed to model social interactions or dependencies between behavioral signals. The integration of these techniques allows for both predictive modeling and explanatory insights, helping clinicians understand not only what risk exists but also why certain behavioral patterns indicate vulnerability.

Despite its promise, multimodal digital phenotyping faces several technical and practical challenges \cite{simon2026digital}. Data heterogeneity, temporal misalignment, missing or noisy sensor readings, and privacy-sensitive information complicate model development and deployment. Ensuring generalizability across populations, cultural contexts, and device types requires careful validation and potentially domain adaptation. Moreover, maintaining user trust demands transparent handling of sensitive behavioral and physiological data, robust security measures, and clear communication of how data are used for assessment and intervention.

When implemented thoughtfully, multimodal digital phenotyping can transform behavioral addiction care by enabling proactive, context-aware, and personalized interventions \cite{alfalahi2023scoping}. It facilitates early detection of risk states, dynamic adaptation of treatment strategies, and continuous monitoring of recovery progress. By combining fine-grained behavioral insights with AI-driven analysis, this approach supports precision mental health, enhancing both the effectiveness and scalability of DTx for addiction management.

\subsubsection{Hybrid human-AI care models}
Hybrid human-AI care models represent an emerging paradigm in behavioral addiction management, combining the strengths of artificial intelligence with the expertise and empathy of human clinicians \cite{kumar2025distributing}. While AI systems excel at processing large-scale, high-frequency data, identifying subtle behavioral patterns, and delivering scalable, personalized interventions, human professionals provide contextual judgment, ethical oversight, and therapeutic rapport that machines cannot replicate \cite{nnubia2025artificial}. By integrating these complementary capabilities, hybrid models aim to optimize treatment effectiveness, engagement, and safety in digital addiction care.

In a typical hybrid framework, AI algorithms continuously monitor multimodal data streams such as app usage logs, physiological signals, conversational text, and contextual indicators to generate risk assessments, personalized recommendations, and timely intervention prompts. These AI-driven insights are then reviewed, validated, or adjusted by clinicians who can incorporate nuanced understanding of the individual’s history, comorbidities, cultural context, and psychosocial environment \cite{shin2025toward}. For example, an AI system may detect early signs of relapse and suggest a specific coping exercise, while a human clinician evaluates the appropriateness of escalation or adapts the strategy to the patient’s unique circumstances.

Hybrid models also enhance scalability while preserving safety. AI can handle routine monitoring, low-risk interventions, and adherence nudges across large user populations, reducing the burden on healthcare providers \cite{wang2025position}. Meanwhile, human intervention is reserved for high-risk or complex cases, such as severe relapses, co-occurring psychiatric conditions, or ethical dilemmas in therapy. This allocation of responsibilities allows healthcare systems to expand access to care without compromising clinical quality or personalized attention.

Another advantage of hybrid approaches is the iterative feedback loop between humans and AI. Clinician input can be used to fine-tune machine learning models, improving predictive accuracy, reducing bias, and enhancing interpretability \cite{tseng2025designing}. Conversely, AI can provide objective, data-driven insights that augment human decision-making, highlight overlooked behavioral trends, and suggest alternative intervention strategies that may not have been considered. Over time, this collaboration fosters adaptive, continuously learning systems that optimize both algorithmic performance and clinical outcomes \cite{fragiadakis2024evaluating}.

Challenges in implementing hybrid models include ensuring seamless integration of AI outputs into clinical workflows, maintaining trust and transparency for both clinicians and users, and establishing clear accountability for decisions made jointly by human and machine \cite{dai2023incorporating}. Privacy and ethical considerations are amplified, as sensitive behavioral and physiological data are shared between AI systems and human professionals. Regulatory compliance, standardized protocols, and explainable AI techniques are essential to address these concerns.

\subsubsection{Preventive \& proactive public health strategies}
Preventive and proactive public health strategies in the context of behavioral addiction leverage AI and digital technologies to shift the focus from reactive treatment toward early intervention, risk mitigation, and sustained population health \cite{hendriks2023proactive}. Rather than waiting for individuals to develop severe addiction symptoms, these strategies aim to identify at-risk populations, anticipate emerging trends, and implement timely measures to reduce incidence, severity, and societal burden. Digital systems enable continuous monitoring, data-driven risk stratification, and personalized outreach at both individual and community levels, making prevention scalable and evidence-based.

AI-driven analytics play a central role by detecting patterns and early warning signals across large populations. Multimodal data from app usage, wearable devices, social media activity, and environmental context can reveal early signs of maladaptive behaviors, stress responses, or social influences that predispose individuals to addiction. Public health authorities and digital platforms can use these insights to design targeted interventions, such as educational campaigns, micro-interventions delivered via mobile applications, or community-level awareness initiatives. This proactive approach allows resources to be allocated efficiently, focusing on high-risk groups before adverse outcomes manifest \cite{sorter2024addressing}.

At the individual level, personalized digital interventions can promote protective behaviors, encourage healthy routines, and provide real-time guidance to mitigate triggers \cite{deshpande202524}. For example, AI systems can nudge users toward alternative activities, suggest stress-reduction exercises during periods of high vulnerability, or connect them with peer-support networks. By continuously adapting to the user’s evolving behavioral and emotional state, these interventions enhance adherence and effectiveness, turning prevention into a dynamic, ongoing process rather than a one-time effort.

Proactive public health strategies also extend to policy and ecosystem-level actions. Aggregated and anonymized data can inform policymakers about population-level risk trends, emerging addictive behaviors, or the effectiveness of regulatory measures \cite{castonguay2024ai}. Insights from AI-driven monitoring can guide the design of school-based prevention programs, workplace wellness initiatives, or public awareness campaigns. Additionally, digital platforms can integrate harm-reduction features, such as self-limiting tools in gaming or gambling apps, to minimize exposure to high-risk behaviors across the community.

However, implementing preventive and proactive strategies requires careful consideration of ethical, privacy, and equity concerns. Data collection and monitoring must respect user consent and confidentiality, and interventions should avoid stigmatization or unintended consequences. Efforts must also address the digital divide to ensure that preventive measures reach diverse populations, including those with limited technological access.

\subsubsection{Incorporation of augmented social support systems}
Augmented social support systems represent a critical component of modern digital addiction management, enhancing traditional human support networks with AI-driven, technology-mediated interventions. Behavioral addictions are strongly influenced by social factors, including peer pressure, isolation, and community engagement \cite{tomei2025perceived}. By integrating digital platforms with real-time social support mechanisms, augmented systems provide both emotional and practical guidance, improving adherence, motivation, and long-term recovery outcomes.

These systems leverage AI to facilitate personalized connections with peers \cite{vctvrtevckova2024role}, mentors, or professional counselors based on shared experiences, behavioral profiles, and current risk states. For instance, digital platforms can suggest peer-support groups for individuals exhibiting early signs of relapse, connect users with mentors who have successfully managed similar addictions, or prioritize interactions with counselors trained to address specific behavioral patterns \cite{smit2023effectiveness}. AI algorithms can monitor engagement, detect changes in sentiment or social behavior, and recommend timely interventions, ensuring that support is delivered when it is most needed.

In addition to peer and professional interactions, augmented social support systems incorporate digital nudges, gamified reinforcement, and community-driven motivation \cite{rokiyah2024role}. Users may receive encouragement messages, recognition for adherence milestones, or invitations to participate in community challenges designed to reinforce positive behaviors. These mechanisms enhance the social reinforcement of healthy habits, creating an ecosystem where adaptive behaviors are continuously encouraged and maladaptive patterns are mitigated.

Context-aware personalization further amplifies the effectiveness of augmented social support \cite{li2026context}. AI models can identify when users are likely to disengage or experience stress, delivering timely micro-interventions that combine social, motivational, and therapeutic components. For example, an individual experiencing a high-risk craving period might be prompted to reach out to a peer-support contact, join a live discussion session, or access a counselor via chat or video call. This proactive and context-sensitive integration of social support complements traditional therapeutic interventions and reinforces resilience.

Despite its potential, incorporating augmented social support systems requires careful attention to privacy, safety, and inclusivity. Sensitive behavioral and emotional data must be securely managed, while interactions must be moderated to prevent harm, misinformation, or stigmatization. Ensuring equitable access across diverse populations and avoiding digital exclusion are also essential for system effectiveness and fairness.

\subsubsection{Integration with broader digital mental health ecosystem}
Integration with the broader digital mental health ecosystem is essential for maximizing the impact, scalability, and continuity of AI-driven addiction management interventions \cite{lim2024integrated}. Behavioral addictions rarely occur in isolation; they often coexist with other mental health challenges, such as depression, anxiety, or stress-related disorders. Seamlessly connecting digital addiction tools with existing mental health platforms, electronic health records (EHRs), telepsychiatry services, and public health initiatives enables comprehensive, coordinated, and person-centered care.

Effective integration allows digital addiction interventions to exchange data and insights with complementary systems, enhancing both personalization and clinical oversight. For example, risk assessments, behavioral logs, or intervention outcomes generated by an addiction management platform can inform broader mental health treatment plans, alert clinicians to emerging comorbid conditions, or guide adaptive therapy sequencing. Similarly, mental health monitoring apps can feed relevant contextual data—such as mood fluctuations, sleep quality, or stress levels—into addiction management systems, supporting more accurate predictions and timely interventions. This interconnected approach ensures that AI-driven care considers the whole person rather than addressing behavioral addiction in isolation.

Technical integration requires adherence to standardized protocols, interoperability frameworks, and privacy regulations. Platforms must support secure data sharing while maintaining patient confidentiality and compliance with healthcare standards such as HIPAA or GDPR. Interoperable APIs, standardized data formats, and semantic mapping of behavioral and clinical indicators facilitate smooth communication across systems, enabling real-time analytics and coordinated care workflows \cite{bond2023digital}. In addition, integration with telehealth services allows for seamless escalation to human clinicians when AI systems detect high-risk events or complex clinical needs.

Beyond clinical coordination, ecosystem integration fosters population-level insights and preventive public health strategies. Aggregated, anonymized data from multiple platforms can reveal emerging behavioral trends, high-risk demographics, or effective intervention patterns, informing policy decisions, community-based programs, and large-scale digital therapeutic initiatives. By aligning digital addiction management tools with the broader mental health infrastructure, health systems can improve resource allocation, reduce fragmentation of care, and enhance both individual and societal outcomes.

However, challenges remain in ensuring equitable access, maintaining data quality across heterogeneous systems, and avoiding technological silos. Ethical and legal considerations, such as informed consent, data ownership, and transparency in algorithmic decision-making, are crucial to preserve trust among users, clinicians, and policymakers. Continuous monitoring, evaluation, and adaptive governance are required to ensure that integration delivers both technical interoperability and clinically meaningful outcomes.

\newpage
\section{Conclusion}\label{sec10}
Behavioral addictions represent a complex and rapidly evolving challenge that requires interdisciplinary, scalable, and adaptive solutions. This book has provided a comprehensive overview of how digital technologies and artificial intelligence are reshaping the landscape of addiction management, spanning clinical foundations, technological enablers, data sources, analytical methods, and broader societal implications. By bridging insights from psychology, neuroscience, and computer science, the book highlights the transition from traditional, reactive treatment models toward proactive, data-driven, and personalized care paradigms.

The integration of digital technologies such as wearables, mobile platforms, AI-powered therapeutics, conversational agents, and telehealth systems enables continuous monitoring, real-time intervention, and enhanced accessibility. These systems support a wide range of functions, including early risk detection, relapse prediction, personalized intervention delivery, and long-term recovery management. At the core of these advancements lie machine learning and AI techniques, which leverage multimodal data to generate actionable insights and optimize intervention strategies. However, challenges related to data heterogeneity, scarcity, bias, and generalization continue to limit the robustness and transferability of current models.

The book also emphasizes the importance of explainability, ethical safeguards, and regulatory compliance in ensuring trustworthy and responsible deployment of AI-driven systems. Issues such as privacy, fairness, digital divide, and user autonomy must be carefully addressed to avoid unintended consequences and to foster widespread adoption. Furthermore, economic and ecosystem-level considerations including interoperability with healthcare systems and sustainable business models play a crucial role in translating research innovations into real-world impact.

Looking forward, emerging directions such as AI-driven personalized health, generative AI-based conversational therapy, multimodal digital phenotyping, and hybrid human-AI care models offer significant potential to enhance effectiveness and scalability. Preventive and proactive public health strategies, along with the integration of augmented social support systems and broader digital mental health ecosystems, further extend the scope of impact beyond individual treatment.

In conclusion, digital technologies and AI hold transformative potential for behavioral addiction management. Realizing this potential requires not only technical innovation but also interdisciplinary collaboration, rigorous clinical validation, and ethical responsibility. By aligning technological capabilities with clinical needs and societal values, future systems can enable more personalized, accessible, and sustainable approaches to addiction care.

\newpage
\section*{Acknowledgments}
This study was supported by The Scientific and Technological Research Council of Turkey (TUBITAK) under Project No. 125K046, and by the Kocaeli University Scientific Research Projects Coordination Unit (BAP) under Grant No. FBA-2026-5065.

\section*{Declaration of generative AI-assisted technologies in the writing process}
During the preparation of this work the authors used ChatGPT to improve language and readability. After using this tool/service, the authors reviewed and edited the content as needed and take(s) full responsibility for the content of the publication. The authors also used GenAI tools to assist in the creation and refinement of figures in this manuscript.

\section*{Competing Interests}
The author declares that he has no known competing financial interests or personal relationships that could have appeared to influence the work reported in this book.

\section*{Ethical approval}
Not applicable

\section*{Consent to participate}
Not applicable

\section*{Consent to publication}
Not applicable

\section*{Availability of data and material}
Not applicable

\section*{Authors' contributions}
S.E. and S.S. contributed to the conceptualization and design of the study. M.H.B.U.  and S.E. supervised the overall research process and contributed to the structuring and critical revision of the manuscript. S.E., S.S., M.H.B.U., and B.G. conducted the literature review and drafted the initial manuscript. B.G., Ö.K., and N.S. contributed to the clinical and psychiatric content and interpretation. H.A. contributed to the social and behavioral aspects of the study. İ.E.A. contributed to clinical insights and real-world applicability. All authors contributed to the writing, review, and editing of the manuscript. All authors read and approved the final version of the manuscript.
\newpage

\bibliographystyle{unsrtnat}  
\bibliography{sn-bibliography}

\clearpage
\pagenumbering{gobble}
\kapaksayfasi{Figures/arka_kapak.png}
\end{document}